\documentclass[a4paper,11pt]{article}
\usepackage{dsfont}
\usepackage{amssymb}
\usepackage{latexsym}
\usepackage{amsmath}
\usepackage{color}
\usepackage{comment}
\usepackage{amsthm}
\usepackage[dvips]{graphicx}
\usepackage{layout} 
\usepackage{ulem}
\usepackage{enumerate}
\usepackage{bm}
\usepackage{bbm} 
\usepackage[bbgreekl]{mathbbol} 
\usepackage{authblk}
\usepackage{setspace} 
\newif\ifrs
\rstrue
\ifrs \usepackage{mathrsfs} \fi  
\usepackage[top=30truemm,bottom=30truemm,left=25truemm,right=25truemm]{geometry}
\usepackage[dvips]{graphicx}
\usepackage{here}
\usepackage[hang,small,bf]{caption}
\usepackage[subrefformat=parens]{subcaption}
\newcommand{\rev}{\color{black}}
\newif\ifcol
\coltrue 

\newtheorem{theorem}{Theorem}[section]

\newtheorem{lemma}[theorem]{Lemma}

\newtheorem{proposition}[theorem]{Proposition}

\newtheorem{remark}[theorem]{Remark}
\newtheorem{example}[theorem]{Example}
\numberwithin{equation}{section}
\newtheorem{theorem*}{Theorem}
\newtheorem{ass*}[theorem*]{Assumption}
\newtheorem{note*}[theorem*]{Note}
\newtheorem{lemma*}[theorem*]{Lemma}
\newtheorem{definition*}[theorem*]{Definition}
\newtheorem{proposition*}[theorem*]{Proposition}
\newtheorem{corollary*}[theorem*]{Corollary}
\newtheorem{remark*}[theorem*]{Remark}
\newtheorem{example*}[theorem*]{Example}
\numberwithin{equation}{section}
\newif\ifcol
\colfalse
\ifcol
\newcommand{\colorm}{\color{magenta}}
\newcommand{\colorr}{\color[rgb]{0.8,0,0}}
\newcommand{\colorg}{\color[rgb]{0,0.5,0}}
\newcommand{\colorb}{\color[rgb]{0,0,0.8}}

\newcommand{\colorn}{\color[rgb]{1,1,1}}

\newcommand{\coloroy}{\color[rgb]{1,0.95,0}}

\else
\newcommand{\colorm}{\color{black}}
\newcommand{\colorb}{\color{black}}
\newcommand{\colorr}{\color{black}}
\newcommand{\colorg}{\color{black}}
\newcommand{\colorn}{\color{black}}
\newcommand{\coloroy}{\color{black}}
\fi
\newif\ifcol
\colfalse 
\ifcol
\newcommand{\colred}{\color[rgb]{0.8,0,0}}
\newcommand{\colblue}{\color[rgb]{0,0,0.8}}
\else
\newcommand{\colred}{\color{black}}
\newcommand{\colblue}{\color{black}}
\fi
\newif\ifcol
\colfalse 
\ifcol
\newcommand{\tred}{\color[rgb]{0.8,0,0}}

\else
\newcommand{\tred}{\color{black}}
\fi
\newif\ifcol
\colfalse 
\ifcol
\newcommand{\fred}{\color[rgb]{0.8,0,0}}

\else
\newcommand{\fred}{\color{black}}
\fi

\excludecomment{en-text}
\includecomment{jp-text}
\includecomment{comment}
\def\tX{\widetilde{X}}
\def\tY{\widetilde{Y}}
\def\infm{{\infty\text{--}}}
\def\inftym{\infm}
\def\koko{{\coloroy{koko}}}
\def\bd{\begin{description}}
\def\ed{\end{description}}

\def\D2{\bbD_{2,\infty-}}

\def\tj{{t_j}}
\def\tjm{{t_{j-1}}}

\def\D{{\bf D}}

\def\F{{\bf F}}

\def\cala{{\cal A}}

\def\cald{{\cal D}}

\def\calf{{\cal F}}

\def\calj{{\cal J}}
\def\calk{{\cal K}}

\def\calm{{\cal M}}

\def\cals{{\cal S}}

\def\ds{\displaystyle}
\def\yeq{\>=\>}
\def\yleq{\>\leq\>}
\def\ygeq{\>\geq\>}

\def\sfk{{\sf k}}
\def\sfm{{\sf m}}

\def\sfd{{\sf d}}
\def\sfp{{\sf p}}
\def\sfr{{\sf r}}

\def\simleq{\ \raisebox{-.7ex}{$\stackrel{{\textstyle <}}{\sim}$}\ }

\def\ep{\epsilon}
\def\half{\frac{1}{2}}

\def\cadlag{c\`adl\`ag\ }
\def\up{\uparrow}
\def\down{\downarrow}

\def\y{\vspace*{3mm}\\}
\def\halflineskip{\vspace*{3mm}}
\def\nn{\nonumber}
\def\be{\begin{equation}}
\def\ee{\end{equation}}
\def\bea{\begin{eqnarray}}
\def\eea{\end{eqnarray}}
\def\beas{\begin{eqnarray*}}
\def\eeas{\end{eqnarray*}}
\def\bi{\begin{itemize}}
\def\ei{\end{itemize}}
\def\im{\item}
\def\bd{\begin{description}}
\def\ed{\end{description}}
\def\l{\left}
\def\r{\right}
\def\bars{\bar{S}}

\newcommand{\bbD}{{\mathbb D}}
\newcommand{\bbE}{{\mathbb E}}

\newcommand{\bbH}{{\mathbb H}}

\newcommand{\bbM}{{\mathbb M}}
\newcommand{\bbN}{{\mathbb N}}

\newcommand{\bbR}{{\mathbb R}}

\newcommand{\bbU}{{\mathbb U}}
\newcommand{\bbV}{{\mathbb V}}

\newcommand{\bbY}{{\mathbb Y}}
\newcommand{\bbZ}{{\mathbb Z}}

\begin{document}

\title{
Global 
jump filters 
and quasi-likelihood analysis 
for 
volatility 
\footnote{
This work was in part supported by 
CREST JPMJCR14D7 Japan Science and Technology Agency; 
Japan Society for the Promotion of Science Grants-in-Aid for Scientific Research 
No. 17H01702 (Scientific Research) 
and by a Cooperative Research Program of the Institute of Statistical Mathematics. 
%
}
}
\author[1]{Haruhiko Inatsugu}
\author[1,2]{Nakahiro Yoshida}
\affil[1]{Graduate School of Mathematical Sciences, University of Tokyo
\footnote{Graduate School of Mathematical Sciences, University of Tokyo: 3-8-1 Komaba, Meguro-ku, Tokyo 153-8914, Japan. e-mail: nakahiro@ms.u-tokyo.ac.jp}
        }
\affil[2]{Japan Science and Technology Agency CREST
        }
\date{June 29, 2018 \y
Revised version: September 17, 2020\\
Revised version 2: February 14, 2021
\footnote{{\fred The theoretical part has been extended by relaxing Condition $[F2]$.}}
}
\maketitle
\ \\
{\it Summary.} 
{\colorb We propose a new estimation scheme for 
{\colorg estimation of the}
volatility parameters of {\colorg a semimartingale} with jumps based on a
jump-detection filter. 
Our filter uses all of data to analyze the relative size of increments and to discriminate jumps more precisely. 
{\colorg 
We construct quasi-maximum likelihood estimators and quasi-Bayesian estimators, 
and show limit theorems for them including 
$L^p$-estimates of the error} 
and asymptotic mixed normality based on the framework of the quasi-likelihood analysis.
{\colorg 
The global jump filters do not need a restrictive condition for the distribution of 
the small jumps.}
By numerical simulation we show that our ``global" method obtains better estimates of the volatility parameter than the previous ``local" methods.}


\section{Introduction}

We consider an $\sfm$-dimensional semimartingale $Y=(Y_t)_{t\in[0,T]}$ 
admitting a decomposition 
\bea\label{300615-1} 
Y_t &=& Y_0+\int_0^tb_sds+\int_0^t\sigma(X_s,\theta)dw_s+J_t,\quad t\in[0,T]
\eea
on a stochastic basis $(\Omega,\calf,{\bf F},P)$ with a filtration 
${\bf F}=(\calf_t)_{t\in[0,T]}$. 
Here $b=(b_t)_{t\in[0,T]}$ is an $\sfm$-dimensional 
{\tred\cadlag adapted process,} 
$X=(X_t)_{t\in[0,T]}$ is {\tred a} $\sfd$-dimensional \cadlag adapted process, 
$w=(w_t)_{t\in[0,T]}$ is an $\sfr$-dimensional standard ${\bf F}$-Wiener process, 
$\theta$ is a parameter in the closure of an open set $\Theta$ in $\bbR^\sfp$, 
and $\sigma:\bbR^\sfd\times\bar{\Theta}\to\bbR^\sfm\otimes\bbR^\sfr$ is a continuous function. 
$J=(J_t)_{t\in[0,T]}$ is the jump part of $Y$, i.e., 
$J_t=\sum_{s\in[0,t]}\Delta Y_s$, where $\Delta Y_s=Y_s-Y_{s-}$ and 
$\Delta Y_0=0$. 
We assume $J_0=0$ and 
{\rev$\sum_{t\in[0,T]}1_{\{\Delta J_t\not=0\}}<\infty$ {\colred a.s.}}
{\colorg 
Model (\ref{300615-1}) is a stochastic regression model, but for example, it 
can express a diffusion type process 
with jumps $\Delta J^X$ 
contaminated by exogenous jump noise $J^Y$:
\beas \left\{\begin{array}{ccl}
Y_t &=& X_t+J^Y_t,\y
X_t &=& X_0+\int_0^tb_sds+\int_0^t\sigma(X_s,\theta)dw_s+J^X_t, 
\end{array}\right.
\eeas
with $J=J^X+J^Y$, and as a special case, a jump-diffusion process.
}
We want to estimate the true value $\theta^*\in\Theta$ of $\theta$ based on the data 
$(X_\tj,Y_\tj)_{j=0,1,...,n}$, where $t_j=t^n_j=jT/n$. 
Asymptotic properties of estimators will be discussed when $n\to\infty$. 
That is, the observations are {\colorb high} frequency data. 
The data of the processes $b$ and $J$ are not available since 
they are not directly observed. 

{\colorm 
Today a substantial amount of literature is available on parametric estimation of 
the diffusion parameter $\theta$ 
of diffusion type processes with/without jumps. 
In the ergodic diffusion case of $J=0$ and 
$T\to\infty$, the drift coefficient is parameterized as well as 
the diffusion coefficient. 
Certain asymptotic properties of estimators 
are found in 
Prakasa Rao 
\cite{PrakasaRao1983,prakasa1988statistical}. 
The 
joint asymptotic normality of estimators was given in Yoshida 
\cite{Yoshida1992b} 
and later generalized in Kessler 
\cite{Kessler1997}. 
The quasi-likelihood analysis (QLA, Yoshida \cite{yoshida2011}) ensures 
not only limit theorems but also moment convergence of 
the QLA estimators, i.e., 
the quasi-maximum likelihood estimator ({\rev QMLE}) and the quasi-Bayesian estimator (QBE). 
\begin{en-text}
Relaxation of the balance condition $nh_n^p\to0$ as $n\to\infty$ 
is practically important, $h_n$ being the sampling step size. 
This problem 
has been pursued by 
these papers and also, 
within the framework of the theory of quasi-likelihood analysis, 
by 
Uchida and Yoshida 
\cite{UchidaYoshida2012Adaptive,uchida2014adaptive} for adaptive estimators and by 
Kamatani and Uchida \cite{KamataniUchida2014} for hybrid multi-step estimators. 
\end{en-text}
The adaptive estimators 
(Uchida and Yoshida 
\cite{UchidaYoshida2012Adaptive,uchida2014adaptive})
and the hybrid multi-step estimators (Kamatani and Uchida \cite{KamataniUchida2014}) 
are of practical importance from computational aspects. 
Statistics becomes non-ergodic under a finite time horizon $T<\infty$. 
Dohnal 
\cite{Dohnal1987} discussed estimation of the diffusion parameter 
based on high frequency data. 
Stable convergence of the quasi-maximum likelihood estimator 
was given by Genon-Catalot and Jacod 
\cite{Genon-CatalotJacod1993}. 
Uchida and Yoshida 
{\tred\cite{uchida2013quasi}} showed
stable convergence of the quasi-Bayesian estimator 
and moment convergence of the QLA estimators. 
The methods of the QLA were essential there and 
will be applied in this article. 
{\rev The} non-synchronous case is addressed by  
Ogihara and Yoshida 
\cite{ogihara2014quasi} within QLA. 
As for inference for jump-diffusion processes, 
under ergodicity, Ogihara and Yoshida 
\cite{OgiharaYoshida2011} 
showed asymptotic normality of the QLA estimators and
moment convergence of their error. 
They used a type of optimal jump-filtered quasi-likelihood function 
in Shimizu and Yoshida 
\cite{ShimizuYoshida2006}. 

The filter in the quasi-likelihood functions of 
Shimizu and Yoshida 
\cite{ShimizuYoshida2006} 
is based on the magnitude of the absolute value of the increment: $\{ |\Delta_i Y| > Ch_n^{\rho} \}$, where 
$\Delta_i Y = Y_{t_i} - Y_{t_{i-1}}$, $\rho \in [0,1/2)$ and $C>0$. 
If an increment is sufficiently large relative to the threshold, 
then 
it is classified as a jump. 
If, on the other hand, the size of the increment is ``moderate'', it is regarded as {\rev coming} from the continuous part. 
Then the parameters in the continuous and jump parts can optimally be estimated by respective data sets obtained by classification of increments. 
{\colorg 
This threshold is natural and in fact, historically,  the idea goes back to studies of limit theorems for semimartingales, even further back to L\'evy processes. 
}

However, this jump detection filter has a caveat. 
Though the efficiency of the estimators has been established theoretically, 
it is known that their real performance strongly depends on a choice of 
tuning parameters; see, e.g., Shimizu 
\cite{Shimizu2008}, Iacus and Yoshida 
\cite{iacus2017simulation}. 
The filter is each time based on only one increment of the data. 
In this sense, this filter can be regarded as a {\it local} method. 
This localism would cause misclassification of increments 
in practice, even though it should not occur mathematically by the large deviation principle 
in the limit, 
and estimated values' instability and strong dependency on the tuning parameters. 
To overcome these problems, we introduce a {\it global} filtering method, which we call 
{\tred the} $\alpha$-{\it threshold method}. 
It uses all of the data to more accurately 
detect increments having jumps, based on the order statistics 
associated with all increments. 
{\colorg 
Another advantage of the global filter is that 
it does not need any restrictive condition {\rev on} the distribution of small jumps.}
This paper provides efficient parametric estimators 
for the model (\ref{300615-1}) under a finite time horizon $T<\infty$ 
by using the $\alpha$-threshold method, while applications of this method 
to the realized volatility and other related problems are straightforward. 
Additionally, it should be remarked that though the $\alpha$-threshold method involves 
the tuning parameter $\alpha$ to determine a selection rule for increments,  
it is robust against the choice of $\alpha$ as we will see later. 

The organization of this paper is as follows. 
In Section \ref{300615-11}, we introduce the $\alpha$-quasi-log likelihood function $\bbH_n(\theta;\alpha)$, 
that is a truncated version of the quasi-log likelihood function made from local Gaussian approximation, 
based on the global filter for the tuning parameter $\alpha$. 
The $\alpha$-quasi-maximum likelihood estimator ($\alpha$-QMLE) $\hat{\theta}^{M,\alpha}_n$ is defined with respect to 
$\bbH_n(\theta;\alpha)$. 
Since the truncation is formulated by the order statistics of the increments, 
this filter destroys adaptivity and martingale structure. 
However, the global filtering lemmas in Section \ref{300615-12} enable us to recover these properties. 
Section \ref{300615-13} gives a rate of convergence of the $\alpha$-QMLE $\hat{\theta}^{M,\alpha}_n$ 
in $L^p$ sense. In order to prove it, with the help of the QLA theory (Yoshida \cite{yoshida2011}), 
the so-called polynomial type large deviation inequality is derived in Theorem \ref{300217-10} for  
an annealed version of the quasi-log likelihood $\bbH_n^\beta(\theta;\alpha)$ of (\ref{300615-14}), 
where $\beta$ is {\tred the} annealing index. 
Moreover, the $(\alpha,\beta)$-quasi-Bayesian estimator ($(\alpha,\beta)$-QBE) 
$\hat{\theta}^{B,\alpha,\beta}_n$ 
can be defined as the Bayesian estimator 
with respect to $\bbH_n^\beta(\theta;\alpha)$ as (\ref{300615-5}). 
Then the polynomial type large deviation inequality makes it possible to prove 
$L^p$-boundedness of the error of the $(\alpha,\beta)$-QBE $\hat{\theta}^{B,\alpha,\beta}_n$ (Proposition \ref{300226-1}). 
The $\alpha$-QMLE and $(\alpha,\beta)$-QBE do not attain the optimal rate of convergence when the parameter $\alpha$ is fixed 
though the fixed $\alpha$-method surely removes jumps as a matter of fact. 
In Section \ref{300615-3}, we introduce a quasi-likelihood function $\bbH_n(\theta)$ depending on a moving level $\alpha_n$. 
The random field $\bbH_n(\theta)$ is more aggressive than $\bbH_n(\theta;\alpha)$ with a fixed $\alpha$. 
Then a polynomial type large deviation inequality is obtained in Theorem \ref{300224-10} but 
the scaling factor is $n^{-1/2}$ in this case so that we can prove $\sqrt{n}$-consistency in $L^p$ sense 
for both QMLE $\hat{\theta}^{M,\alpha_n}_n$ and QBE $\hat{\theta}^{B,\alpha_n}_n$ 
associated with the random field $\bbH_n(\theta)$ (Proposition \ref{300615-16}). 
Stable convergence of these estimators and moment convergence are validated by Theorem \ref{300430-8}. 
The moving threshold method attains the optimal rate of convergence in contrast to the fixed-$\alpha$ method. 
However, 
the theory requires the sequence $\alpha_n$ should keep a certain balance: 
too large $\alpha_n$ causes deficiency and too small $\alpha_n$ may fail to filter out jumps. 
To balance {\colorg efficiency of estimation and precision in filtering}
by taking advantage of the stability of the fixed-$\alpha$ scheme, 
in Section \ref{300430-10}, 
we construct a one-step estimator $\check{\theta}^{M,\alpha}_n$ 
for a fixed $\alpha$ 
{\colorg and the aggressive $\bbH_n(\theta)$}
with the $\alpha$-QMLE $\hat{\theta}^{M,\alpha}_n$ as the initial estimator. 
{\colorg Similarly, the one-step estimator $\check{\theta}^{B,\alpha,\beta}_n$ is constructed 
for fixed $(\alpha,\beta)$ and $\bbH_n(\theta)$ 
with the $(\alpha,\beta)$-quasi-Bayesian estimator $\hat{\theta}^{B,\alpha,\beta}_n$ for the initial estimator. }
By combining the results in Sections \ref{300615-2} and \ref{300615-3}, 
we show that these estimators enjoy  
the same stable convergence and moment convergence as 
QMLE $\hat{\theta}^{M,\alpha_n}_n$ and QBE $\hat{\theta}^{B,\alpha_n}_n$. 
It turns out {\colorg in Section \ref{300615-6}} 
that the so-constructed estimators are accurate and quite stable against $\alpha$, {\colorg in practice}. 
In Section \ref{300228-1}, we relax the conditions for stable convergence 
by a localization argument. 
Section \ref{300615-6} 
presents some simulation results and shows that the global filter can detect jumps more precisely than the local threshold {\tred methods}.  
\begin{en-text}
In Section \ref{300615-2}, we construct the quasi-maximum likelihood estimator (QMLE) and quasi-Bayesian estimator 
based on the global filter when the tuning parameter $\alpha$ is given. 
Consistency is shown by the quasi likelihood analysis, following Yoshida 
\cite{yoshida2011}. 
We see that the estimator does not attain the optimal rate of convergence when the parameter $\alpha$ is fixed 
though the fixed $\alpha$-method surely removes jumps as a matter of fact. 
Section \ref{300615-3} discusses certain estimators with $\alpha$ varying according to the sample size.  
The moving threshold method attains the optimal rate of convergence in contrast to the fixed-$\alpha$ case. 
Section \ref{300430-10} presents one-step estimators 
by taking advantage of the stability of the fixed-$\alpha$ estimator 
and proves their efficiency by combining the results of Sections \ref{300615-2} and 
\ref{300615-3}. 
It turns out that the so-constructed estimators are accurate and quite stable 
against $\alpha$. 
In Section \ref{300228-1}, we relax the conditions for stable convergence 
by a localization argument. 
Finally, Section \ref{300615-6}  
presents some simulation results of the quasi QMLE and show that the global filter can detect jumps more precisely than the local threshold method 
and that the estimator is robust against the choice of the tuning parameter. 
\end{en-text}
}

\section{Global filter: $\alpha$-threshold method}\label{300615-2}
\subsection{Model structure}
{\colorm 
We will work with the model (\ref{300615-1}). 
To structure the model suitably, we begin with an example. 
}
\begin{example}\rm 
Consider a two-dimensional stochastic differential equation 
partly having jumps: 
\beas\left\{\begin{array}{ccl}
d\xi_t &=& b_t^\xi dt+\sigma^\xi(\xi_t,\eta_t,\zeta_t,\theta)dw^{\xi}_t+dJ^{\xi}_t\y
d\eta_t &=& b_t^\eta dt+\sigma^\eta(\xi_t,\eta_t,\zeta_t,\theta)dw^{\eta}_t.
\end{array}\right.
\eeas
We can set $Y=(\xi,\eta)$, $X=(\xi,\eta,\zeta)$ and $J=(J^{\xi},0)$. 
No jump filter is necessary for the component $\eta$. 
\end{example}
This example suggests that different treatments should be given component-wise. 
{\colorb
We assume that 
\beas
\sigma = \text{diag}[\sigma^{({\rev 1})}(x, \theta), \ldots, \sigma^{(\sfk)}(x, \theta)]
\eeas
for some $\sfm_k\times\sfm_k$ {\tred nonnegative symmetric} matrices $\sigma^{(k)}(x,\theta)$, $k=1,...,\sfk$, and we 
further assume that $w=(w^{(k)})_{k=1,...,\sfk}$ with $\sfr={\tred \sum_{k=1}^\sfm } \sfm_k = \sfm$. 
Let $S=\sigma^{\otimes2}=\sigma\sigma^\star$.  
Then $S(x,\theta)$ has the form of }
\beas 
S(x,\theta)=\text{diag}\big[S^{(1)}(x,\theta),...,S^{(\sfk)}(x,\theta)\big]
\eeas
for $\sfm_k\times\sfm_k$ matrices $S^{(k)}(x,\theta)
{\colorg = \sigma^{(k)}(\sigma^{(k)})^\star(x,\theta)}
$, $k=1,...,\sfk$. 
According to the blocks of $S$, we write 
\beas
Y_t\yeq
\begin{bmatrix}
Y^{(1)}_t  \\
\vdots\\
Y^{(\sfk)}_t
\end{bmatrix}, \qquad 
b_t\yeq
\begin{bmatrix}
b^{(1)}_t  \\
\vdots\\
b^{(\sfk)}_t
\end{bmatrix}, \qquad
w_t\yeq
\begin{bmatrix}
w^{(1)}_t  \\
\vdots\\
w^{(\sfk)}_t
\end{bmatrix}, \qquad
J_t\yeq
\begin{bmatrix}
J^{(1)}_t  \\
\vdots\\
J^{(\sfk)}_t
\end{bmatrix}.
\eeas

Let $N^X_t=\sum_{s\leq t}1_{\{\Delta X_s\not=0\}}$. 
{\tred We} will pose a condition that $N^X_T<\infty$ a.s.
The jump part $J^X$ of $X$ is defined by $J^X_t=\sum_{s\leq t}\Delta X_s$.

\subsection{Quasi likelihood function by order statistics}\label{300615-11}
In this section, we will give a filter that removes $\Delta J$. 
\cite{ShimizuYoshida2006} and 
\cite{OgiharaYoshida2011} 
used certain jump detecting filters that cut large increments $\Delta_jY$ by a threshold 
comparable to diffusion increments. It is a {\it local} filter 
because the classification is done for each increment 
{\tred without using other increments}. 
Contrarily, in this paper, we propose a {\it global} filter that removes 
increments $\Delta_jY$ when $|\Delta_jY|$ is in an upper class among 
all data $\{|\Delta_iY|\}_{i=1,...,n}$.

We prepare statistics $\bars^{(k)}_{n,j-1}$ ($k=1,...,\sfk$; $j=1,...,n$; $n\in\bbN$) 
such that each $\bars^{(k)}_{n,j-1}$ is an 
initial estimator of $S^{(k)}(X_{t_{j-1}},\theta^*)$ up to a scaling constant, that is, 
there exists a (possibly unknown) positive constant $c^{(k)}$ 
such that every $S^{(k)}(X_{t_{j-1}},\theta^*)$ is approximated by $c^{(k)}\bars^{(k)}_{n,j-1}$, 
as precisely stated later. 
We do not assume that $\bars^{(k)}_{n,j-1}$ is $\calf_\tjm$-measurable. 

\begin{example}\label{201902031722}\rm Let $K$ be a positive integer. 
Let $(\bar{i}_n)$ be a diverging sequence of positive integers, e.g., 
$\bar{i}_n\sim h^{-1/2}$. 
Let 
\beas 
\hat{S}^{(k)}_{n,j-1} &=& 
\frac{
\sum_{i=-\bar{i}_n}^{\bar{i}_n}\big(\Delta_{j-i}Y^{(k)}\big)^{\otimes2}
1_{\big\{
|\Delta_{j-i-K+1}Y^{(k)}|\wedge\cdots\wedge|\Delta_{j-i-1}Y^{(k)}|
\geq
|\Delta_{j-i}Y^{(k)}|\big\}} 
}
{
h\max\bigg\{1,\sum_{i=-\bar{i}_n}^{\bar{i}_n}
1_{\big\{
|\Delta_{j-i-K+1}Y^{(k)}|\wedge\cdots\wedge|\Delta_{j-i-1}Y^{(k)}|
\geq
|\Delta_{j-i}Y^{(k)}|\big\}} 
\bigg\}
}.
\eeas
Here $\Delta_jY^{(k)}$ reads $0$ when $j\leq0$ or $j>n$. 
An example of $\bars^{(k)}_{n,j-1}$ is 
\bea\label{300211-1} 
\bars^{(k)}_{n,j-1} &=& 
\hat{S}^{(k)}_{n,j-1}1_{\{\lambda_{min}( \hat{S}^{(k)}_{n,j-1})>2^{-1}\ep_0\}}
+2^{-1} \ep_0 {\colorb I_{\sfm_k}} 1_{\{\lambda_{min}( \hat{S}^{(k)}_{n,j-1})\leq2^{-1}\ep_0\}}, 
\eea
suppose that $\ds \inf_{x,\theta}\lambda_{min}(S^{(k)}(x,\theta))\geq\ep_0$ 
for some positive constant $\ep_0$, where 
$\lambda_{min}$ is the minimum eigenvalue of the matrix. 
\end{example}
\halflineskip

Let $\alpha=(\alpha^{(k)})_{k\in{\colorb \{1, \ldots, \sfk \}}}\in[0,1)^\sfk$. 
Our global jump filter is constructed as follows. 
Denote by $\calj_n^{(k)}(\alpha^{(k)})$ the set of $j\in\{1,...,n\}$ such that
\beas 
\#\big\{j'\in\{1,...,n\};\>|(\bars^{(k)}_{n,j'-1})^{-1/2}\Delta_{j'}Y^{(k)}|{\colorg >}|(\bars^{(k)}_{n,j-1})^{-1/2}\Delta_jY^{(k)}|\big\}&\geq&\alpha^{(k)} n
\eeas
for $k=1,...,\sfk$ and  $n\in\bbN$.
{\colorg 
If $\alpha^{(k)}=0$, 
then $\calj_n^{(k)}(\alpha^{(k)})=\{1,....,n\}$, 
that is, there is no filter for the $k$-th component.}
The density function of the multi-dimensional normal distribution with mean vector $\mu$ and 
covariance matrix $C$ is denoted by $\phi(z;\mu,C)$. 
Let
\beas 
q^{(k)}(\alpha^{(k)}) &=& \frac{
\text{Tr}\bigg(\int_{\{|z|\leq c(\alpha^{(k)})^{1/2}\}} z^{\otimes2}\phi(z;0,{\colorb I_{\sfm_k}})dz\bigg)
}{
\text{Tr}\bigg(\int_{\bbR^{{\tred\sfm_k}}}z^{\otimes2}\phi(z;0, {\colorb I_{\sfm_k}})dz\bigg)
},
\eeas
{\tred equivalently}, 
\beas 
q^{(k)}(\alpha^{(k)}) 
&=&
({\colorb \sfm_k})^{-1}\text{Tr}\bigg(\int_{\{|z|\leq c(\alpha^{(k)})^{1/2}\}} z^{\otimes2}\phi(z;0, {\colorb I_{\sfm_k}})dz\bigg)
\\&=&
(\sfm_k)^{-1}E[V1_{\{V\leq c(\alpha^{(k)})\}}],\quad 
\eeas
for a random variable $V\sim \chi^2({\sfm_k})$, the chi-squared distribution with ${\colorb \sfm_k}$ degrees of freedom, 
{\tred where} $c(\alpha^{(k)})$ is determined by
\beas 
P[V\leq c(\alpha^{(k)})]&=&1-\alpha^{(k)}. 
\eeas
Let
$p(\alpha^{(k)})=1-\alpha^{(k)}$. 
Now the ${\bm \alpha}${\bf-quasi-log likelihood function} $\bbH_{n}(\theta;\alpha)$ is defined by 
\beas 
\bbH_{n}(\theta;\alpha) 
&=& 
-\half\sum_{k=1}^\sfk\sum_{j\in\calj^{(k)}_n(\alpha^{(k)})}\bigg\{
q^{(k)}(\alpha^{(k)})^{-1}h^{-1}S^{(k)}(X_\tjm,\theta)^{-1}\big[\big(\Delta_jY^{(k)}\big)^{\otimes2}\big]K^{(k)}_{n,j}
\\&&
+p(\alpha^{(k)})^{-1}\log\det S^{(k)}(X_\tjm,\theta)\bigg\}
\eeas
where  
\bea\label{300211-5} 
K^{(k)}_{n,j} &=& 1_{\big\{|\Delta_jY^{(k)}|<C^{(k)}_*n^{-\frac{1}{4}}\big\}}
\eea
and $C^{(k)}_*$ are arbitrarily given positive constants. 
{\colorg 
For a tensor $T=(T_{i_1,...,i_k})_{i_1,...,i_k}$, we write 
\beas 
T[x_1,...,x_k]
&=&
T[x_1\otimes\cdots\otimes x_k]
\>=\>
\sum_{i_1,...,i_k}T_{i_1,...,i_k}
x_1^{i_1}\cdots x_k^{i_k}
\eeas
for $x_1=(x_1^{i_1})_{i_1}$, ..., $x_k=(x_k^{i_k})_{i_k}$. 
We denote $u^{\otimes r}=u\otimes\cdots\otimes u$ ($r$ times). 
Brackets $[\ \ ]$ stand for a multilinear mapping. 
{\tred 
This notation also applies to tensor-valued tensors. 
}

If $\alpha^{(k)}=0$, then 
$\calj^{(k)}_n(\alpha^{(k)}{\rev)}=\{1,...,n\}$, 
$c(\alpha^{(k)})=+\infty$, $p^{(k)}(\alpha^{(k)})=	1$ and $q^{(k)}(\alpha^{(k)})=1$, 
so the $k$-th component of 
$\bbH_n(\theta;\alpha)$ essentially becomes the ordinary quasi-log likelihood function 
by local Gaussian approximation. 
}

\begin{remark}\rm 
{\tred (i)} 
The cap $K^{(k)}_{n,j}$ can be removed if a suitable condition is assumed for the {\rev big} jump sizes of $J$, 
{\rev
e.g., $\sup_{t\in[0,T]}|\Delta J_t|\in L^\inftym{\colred =\cap_{p>1}L^p}$. 
} 
It is also reasonable to use  
\beas 
K^{(k)}_{n,j} &=& 1_{\big\{|\bars_{n,j-1}^{-1/2}\Delta_jY^{(k)}|<C^{(k)}_*n^{-\frac{1}{4}}\big\}} 
\eeas
if $\bars_{n,j-1}$ is uniformly $L^\inftym$-bounded. 
{\rev
In any case, the factor $K^{(k)}_{n,j}$ only serves for removing the effects of too big jumps and 
the classification is practically never affected by it since the global filter puts a threshold of the order less than $n^{-1/2}\log n$. 
As a matter of fact, the threshold of $K^{(k)}_{n,j}$ is of order $O(n^{-1/4})$, 
that is far looser than the ordinary local filters, and 
the truncation is exercised only with exponentially small probability. 
On the other hand, 
the global filter puts no restrictive condition on the distribution of the size of small jumps, 
like vanishing at the origin or boundedness of the density of the jump sizes, as assumed for the local filters so far. 
It should be emphasized that the difficulties in jump filtering are focused on the treatments of small jumps that look like the Brownian increments. 
{\tred (ii) 
The symmetry of $\sigma^{(k)}(x,\theta)$ is not restrictive because 
$\sigma^{(k)}(X_t,\theta)dw^{(k)}_t=S^{(k)}(X_t,\theta)^{1/2}\cdot\big(S^{(k)}(X_t,\theta)^{-1/2}\sigma^{(k)}(X_t,\theta)dw^{(k)}_t\big)$. On the other hand, 
we could introduce an $\sfm_k\times\sfm_k$ random matrix $\bar{\sigma}^{(k)}_{n,j-1}$ 
approximating $\sigma^{(k)}(X_\tjm,\theta^*)$ up to scaling, and use 
$\big(\bar{\sigma}^{(k)}_{n,j-1}\big)^{-1}\Delta_jY^{(k)}$ for 
$\big(\bar{S}^{(k)}_{n,j-1}\big)^{-1/2}\Delta_jY^{(k)}$, in order to remove the assumption of symmetry. 
}
}
\end{remark}

The {\bf $\boldsymbol{\alpha}$-quasi-maximum likelihood estimator} of $\theta$ ($\alpha$-QMLE) is any measurable mapping 
$\hat{\theta}^{M,\alpha}_n$ characterized by 
\beas 
\bbH_n(\hat{\theta}^{M,\alpha}_n;\alpha)
&=& 
\max_{\theta\in\bar{\Theta}}\bbH_n(\theta{\colorb ;} \alpha).
\eeas
We will identify an estimator of $\theta$, that is a measurable mapping of the data, 
with the pull-back of it to $\Omega$ since the aim of discussion here is to obtain 
asymptotic properties of the estimators' distribution. 

\begin{en-text}
The {\bf $\boldsymbol{\alpha}$-Bayesian estimator} $\hat{\theta}^{B,\alpha}_n$ of $\theta$ ($\alpha$-QBE) is defined by 
\beas 
\hat{\theta}^{B,\alpha}_n
&=& 
\bigg[\int_\Theta\exp\big(\bbH_n(\theta;\alpha)\big)\varpi(\theta)d\theta\bigg]^{-1}
\int_\Theta\theta\exp\big(\bbH_n(\theta;\alpha)\big)\varpi(\theta)d\theta,
\eeas
where $\varpi$ is a continuous function on $\Theta$ satisfying 
$0<\inf_{\theta\in\Theta}\varpi(\theta)\leq\sup_{\theta\in\Theta}\varpi(\theta)<\infty$. 
Set $\hat{u}^{{\sf A},\alpha}_n=\sqrt{n}\big(\hat{\theta}^{{\sf A},\alpha}_n-\theta^*\big)$ for ${\sf A}\in\{M,B\}$. 
\end{en-text}


\subsection{Assumptions}

We assume Sobolev's embedding inequality
\beas 
\sup_{\theta\in\Theta}\big|f(\theta)\big|
&\leq& 
C_{\Theta,p}\bigg\{\sum_{i=0}^1\int_\Theta \big|\partial_\theta^if(\theta)\big|^p {\colorb d \theta} \bigg\}^{1/p}
\qquad(f\in C^1(\Theta))
\eeas
for a bounded open set $\Theta$ in $\bbR^\sfp$, where $C_{\Theta,p}$ is a constant, $p>\sfp$. 
This inequality is valid, e.g., if $\Theta$ has a Lipschitz boundary. 
Denote by $C_\up^{a,b}(\bbR^\sfd\times\Theta;\bbR^\sfm\otimes\bbR^\sfr)$ the set of continuous functions 
$f:\bbR^\sfd\times\Theta\to\bbR^\sfm\otimes\bbR^\sfr$ that have continuous derivatives 
$\partial_{s_1}\cdots\partial_{s_\ell}f$ for all $(s_1,...,s_\ell)\in\{\theta,x\}^\ell$ such that 
$\#\{i\in\{1,...,\ell\};\>s_i=x\}\leq a$ and $\#\{i\in\{1,...,\ell\};\>s_i=\theta\}\leq b$, 
{\colorm 
and each of these derivatives satisfies
\beas 
\sup_{\theta\in\Theta}\big|\partial_{s_1}\cdots\partial_{s_\ell}f(x,\theta)\big|
&\leq& 
C(s_1,...,s_\ell)\big(1+|x|^{C(s_1,...,s_\ell)}\big)\quad(x\in\bbR^\sfd)
\eeas
for some positive constant $C(s_1,...,s_\ell)$.
}
%
Let $\|V\|_p=\big(E[|V|^p])^{1/p}$ for a vector-valued random variable $V$ and $p>0$. 
Let $N_t^{(k)}=\sum_{s\leq t}1_{\{\Delta J^{(k)}_s\not=0\}}$ 
and $N_t=\sum_{s\leq t}1_{\{\Delta J_s\not=0\}}$
We shall consider the following conditions. 
{\colorg Let $\tX=X-J^X$ for $J^X=\sum_{s\in[0,\cdot]}\Delta X_s$. }
\bd
\im[[F1\!\!]]$_\kappa$
{(i)} For every $p>1$, $\sup_{t\in[0,T]}\|X_t\|_p<\infty$ and there exists a constant $C(p)$ such that 
\beas 
\| {\colorg \tX_t-\tX_s}
\|_p\leq C(p)|t-s|^{1/2} \qquad(t,s\in[0,T]). 
\eeas
\bd\im[(ii)] $\sup_{t\in[0,T]}\|b_t\|_p<\infty$ for every $p>1$. 
\im[(iii)] $\sigma\in C^{2,\kappa}_\up(\bbR^\sfd\times\Theta;\bbR^{\sfm}\otimes\bbR^\sfr)$, 
$S(X_t,\theta)$ is invertible a.s. for every $\theta\in\Theta$, and \\
$\sup_{t\in[0,T],\theta\in\Theta}\|S(X_t,\theta)^{-1}\|_p<\infty$ for every $p>1$. 
\im[(iv)] 
$N_T\in L^\infm$ 
and $N^X_T\in L^\inftym$. 
\ed
\ed

{\fred 
Let $(\kappa_n)_{n\in\bbN}$ be a sequence of positive integers satisfying 
$\kappa_n=O(n^{1/2})$ as $n\to\infty$. 
For $j\in \{1,...,n\}$, let  $I_{n,j}=\big\{i\in\{1,...,n\};\>|i-j|\leq\kappa_n\big\}$. 
Let ${\tt I}_{n,j}=\cup_{i\in I_{n,j}}[t_{i-1},t_i]$. 
Define the index set ${\sf L}_n^{(k)}$ by 
\beas 
{\sf L}_n^{(k)} 
&=& 
\big\{j\in\{1,...,n\};\>N^{(k)}({\tt I}_{n,j})+N^X({\tt I}_{n,j})\not=0\big\}.
\eeas

}
\bd
\im[[F2\!\!]] 
{\bf (i)} 
$\bars^{(k)}_{n,j-1}$ are {\tred symmetric,} invertible and 
$\ds \sup_{n\in\bbN}\max_{j=1,...,n}\big\|(\bars^{(k)}_{n,j-1})^{-1}\big\|_p<\infty$ for every $p>1$ and $k=1,...,\sfk$. 
\bd
\im[(ii)] 
There exist positive constants $\gamma_0$ and $c^{(k)}$ ($k=1,...,\sfk$) such that 
\beas 
\sup_{n\in\bbN}\max_{j=1,...,n}n^{\gamma_0}
{\fred 
\bigg\|\bigg(S^{(k)}(X_\tjm,\theta^*)-c^{(k)}\bars^{(k)}_{n,j-1}\bigg)1_{\{j\in({\sf L}_n^{(k)})^c\}}\bigg\|_p
}
&<& \infty
\eeas
for every $p>1$ and $k=1,...,\sfk$. 
\ed
\ed

\begin{remark}\rm 
In {\fred $[F2]$ (ii)}, we assumed that there exists a positive constant $c^{(k)}$ 
such that every $S^{(k)}(X_{t_{j-1}},\theta^*)$ is approximated by 
$c^{(k)}\bars_{n,j-1}^{(k)}$. 
In estimation of $\theta$, we only assume positivity of $c^{(k)}$ but the values of them can be unknown 
since the function $\bbH_n$ does not involve $c^{(k)}$. 
When $S^{(k)}(X_{t_{j-1}},\theta^*)$ is a {\rev scalar} matrix, Condition ${\fred[F2]}$ is satisfied 
simply by $\bars^{(k)}_{n,j-1}=I_{{\rev \sfm_k}}$. 
\end{remark}

{\rev
\begin{remark}\rm
The $\bar{S}^{(k)}_{n,j-1}$ given by (\ref{300211-1}) in Example \ref{201902031722} 
satisfies Condition ${\fred[F2]}$ with $\gamma_0=1/4$ if one takes $\bar{i}_n\sim h^{-1/2}$. 
The constant $c^{(k)}$ depends on the depth $K$ of the threshold. 
It is possible to give an explicit expression of $c^{(k)}$ but not required by the condition. 
\end{remark}
}

\subsection{Global filtering lemmas}\label{300615-12} 
{\rev
The $\alpha$-quasi-log likelihood function $\bbH_n(\theta;\alpha)$ involves the summation 
regarding the index set $\calj^{(k)}_n(\alpha^{(k)})$. 
The global jump filter $\calj^{(k)}_n(\alpha^{(k)})$ avoids taking jumps but 
it completely destroys the martingale structure that the ordinary quasi-log likelihood function originally possessed, 
and without the martingale structure, we cannot follow a standard way to validate desirable asymptotic properties 
the estimator should have. 
However, it is possible to recover the martingale structure to some extent by deforming 
the global jump filter to a suitable deterministic filter. 
In this section, we will give several lemmas that enable such a deformation. 
}

As before, $\alpha=(\alpha^{(k)})_{k=1,...,\sfk}$ is a fixed vector in $[0,1)^\sfk$. 
We may assume that $\gamma_0\in(0,1/2]$ in ${\fred[F2]}$. 
Let 
\beas 
U^{(k)}_j &=&(c^{(k)})^{-1/2}h^{-1/2}(\bars^{(k)}_{n,j-1})^{-1/2}\Delta_jY^{(k)}\quad\text{and}\quad
W^{(k)}_j \yeq h^{-1/2}\Delta_jw^{(k)}.
\eeas
By $[F1]_0$ and ${\fred[F2]}$, we have 
\beas 
\sup_{n\in\bbN}\sup_{j=1,...,n}\big\|R^{(k)}_j
{\fred 1_{\{j\in({\sf L}_n^{(k)})^c\}}}
\big\|_p \yeq O(n^{-\gamma_0})
\eeas
for every $p>1$, 
where 
\beas 
R^{(k)}_j = U^{(k)}_j-W^{(k)}_j-(c^{(k)})^{-1/2}h^{-1/2}(\bars^{(k)}_{n,j-1})^{-1/2}\Delta_jJ^{(k)}.
\eeas 
{\tred Remark that 
$A^{1/2}=\frac{1}{\pi}\int_0^\infty \lambda^{-1/2}A(\lambda+A)^{-1}d\lambda$ 
for a positive-definite matrix $A$. }

\def\bW{\overline{W}^{(k)}}
\def\bU{\overline{U}^{(k)}}
Denote $|W^{(k)}_j|$ and $|U^{(k)}_j|$ 
by $\bW_j$ and $\bU_j$, respectively. 
$\bW_{(j)}$ denotes the $j$-th ordered statistic of $\{\bW_1,...,\bW_n\}$, and 
$\bU_{(j)}$ denotes {\colorb the} $j$-th ordered statistic of $\{\bU_1,...,\bU_n\}$. 
The rank of $\bW_j$ is denoted by $r({\colorb \bW_{j}})$. 
Denote by $q_{\alpha^{(k)}}$ the $\alpha^{(k)}$-quantile of the distribution of $\bW_1$. 
{\colorg The number $q_{\alpha^{(k)}}$ depends on $\sfm_k$. 
}

Let $0<\gamma_2<\gamma_1<\gamma_0$. 
Let $a_n^{(k)}=\lfloor\bar{\alpha}^{(k)}n-n^{1-\gamma_2}\rfloor$, where $\bar{\alpha}^{(k)}=1-\alpha^{(k)}=p(\alpha^{(k)})$. 
Define the event $N_{n,j}^{(k)}$ by 
\beas 
N_{n,j}^{(k)} &=& \big\{r(\bW_j)\leq a_n^{(k)}-n^{1-\gamma_2}\big\}
{\colorb \cap} \big\{\bW_{ {\colorb (a_n^{(k)})} }-\bW_j<n^{-\gamma_1}\big\}.
\eeas
\begin{lemma}\label{300618-1}
{\colorg Suppose that $\alpha^{(k)}\in(0,1)$. Then} 
$\ds 
P {\colorb \Big[ \bigcup}_{j=1,..,,n}N_{n,j}^{(k)} {\colorb \Big] } \yeq O(n^{-L})$ as $n\to\infty$ for every $L>0$. 
\end{lemma}
\proof 
We have 
\beas &&
P {\colorb \Big[ }\bW_{{\colorb (a_n^{(k)})}}>q_{\bar{\alpha}^{(k)}}+n^{-\gamma_1} {\colorb \Big]}
\\&=&
P\bigg[\sum_{j=1}^n 1_{\{\bW_j\leq q_{\bar{\alpha}^{(k)}}+n^{-\gamma_1}\}}<a_n^{(k)}\bigg]
\\&=&
P\bigg[n^{-1/2}\sum_{j=1}^n\bigg\{1_{\{\bW_j\leq q_{\bar{\alpha}^{(k)}}+n^{-\gamma_1}\}}
-P\big[\bW_j\leq q_{\bar{\alpha}^{(k)}}+n^{-\gamma_1}\big]\bigg\}
<-n^{\half-\gamma_1}c(n)\bigg]
\\&=&
O(n^{-L}) 
\eeas
for every $L>0$, where $(c(n))_{n\in\bbN}$ is a sequence of numbers such that $\inf_{n\in\bbN}c(n)>0$ 
{\colorb (the existence of such $c(n)$ can be proved by the mean value theorem)}. 
{\colorb The last equality in the above estimates is obtained by the following argument. 
For $A_j = \{\bW_j\leq q_{\bar{\alpha}^{(k)}}+n^{-\gamma_1}\}$ and 
$Z_j = 1_{A_j}-P[A_1]$, 
by the Burkholder-Davis-Gundy inequality, {\colorm Jensen's inequality and $|Z_j|\leq1$,} 
we obtain   
\beas
P\bigg[n^{-1/2}\sum_{j=1}^n Z_j < -n^{\half-\gamma_1}c(n)\bigg] 
&{\colorm \simleq}& 
{\colorm 
n^{-2p(\frac{1}{2} - \gamma_1)}c(n)^{-2p}  E\Bigg[ n^{-1}\sum_{j=1}^n |Z_j|^{2p}  \Bigg] 
}
\\&=& 
O(n^{-{\colorm p(1-2\gamma_1)}})
\eeas
for every $p>1$. 
}

Let 
\beas
B_n^{(k)}=\big\{\big|\bW_{{\colorb (a_n^{(k)})}}-q_{\bar{\alpha}^{(k)}}\big|>n^{-\gamma_1}\big\}.
\eeas 
We can estimate $P\big[\bW_{{\colorb (a_n^{(k)})}}<q_{\bar{\alpha}^{(k)}}-n^{-\gamma_1}\big]$, and so we have 
\bea\label{300213-1}
P\big[B_n^{(k)}\big]
&=& O(n^{-L})
\eea
for every $L>0$. 

By definition, on the event $N^{{\colorb (k)}}_{n,j}\cap (B_n^{(k)})^c$, 
the number of data $\bW_{j'}$ on the interval 
$\big[q_{\bar{\alpha}^{(k)}}-2n^{-\gamma_1},q_{\bar{\alpha}^{(k)}}+2n^{-\gamma_1}\big]$ is not less than 
$n^{1-\gamma_2}$. 
However, 
\bea\label{300214-1} &&
P\bigg[\sum_{j'=1}^n1_{\big\{\bW_{j'}\in\big[q_{\bar{\alpha}^{(k)}}-2n^{-\gamma_1},q_{\bar{\alpha}^{(k)}}+2n^{-\gamma_1}\big]\big\}}
\geq n^{1-\gamma_2}\bigg]
\nn\\&=&
P\bigg[n^{-1+\gamma_1}\sum_{j'=1}^n1_{\big\{\bW_{j'}\in\big[q_{\bar{\alpha}^{(k)}}-2n^{-\gamma_1},q_{\bar{\alpha}^{(k)}}+2n^{-\gamma_1}\big]\big\}}
\geq n^{\gamma_1-\gamma_2}\bigg]
\nn\\&=&
O(n^{-L})
\eea
for every $L>0$. 
Indeed, the family 
\beas 
\bigg\{
n^{-1/2}\sum_{j'=1}^n
\bigg(
1_{\big\{\bW_{j'}\in\big[q_{\bar{\alpha}^{(k)}}-2n^{-\gamma_1},q_{\bar{\alpha}^{(k)}}+2n^{-\gamma_1}\big]\big\}}
-E\bigg[1_{\big\{\bW_{j'}\in\big[q_{\bar{\alpha}^{(k)}}-2n^{-\gamma_1},q_{\bar{\alpha}^{(k)}}+2n^{-\gamma_1}\big]\big\}}
\bigg]\bigg)
\bigg\}_{n\in\bbN}
\eeas
is bounded in $L^\inftym$ {\colorb (this can be proved by the same argument as above)}. 
Since the estimate (\ref{300214-1}) is independent of $j\in\{1,...,n\}$, combining it with (\ref{300213-1}), we obtain
\beas \max_{j=1,..,,n}
P\big[N_{n,j}^{(k)}\big] \yeq O(n^{-L})
\eeas 
as $n\to\infty$ for every $L>0$. Now the desired inequality of the lemma is obvious. 
\qed\halflineskip

Let 
\beas 
\hat{\calj}_n^{(k)}(\alpha^{(k)}) &=& 
\bigg\{j\in\{1,...,n\};\>r(\bW_j)\leq \hat{a}^{(k)}_n\bigg\},
\eeas
where 
\beas
\hat{a}^{(k)}_n &=& \lfloor a^{(k)}_n-n^{1-\gamma_2}\rfloor.
\eeas
%
%
%
%
Let 
\beas 
\Omega_n &=& 
{\fred\bigg\{\sum_k\#{\sf L}_n^{(k)} < n^{1-\gamma_2}\bigg\}} 
{\colorg \bigcap}
\bigg(\bigcap_{k=1,...,\sfk}\bigcap_{j=1,...,n}
\bigg[\big\{|R^{(k)}_j|{\fred 1_{\{j\in({\sf L}_n^{(k)})^c\}}}
<2^{-1}n^{-\gamma_1}\big\}\cap(N^{(k)}_{n,j})^c\bigg]
\bigg). 
\eeas

\begin{lemma} 
\bea\label{300217-1} 
\hat{\calj}_n^{(k)}(\alpha^{(k)})\cap({\fred{\sf L}^{(k)}_n})^c
&\subset& 
\calj^{(k)}_n(\alpha^{(k)}) 
\eea
on $\Omega_n$. 
In particular 
\bea\label{300217-2} 
\# \big[\calj^{(k)}_n(\alpha^{(k)})\ominus\hat{\calj}_n^{(k)}(\alpha^{(k)})\big] &\leq& 
c_*n^{1-\gamma_2}
+{\fred \#\>{\sf L}_n^{(k)}}
\eea
on $\Omega_n$, where $c_*$ is a positive constant.  
{\colorb Here $\ominus$ denotes the symmetric difference operator of sets.  }
\end{lemma}
\proof 
%
On $\Omega_n$, 
{\colorg 
if a pair $(j_1,j_2)\in({\fred{\sf L}^{(k)}_n})^c\times({\fred{\sf L}^{(k)}_n})^c$ satisfies 
$r(\bW_{j_1})\leq\hat{a}^{(k)}_n$ and $r(\bW_{j_2})\geq a^{(k)}_n$, 
then 
$\bU_{j_1} <  \bW_{j_1} + 2^{-1}n^{-\gamma_1} 
\leq \bW_{(a^{(k)}_n)} - 2^{-1}n^{-\gamma_1} \leq 
\bW_{j_2} - 2^{-1}n^{-\gamma_1} 
< \bU_{j_2}$. 
Therefore, if $j\in\hat{\calj}^{(k)}_n(\alpha^{(k)})\cap({\fred{\sf L}^{(k)}_n})^c$, then 
$j\in\calj^{(k)}_n(\alpha^{(k)})$
}
since one can find 
at least 
$\lceil\alpha^{(k)} n\rceil {\colorg (\leq (n - a^{(k)}_n + 1)-n^{1-\gamma_2})}$ 
variables among 
$\bU_{(a_n^{(k)})},.....,\bU_{(n)}$ that are {\colorg larger} than $\bU_j$. 
Therefore (\ref{300217-1}) holds, and so does (\ref{300217-2}) 
{\colorg as follows. From (\ref{300217-1}), we have 
$
 \# \big[\calj^{(k)}_n(\alpha^{(k)})\ominus\hat{\calj}_n^{(k)}(\alpha^{(k)})\big] 
\leq
{\sf N}+ {\fred\#\> {\sf L}^{(k)}_n} 
$ 
for 
\beas
{\sf N}&=&\# \big[ \calj^{(k)}_n(\alpha^{(k)}) \cap \hat{\calj}_n^{(k)}(\alpha^{(k)})^{c} 
\cap ({\fred{\sf L}_n^{(k)}})^c\big].
\eeas 
Suppose that 
$j \in \calj^{(k)}_n(\alpha^{(k)}) \cap \hat{\calj}_n^{(k)}(\alpha^{(k)})^{c}
\cap ({\fred{\sf L}_n^{(k)}})^c$. 
{\tred In Case $r(\bW_j)< a^{(k)}_n$, 
since $\hat{a}^{(k)}_n<r(\bW_j)< a^{(k)}_n$, we know 
the number of such $j$ is less than or equal to $n^{1-\gamma_2}$. 
}
{\tred In Case $r(\bW_j)\geq a^{(k)}_n$, as seen above, } 
$\bU_{j_1}<\bU_j$ on $\Omega_n$ 
for all $j_1\in({\fred{\sf L}_n^{(k)}})^c$ satisfying $r(\bW_{j_1})\leq\hat{a}^{(k)}_n$, 
since $j\in({\fred{\sf L}_n^{(k)}})^c$ and $r(\bW_j){\tred\>\geq\>} a^{(k)}_n$.  
The number of such $j_1$s is at least $\hat{a}^{(k)}_n-\lfloor n^{1-\gamma_2}\rfloor$. 
On the other hand, $j \in \calj^{(k)}_n(\alpha^{(k)})$ gives 
$\#\{j'\in\{1,...,n\};\>\bU_j<\bU_{j'}\}\geq \lceil \alpha^{(k)}n\rceil$. 
Therefore
\beas 
{\sf N} 
&\leq& 
{\tred n^{1-\gamma_2}+}
n-\big(\hat{a}^{(k)}_n-\lfloor n^{1-\gamma_2}\rfloor\big)-\lceil \alpha^{(k)}n\rceil
\>\leq\>
{\tred 4}n^{1-\gamma_2}+{\tred 2}
\eeas
on $\Omega_n$. 
{\fred We} obtain (\ref{300217-2}) on $\Omega_n$ with 
$c_*={\tred 6}$ 
if we use the inequality 
${\tred4}n^{1-\gamma_2}+{\tred2}\leq {\tred6}n^{1-\gamma_2}$. 
}
\qed\halflineskip

Let $\gamma_3>0$. 
For random variables $(V_j)_{j=1,...,n}$, let 
\beas 
\cald_n^{(k)} &=& n^{\gamma_3}
{\colorb \Bigg|} \frac{1}{n}\sum_{j\in\calj^{(k)}_n(\alpha^{(k)})}V_j-\frac{1}{n}\sum_{j\in\hat{\calj}^{(k)}_n(\alpha^{(k)})}V_j {\colorb \Bigg|}.
\eeas

\begin{lemma}\label{300217-5}
{\bf (i)} Let $p_1>1$. Then
\beas 
\|\cald_n^{(k)}\|_p&\leq& 
\big(c_*n^{\gamma_3-\gamma_2}+n^{-1+\gamma_3}
\|{\fred\#\>{\sf L}_n^{(k)}}\|_{p_1}
\big)
\bigg\|\max_{j=1,...,n}\big|V_j\big|\bigg\|_{pp_1(p_1-p)^{-1}}
\\&&
+n^{\gamma_3}\bigg\|\max_{j=1,...,n}\big|V_j\big|1_{\Omega_n^c}\bigg\|_{p}
\eeas
for $p\in(1,p_1)$. 
\bd
\im[(ii)] Let $\gamma_4>0$ and $p_1>1$. Then
\beas 
\|\cald_n^{(k)}\|_p&\leq& 
\big(c_*n^{\gamma_3-\gamma_2}+n^{-1+\gamma_3}
\|{\fred\#\>{\sf L}_n^{(k)}}\|_{p_1}
\big)
\nn\\&&\times
\bigg(n^{\gamma_4}
+n\max_{j=1,...,n}\bigg\|\big|V_j\big|1_{\{|V_j|>n^{\gamma_4}\}}\bigg\|_{pp_1(p_1-p)^{-1}}
\bigg)
\\&&
+n^{\gamma_3}\bigg\|\max_{j=1,...,n}\big|V_j\big|1_{\Omega_n^c}\bigg\|_{p}
\eeas
for $p\in(1,p_1)$. 
\ed
\end{lemma}
\proof The estimate in (i) is obvious from (\ref{300217-2}). (ii) follows from (i). 
\qed\halflineskip

\begin{en-text}
We denote by 
$R^{(k)}(\alpha^{(k)})$ the region 
\beas &&
R^{(k)}(\alpha^{(k)}) \yeq 
\big[-c^{(k)}(\alpha^{(k)}),c^{(k)}(\alpha^{(k)})\big]^{m^{(k)}},
\eeas
where the constant $c^{(k)}(\alpha^{(k)})$ is specified by 
\beas
\bigg(\int_{-c^{(k)}(\alpha^{(k)})}^{c^{(k)}(\alpha^{(k)})}\phi(u;0,1)du\bigg)^{m^{(k)}} \yeq 1-\alpha^{(k)}.
\eeas
\end{en-text}

Let $\widetilde{\calj}^{(k)}_n(\alpha^{(k)})=\big\{j;\>|h^{-1/2}\Delta_jw^{(k)}|\leq q_{\bar{\alpha}^{(k)}}\big\} 
{= \colorb \big\{ j;\> \bW_j \leq q_{\bar{\alpha}^{(k)}} \big\} }$. 
Let 
\beas 
\tilde{\cald}_n^{(k)} &=& n^{\gamma_3}
\bigg|\frac{1}{n}\sum_{j\in\hat{\calj}^{(k)}_n(\alpha^{(k)})}V_j-\frac{1}{n}\sum_{j\in\widetilde{\calj}^{(k)}_n(\alpha^{(k)})}V_j\bigg|.
\eeas
%
\begin{lemma}\label{300217-6}
{\colorg 
Let $\tilde{\Omega}_n=\big\{\big| \bW_{(\hat{a}^{(k)}_n)}-q_{\bar{\alpha}^{(k)}}\big|
< {\tred\check{C}\>}n^{-\gamma_2}\big\}$, {\tred where $\check{C}$ is a positive constant. 
Then}
\bd
\im[(i)] For $p\geq1$, 
\beas 
\|\tilde{\cald}_n^{(k)}\|_p
&\leq& 
n^{\gamma_3}\bigg\|\max_{j'=1,...,n}|V_{j'}|\>
\frac{1}{n}\sum_{j=1}^n
1_{\big\{ \big|\bW_j-q_{\bar{\alpha}^{(k)}}\big|\leq {\tred\check{C}\>}n^{-\gamma_2}\big\}}
\bigg\|_p
+n^{\gamma_3}\bigg\|
1_{\tilde{\Omega}_n^c}
\max_{j'=1,...,n}|V_{j'}|\bigg\|_p.
\eeas

\im[(ii)] For $p_1>p\geq1$, 
\beas 
\|\tilde{\cald}_n^{(k)}\|_p
&\leq& 
n^{\gamma_3}\bigg\|\max_{j=1,...,n}|V_{j}|\>
\bigg\|_p
P\bigg[\big|\bW_1-q_{\bar{\alpha}^{(k)}}\big|\leq {\tred\check{C}\>}n^{-\gamma_2}\bigg]
\\&&
+n^{\gamma_3}\bigg\|\max_{j=1,...,n}|V_{j}|
\bigg\|_{pp_1(p_1-p)^{-1}}
\nn\\&&\hspace{30pt}\times
\bigg\|
\frac{1}{n}\sum_{j=1}^n\bigg(
1_{\big\{ \big|\bW_j-q_{\bar{\alpha}^{(k)}}\big|\leq {\tred\check{C}\>}n^{-\gamma_2}\big\}}
-
P\bigg[\big|\bW_1-q_{\bar{\alpha}^{(k)}}\big|\leq {\tred\check{C}\>}n^{-\gamma_2}\bigg]
\bigg)
\bigg\|_{p_1}
\\&&
+n^{\gamma_3}
P[{\tred\tilde{\Omega}}_n^c]^{1/p_1}
\bigg\|\max_{j=1,...,n}|V_{j}|\bigg\|_{pp_1(p_1-p)^{-1}}.
\eeas
\ed
}
\end{lemma}
{\colorg \proof
(i) follows from 
\beas 
1_{\tilde{\Omega}_n}\bigg|1_{\{\bW_j\leq\bW_{(\hat{a}_n^{(k)})}\}}
-1_{\{\bW_j\leq q_{\bar{\alpha}^{(k)}}\}}\bigg| 
&\leq&
1_{\big\{ \big|\bW_j-q_{\bar{\alpha}^{(k)}}\big|\leq {\tred\check{C}\>}n^{-\gamma_2}\big\}},
\eeas
and (ii) follows from (i). 
\begin{en-text}
\beas 
\tilde{\cald}_n^{(k)}1_{\tilde{\Omega}_n}
&\leq&
n^{\gamma_3-1}
1_{\tilde{\Omega}_n}\max_{j'=1,...,n}|V_{j'}|\>
\sum_{j=1}^n \bigg|1_{\{\bW_j\leq\bW_{(\hat{a}_n^{(k)})}\}}
-1_{\{\bW_j\leq q_{\bar{\alpha}^{(k)}}\}}\bigg| 
\\&\leq&
n^{\gamma_3-\gamma_2}
\max_{j'=1,...,n}|V_{j'}|, 
\eeas
and (ii) follows from (i). 
\end{en-text}
\qed}

\halflineskip
\noindent
{\colorg 
{\tred We take a sufficiently large $\check{C}$. Then}
the term involving $\tilde{\Omega}_n^c$ on the right-hand side of each inequality in Lemma \ref{300217-6} 
can be estimated as the proof of Lemma \ref{300618-1}. 
{\tred For example, $P[\tilde{\Omega}_n^c]=O(n^{-L})$ for any $L>0$. }
\halflineskip}

\begin{en-text}
\beas 
P\big[|U^{(k)}_j|\wedge|W^{(k)}_j|<r(\alpha^{(k)})<r(\alpha^{(k)})+ n^{-\gamma_1}<|U^{(k)}_j|\vee|W^{(k)}_j|]
&\leq& n^{-L}
\eeas
and
\beas 
P\big[|U^{(k)}_j|\wedge|W^{(k)}_j|<r(\alpha^{(k)})- n^{-\gamma_1}<r(\alpha^{(k)})<|U^{(k)}_j|\vee|W^{(k)}_j|]
&\leq& n^{-L}
\eeas
Therefore, essentially, disorder occurs only when $|W^{(k)}_j|\in[r(\alpha^{(k)})-n^{-\gamma_1},r(\alpha^{(k)})+n^{-\gamma_1}]$. 
The contribution of such $W^{(k)}_j$s to $n^{-1}\bbH_n(\theta;\alpha)$  is asymptotically negligible. 
\end{en-text}

\begin{lemma}\label{300217-9}
Let $k\in\{1,...,\sfk\}$ and let $f\in C^{1,1}_\up\big(\bbR^\sfd\times\Theta;\bbR)$. 
Suppose that ${\colorg [F1]_{ 0}}$ is fulfilled. Then   
\beas 
\sup_{n\in\bbN}\left\|\>\sup_{\theta\in\Theta}
n^\ep\left|\frac{1}{n}\sum_{j\in\calj^{(k)}_n(\alpha^{(k)})}p(\alpha^{(k)})^{-1}
f(X_\tjm,\theta)-\frac{1}{T}\int_0^Tf(X_t,\theta)dt\right|\>\right\|_p
&<&\infty
\eeas
for every $p\geq1$ and $\ep < {\colorb \gamma_2}$ .
\end{lemma}
\proof Use Sobolev's inequality and Burkholder's inequality as well as 
{\tred Lemmas \ref{300618-1},  
\ref{300217-5} (ii) and 
\ref{300217-6} (ii).
} 
{\colorb More precisely, we {\colorm have the following decomposition 
\beas
&& \frac{1}{n}\sum_{j\in\calj^{(k)}_n(\alpha^{(k)})}
{\colorg p(\alpha^{(k)})^{-1}}
f(X_\tjm,\theta)-\frac{1}{T}\int_0^Tf(X_t,\theta)dt \\
&=& 
p(\alpha^{(k)})^{-1}\left\{
\frac{1}{n}\sum_{j\in\calj^{(k)}_n(\alpha^{(k)})} f(X_\tjm,\theta) 
- \frac{1}{n}\sum_{j\in \hat{\calj}^{(k)}_n(\alpha^{(k)})}f(X_\tjm,\theta) \right\}
\\&&
+ p(\alpha^{(k)})^{-1}\left\{
\frac{1}{n}\sum_{j\in \hat{\calj}^{(k)}_n(\alpha^{(k)})}f(X_\tjm,\theta) 
- \frac{1}{n}\sum_{j\in \tilde{\calj}^{(k)}_n(\alpha^{(k)})} f(X_\tjm,\theta) \right\}
\\&&
+ \frac{1}{np(\alpha^{(k)})}\sum_{j=1}^n  
f(X_\tjm,\theta) \bigg\{1_{ \big\{ \bW_j \leq {\tred q_{\bar{\alpha}^{(k)}}} \big\}}
-p(\alpha^{(k)})\bigg\}
\\&&+ \frac{1}{nh}\sum_{j=1}^n \int_{t_{j-1}}^{t_j} [ f(X_\tjm,\theta) - f(X_t, \theta) ] dt 
\\&=:& 
I_{1,n}^{(k)}(\theta) + I_{2,n}^{(k)}(\theta) + I_{3,n}^{(k)}(\theta) + I_{4,n}^{(k)}(\theta). 
\eeas
}
{\colorg We may assume $\alpha^{(k)}>0$ since 
only $I^{(k)}_{4,n}(\theta)$ remains when $\alpha^{(k)}=0$ and it will be estimated 
below. }

As for $I_{1,n}^{(k)}(\theta)$, we apply Lemma {\fred\ref{300217-5} (ii)} to obtain 
\beas
\Bigg\| \sup_{\theta \in \Theta} n^\ep 
{\colorm \big|}I_{1,n}^{(k)}(\theta) {\colorm \big|}\Bigg\|_p 
&{\colorm \simleq}&
{\colorm \sum_{i=0,1} \sup_{\theta \in \Theta}
\left \|  n^\ep  \left| \frac{1}{n}\sum_{j\in\calj^{(k)}_n(\alpha^{(k)})}
\partial_\theta^if(X_\tjm,\theta) 
- \frac{1}{n}\sum_{j\in \hat{\calj}^{(k)}_n(\alpha^{(k)})}
\partial_\theta^if(X_\tjm,\theta) \right| \right \|_p} \\
&{\colorm \simleq}&
{\colorm 
\colorm \sum_{i=0,1}\sup_{\theta \in \Theta} \Bigg\{
}
\Big( c_{*} n^{\ep - \gamma_2} + n^{-1+\ep} 
\|{\fred\#\>{\sf L}_n^{(k)}}\|_{p_1}
\Big) \\
&\>& \> 
\times \Bigg( n^{\gamma_4} + 
{\colorm 
n \max_{j} \Bigg\|  |\partial_\theta^if(X_\tjm,\theta)| 
1_{\{ |\partial_\theta^if(X_\tjm,\theta)| \geq n^{\gamma_4} \}} \Bigg\|_{\frac{p p_1}{p_1 - p}} 
}
\Bigg) 
\\&\>&
{\colorm 
+ n^\ep \Bigg \| \max_{j}|\partial_\theta^if(X_\tjm,\theta)| 1_{\Omega_n^c} \Bigg \|_p
\Bigg\}. 
}
\eeas
By taking $\gamma_4 > 0$ small enough, we can verify that the right-hand side is $o(1)$ 
{\fred since 
\beas 
\big\|\#\>{\sf L}_n^{(k)}\big\|_p\simleq \kappa_n\big\|N_T+N_T^X\big\|_p=O(n^{1/2}).
\eeas}\noindent
Note that we have used the fact $P[\Omega_n^c] = O(n^{-L})$ {\tred for any $L>0$}. 
A similar argument {\colorm with Lemma \ref{300217-6} (ii)} yields 
$\big\| \sup_{\theta \in \Theta} n^\ep {\colorm \big|}I_{2,n}^{(k)}(\theta){\colorm \big|}
 \big\|_p = o(1)$.

{\colorm 
As for $I_{3,n}^{(k)}(\theta)$, 
applying the Burkholder-Davis-Gundy inequality for the discrete-time martingales 
as well as Jensen's inequality, 
we have
\beas&&
\sup_{\theta \in \Theta}
\left\|
n^\ep\sum_{j=1}^n \frac{1}{n}
\partial_\theta^i f(X_\tjm,\theta) \bigg\{1_{ \big\{ \bW_j \leq {\tred q_{\bar{\alpha}^{(k)}}} \big\}}
-p(\alpha^{(k)})\bigg\}
\right\|_p^p
\\&\simleq&
\sup_{\theta \in \Theta} n^{-p\big(\half- \ep\big)}  E\Bigg[ \Bigg| 
\frac{1}{n}
\sum_{j=1}^n 
\big|\partial_\theta^i f(X_\tjm,\theta)\big|^2
\bigg\{1_{ \big\{ \bW_j \leq {\tred q_{\bar{\alpha}^{(k)}}} \big\}}
-p(\alpha^{(k)})\bigg\}^2 \Bigg|^{\frac{p}{2}} \Bigg] 
\\&=& O\Big( n^{-( \frac{1}{2} - \ep) p} \Big) 
\eeas
for every $p\geq2$ and $i=0,1$. 
Hence, by Sobolev's inequality, we conclude
\beas
\bigg\|\sup_{\theta\in\Theta}n^\ep\big|I_{3,n}^{(k)}(\theta)\big| \bigg\|_p
&=& O\Big(n^{ -\frac{1}{2} + \ep}\Big)
\eeas
for every $p\geq1$. 
}

Finally, 
{\colorm we will estimate $I_{4,n}^{(k)}(\theta)$. 
Since $f\in C^{1,1}_\up(\bbR^\sfd\times\Theta;\bbR)$, 
there exists a positive constant $C$ such that 
\beas 
C_f(x,y) 
&\leq& 
C(1+|x|^C+|y|^C)
\eeas
where 
$
C_f(x,y) = 
\int_0^1 \sup_{\theta \in \Theta}   \big|\partial_{x}f( x + \xi(y - x),\theta)\big| d\xi
$ 
for $x,y\in\bbR^\sfd$. 
Then by $[F1]_{{\rev 0}}$ (i) and (ii), we obtain
\beas&&
\bigg\| n^\ep \sup_{\theta\in\Theta}\big|I_{4,n}^{(k)}(\theta)\big| \bigg\|_p
\nn\\&\leq&
n^\ep\times
\frac{1}{n h} 
 \sum_{j=1}^n  \int_{t_{j-1}}^{t_j}
 \big\|1_{\{\Delta_jN^X=0\}}  C_f(X_{t_{j-1}},X_t) |X_t - X_{t_{j-1}}| \big\|_p \> dt
\\&&
+n^\ep
\Bigg\|
\frac{1}{n h} 
 \sum_{j=1}^n 1_{\{\Delta_jN^X\not=0\}}
 \int_{t_{j-1}}^{t_j}  C_f(X_{t_{j-1}},X_t) |X_t - X_{t_{j-1}}|  \> dt
\Bigg\|_p
\\&\simleq&
n^{-\half+\ep} 
+n^{-\half+\ep}\Bigg\|
(N^X_T)^\half
\bigg\{n^{-1} \sum_{j=1}^n \bigg( h^{-1}\int_{t_{j-1}}^{t_j}  C_f(X_{t_{j-1}},X_t) |X_t - X_{t_{j-1}}|  \> dt\bigg)^2\bigg\}^\half\Bigg\|_p
\\&\simleq&
n^{-\half+\ep} 
+n^{-\half+\ep} 
\big\|N^X_T\big\|_p^\half
\\&=&
O(n^{-\half+\ep})
\eeas
for every $p\geq1$. 
This completes the proof. 
}}
\qed\halflineskip

By $L^p$-estimate, we obtain the following lemma. 
\begin{lemma}\label{300217-8}
Let $k\in\{1,...,\sfk\}$ and let $f\in C^{0,1}_\up\big(\bbR^\sfd\times\Theta; {\colorb \bbR^{\sfm_k} \otimes \bbR^{\sfm_k} })$. 
Suppose that ${\colorg [F1]_{ 0}}$ is fulfilled. Then 
\beas 
\sup_{n\in\bbN}\left\|\>\sup_{\theta\in\Theta}
n^{{\colorg \half-\ep}}
\left|\sum_{j\in\calj^{(k)}_n(\alpha^{(k)})}
f(X_\tjm,\theta)
\bigg[\big({\colorb \Delta_jY^{(k)}}\big)^{\otimes2}K^{(k)}_{n,j}
-\big( {\colorb \sigma^{(k)}(X_\tjm,\theta^*)\Delta_jw^{(k)}} \big)^{\otimes2}\bigg]
\right|\>\right\|_p
&<&\infty
\eeas
for every $p\geq1$ {\colorg and $\ep>0$}.
\end{lemma}

{\colorg 
\proof
Let $\tY^{(k)}=Y^{(k)}-J^{(k)}$. 
Let $\check{N}=N+N^X$. 
Let 
\beas 
Q_j &=& \big(\sigma^{(k)}(X_\tjm,\theta^*)\Delta_jw^{(k)} \big)^{\otimes2}.
\eeas
Then  
\bea\label{300623-1}&&
\sup_{\theta\in\Theta}
\left\|
n^{{\colorg \half-\ep}}
\left|\sum_{j\in\calj^{(k)}_n(\alpha^{(k)})}1_{\{\Delta_j\check{N}>0\}}
f(X_\tjm,\theta)
\bigg[\big(\Delta_jY^{(k)}\big)^{\otimes2}K^{(k)}_{n,j}
-Q_j\bigg]
\right|\>\right\|_p
\nn\\&\leq&
\sup_{\theta\in\Theta}
\left\|
n^{{\colorg \half-\ep}}
\max_{j=1,...,n}\left|
f(X_\tjm,\theta)
\bigg[\big(\Delta_jY^{(k)}\big)^{\otimes2}K^{(k)}_{n,j}
-Q_j\bigg]
\right|\>\right\|_{2p}
\big\|\check{N}_T\|_{2p}
\nn\\&=&
o(1)
\eea
as $n\to\infty$ {\tred thanks to $K^{(k)}_{n,j}$}. 
%

Let $\eta=1-\ep/2$. 
Then, by the Burkholder-Davis-Gundy inequality, for any $L{\tred\>\geq2}$, 
\beas 
P_n&:=&
P\bigg[\max_{j=1,...,n}\bigg|1_{\{\Delta_j\check{N}=0\}}
\int_\tjm^\tj\big\{\sigma(X_t,\theta^*)-\sigma(X_\tjm,\theta^*)\big\}dw_t\bigg|
>n^{-\eta}\bigg]
\\&\leq&
P\bigg[\max_{j=1,...,n}\bigg|
\int_\tjm^\tj\big\{\sigma(\tX_t+J^X_\tjm,\theta^*)-\sigma(X_\tjm,\theta^*)\big\}dw_t\bigg|
>n^{-\eta}\bigg]
\\&\simleq&
\sum_{j=1}^n n^{L\eta}
E\bigg[\bigg(
\int_\tjm^\tj\big|\sigma(\tX_t+J^X_\tjm,\theta^*)-\sigma(X_\tjm,\theta^*)\big|^2dt\bigg)^{L/2}
\bigg]
\\&\leq&
\sum_{j=1}^n n^{L\eta}
h^{L/2-1}
\int_\tjm^\tj E\big[\big|\sigma(\tX_t+J^X_\tjm,\theta^*)-\sigma({\tred\tX_\tjm+J^X_\tjm},\theta^*)\big|^L\big]dt
\\&=&
O\big(n\times n^{L\eta}\times n^{-L/2+1}\times n^{-1}\times n^{-L(1/2-\ep/4)}\big)
\\&=&
O(n^{1-L\ep/4}).
\eeas
In the last part, we used Taylor's formula and H\"older's inequality. 
Therefore, 
$P_n=O(n^{-L})$ for any $L>0$. 

Expand $\Delta_j\tY^{(k)}$ with the formula 
\beas 
\Delta_j\tY^{(k)}
&=& 
\sigma^{(k)}(X_\tjm,\theta^*) \Delta_jw^{(k)}
+\int_\tjm^\tj\big\{\sigma^{(k)}(X_t,\theta^*)-\sigma^{(k)}(X_\tjm,\theta^*)\big\}dw^{(k)}_t
+\int_\tjm^\tj b^{(k)}_tdt
\\&=:&
\xi_{1,j}+\xi_{2,j}+\xi_{3,j}.
\eeas
Then we have
\beas&&
\sup_{\theta\in\Theta}
\Bigg\|
n^{{\colorg \half-\ep}}
\Bigg|\sum_{j\in\calj^{(k)}_n(\alpha^{(k)})}1_{\{\Delta_j\check{N}=0\}}
f(X_\tjm,\theta)
\big[\xi_{1,j}\otimes\xi_{2,j}\big]
\Bigg|\>\Bigg\|_p
\\&\simleq&
n^{\half-\frac{\ep}{2}}
\sup_{j=1,...,n\atop \theta\in\Theta}
\big\|
|f(X_\tjm,\theta)||\xi_{1,j}|
\>\big\|_p
+{\tred n^{1-\ep}}P_n^{\frac{1}{2p}}
\\&=&
o(1).
\eeas
Thus, we can see 
\beas&&
\sup_{\theta\in\Theta}
\Bigg\|
n^{{\colorg \half-\ep}}
\Bigg|\sum_{j\in\calj^{(k)}_n(\alpha^{(k)})}1_{\{\Delta_j\check{N}=0\}}
f(X_\tjm,\theta)
\big[\xi_{i_1,j}\otimes\xi_{i_2,j}\big]
\Bigg|\>\Bigg\|_p
\yeq
o(1)
\eeas
for $(i_1,i_2)\in\{1,2,3\}^2\setminus\{(1,1)\}$. 
Consequently, 
\bea\label{300623-2}&&
\sup_{\theta\in\Theta}
\left\|
n^{{\colorg \half-\ep}}
\left|\sum_{j\in\calj^{(k)}_n(\alpha^{(k)})}1_{\{\Delta_j\check{N}=0\}}
f(X_\tjm,\theta)
\bigg[\big(\Delta_jY^{(k)}\big)^{\otimes2}K^{(k)}_{n,j}
-Q_j\bigg]
\right|\>\right\|_p
\nn\\&\leq&
\sup_{\theta\in\Theta}
\left\|
n^{{\colorg \half-\ep}}
\left|\sum_{j\in\calj^{(k)}_n(\alpha^{(k)})}1_{\{\Delta_j\check{N}=0\}}
f(X_\tjm,\theta)
\bigg[\big(\Delta_j\tY^{(k)}\big)^{\otimes2}
-Q_j\bigg]
\right|\>\right\|_p
+O(n^{-L})
\nn\\&=&
o(1)
\eea
for every $p>1$ and $L>0$. 

From (\ref{300623-1}) and (\ref{300623-2}), we obtain 
\bea\label{300623-3}
\sup_{\theta\in\Theta}
\left\|
n^{{\colorg \half-\ep}}
\left|\sum_{j\in\calj^{(k)}_n(\alpha^{(k)})}
f(X_\tjm,\theta)
\bigg[\big(\Delta_jY^{(k)}\big)^{\otimes2}K^{(k)}_{n,j}
-Q_j\bigg]
\right|\>\right\|_p
&=&
o(1)
\eea
for every $p>1$. 
Applying the same estimate as (\ref{300623-3}) to $\partial_\theta f$ for $f$, 
we conclude the proof by Sobolev's inequality. \qed\halflineskip
}

Lemmas {\fred\ref{300217-5}}, \ref{300217-6} and \ref{300217-8} suggest approximation of $n^{-1}\bbH_{n}(\theta;\alpha)$ by 
\beas&&
-\frac{1}{2n}\sum_{k=1}^\sfk
\sum_{j\in\widetilde{\calj}^{(k)}_n(\alpha^{(k)})}\bigg\{
q^{(k)}(\alpha^{(k)})^{-1}
S^{(k)}(X_\tjm,\theta^*)^{1/2}S^{(k)}(X_\tjm,\theta)^{-1}S^{(k)}(X_\tjm,\theta^*)^{1/2}
\nn\\&&\hspace{150pt}\cdot
\big[\big(h^{-1/2}\Delta_jw^{(k)}\big)^{\otimes2}\big]
+p(\alpha^{(k)})^{-1}\log\det S^{(k)}(X_\tjm,\theta)\bigg\},
\eeas
as we will see {\colorm its validity} below.

\subsection{Polynomial type large deviation inequality and the rate of convergence of 
{\colorg the ${\bm \alpha}$-QMLE and the ${\bm{(\alpha,\beta)}}$-QBE}
}\label{300615-13}
We will show convergence of the $\alpha$-QMLE. 
{\rev
To this end, we will use 
a polynomial type large deviation inequality given in Theorem \ref{300217-10} below 
for a random field associated with $\bbH_n(\theta;\alpha)$. 
Proof of Theorem \ref{300217-10} will be given in Section \ref{300430-2}, based on 
the QLA theory (\cite{yoshida2011}) with the aid of the global filtering lemmas in Section \ref{300615-12}.
}
Though the rate of convergence is less optimal, 
{\rev 
the global filter has the advantage of eliminating jumps with high precision, and}
we can use it as a stable initial estimator to obtain an efficient estimator {\rev later}. 
We do not assume {\rev any} restrictive condition of the distribution of small jumps 
{\colorg though the previous jump filters required such a condition for optimal estimation.} 

We introduce a middle resolution {\colorg (or annealed)} random field. A similar method was used in 
Uchida and Yoshida 
{\tred\cite{UchidaYoshida2012Adaptive}} to relax the so-called balance condition between 
the number of observations and the discretization step for an ergodic diffusion model. 
For $\beta\in(0,\gamma_0)$, 
let 
\bea\label{300615-14}
\bbH_n^\beta(\theta;\alpha)  &=& n^{-1+2\beta}\bbH_n(\theta;\alpha).
\eea
{\rev The random field $\bbH_n^\beta(\theta;\alpha)$ mitigates the sharpness of the contrast $\bbH_n(\theta;\alpha)$.}
Let 
\beas
\bbY_n(\theta;\alpha) &=& n^{-2\beta}\big\{\bbH_n^\beta(\theta;\alpha)-\bbH_n^\beta(\theta^*;\alpha)\big\}
\yeq n^{-1}\big\{\bbH_n(\theta;\alpha)-\bbH_n(\theta^*;\alpha)\big\}. 
\eeas
Let 
\beas 
\bbY(\theta) &=& 
-\frac{1}{2T}\sum_{k=1}^\sfk\int_0^T\bigg\{
\text{Tr}\bigg( {\colorb S^{(k)}(X_t,\theta)^{-1}S^{(k)}(X_t,\theta^*) }-I_{\sfm_k}\bigg)
\\&&
+\log\frac{\det {\colorb S^{(k)}(X_t,\theta)} }{\det {\colorb S^{(k)}(X_t,\theta^*) }}\bigg\}dt.
\eeas

The key index $\chi_0$ is defined by 
\beas 
\chi_0 &=& \inf_{\theta\not=\theta^*}\frac{-\bbY(\theta)}{|\theta-\theta^*|^2}.
\eeas
Non-degeneracy of $\chi_0$ plays an essential role in the QLA. 
\bd\im[[F3\!\!]] For every positive number $L$, there exists a constant $C_L$ such that 
\beas 
P\big[\chi_0<r^{-1}\big] &\leq& C_L\>r^{-L}\qquad(r>0). 
\eeas
\ed
\begin{remark}\rm 
An analytic criterion and a geometric criterion are known to insure Condition $[F3]$ 
when $X$ is a non-degenerate diffusion process. 
See Uchida and Yoshida 
\cite{uchida2013quasi} for details. 
Since the proof of this fact depends on short time asymptotic properties, 
we can modify it by taking the same approach before the first jump even when $X$ has 
finitely active jumps. 
Details will be provided elsewhere. 
On the other hand, those criteria can apply to the jump diffusion $X$ 
without remaking them if we work under localization. 
See Section \ref{300228-1}. 
\end{remark}

Let $\bbU^\beta_n=\{u\in\bbR^\sfp;\>\theta^*+n^{-\beta}u\in\Theta\}$. 
Let $\bbV^\beta_n(r)=\{u\in\bbU^\beta_n;\>|u|\geq r\}$. 
The quasi-likelihood ratio random field $\bbZ^\beta_n(\cdot;\alpha)$ of order $\beta$ is defined by 
\beas 
\bbZ^\beta_n(u;\alpha) &=& 
\exp\bigg\{\bbH^\beta_n\big(\theta^*+n^{-\beta}u;\alpha\big)-\bbH^\beta_n\big(\theta^*;\alpha\big)\bigg\}
\qquad(u\in\bbU^\beta_n).
\eeas
{\rev 
The random field $\bbZ^\beta_n(u;\alpha)$ is ``annealed'' since 
the contrast function $-\bbH^\beta_n(\theta;\alpha)$ becomes a milder penalty 
than $-\bbH_n(\theta;\alpha)$ because $\beta<1/2$. 
}

{\colorr The following theorem will be proved in Section \ref{300430-2}. }
\begin{theorem}\label{300217-10}
Suppose that $[F1]_4$, ${\fred[F2]}$ and $[F3]$ are fulfilled. 
Let $c_0\in(1,2)$. 
Then, for every positive number $L$, 
there exists a constant $C(\alpha,\beta,c_0,L)$ such that 
\beas 
P\bigg[\sup_{u\in\bbV_n(r)}\bbZ^\beta_n(u;\alpha)\geq e^{-r^{c_0}}\bigg] &\leq& \frac{C(\alpha,\beta,c_0,L)}{r^L}
\eeas
for all $r>0$ and $n\in\bbN$. 
\end{theorem}

Obviously, an $\alpha$-QMLE $\hat{\theta}^{M,\alpha}_n$of $\theta$ with respect to $\bbH_n(\cdot;\alpha)$
is a QMLE with respect to $\bbH^\beta_n(\cdot;\alpha)$. 
The following rate of convergence is a consequence of Theorem \ref{300217-10}, 
{\colorg as usual in the QLA theory.}
\begin{proposition}\label{300225-31}
Suppose that $[F1]_4$, ${\fred[F2]}$ and $[F3]$ are satisfied. Then 
$\ds 
\sup_{n\in\bbN}\big\|n^\beta\big(\hat{\theta}^{M,\alpha}_n-\theta^*\big)\big\|_p\><\> \infty
$ for every $p>1$ and every $\beta<\gamma_0$. 
\end{proposition}
\halflineskip

The {\bf $\boldsymbol{(\alpha,\beta)}$-quasi-Bayesian estimator} ($(\alpha,\beta)$-QBE) 
$\hat{\theta}^{B,\alpha,\beta}_n$ 
of $\theta$ is defined by 
\bea\label{300615-5} 
\hat{\theta}^{B,\alpha,\beta}_n
&=& 
\bigg[\int_\Theta\exp\big(\bbH_n^\beta(\theta;\alpha)\big)\varpi(\theta)d\theta\bigg]^{-1}
\int_\Theta\theta\exp\big(\bbH_n^\beta(\theta;\alpha)\big)\varpi(\theta)d\theta,
\eea
where $\varpi$ is a continuous function on $\Theta$ satisfying 
$0<\inf_{\theta\in\Theta}\varpi(\theta)\leq\sup_{\theta\in\Theta}\varpi(\theta)<\infty$. 
%
Once again Theorem \ref{300217-10} ensures $L^\inftym$-boundedness of the error of 
the $(\alpha,\beta)$-QBE: 
\begin{proposition}\label{300226-1} 
Suppose that $[F1]_4$, ${\fred[F2]}$ and $[F3]$ are satisfied. 
Let $\beta\in(0,\gamma_0)$. Then 
\beas 
\sup_{n\in\bbN}\big\|n^\beta\big(\hat{\theta}^{B,\alpha,\beta}_n-\theta^*\big)\big\|_p\><\> \infty
\eeas
for every $p>1$.
\end{proposition}
\proof 
Let $\hat{u}_n^{B,\alpha,\beta}=n^\beta\big(\hat{\theta}_n^{B,\alpha,\beta}-\theta^*\big)$. 
Then 
\beas 
\hat{u}_n^{B,\alpha,\beta} 
&=&
\bigg(\int_{\bbU^\beta_n} {\colorb \bbZ^\beta_n(u; \alpha)} \varpi(\theta^*+n^{-\beta}u)du\bigg)^{-1}
\int_{\bbU^\beta_n} u\> {\colorb \bbZ^\beta_n(u; \alpha)} \varpi(\theta^*+n^{-\beta}u)du;
\eeas
{\tred recall} 
$\bbU^\beta_n=\{u\in\bbR^\sfp;\>\theta^*+n^{-\beta}u\in\Theta\}$. 

{\colorr
Let {\tred$C_1>0$,} $p>1$, $L>p+1$ {\tred and} $D>\sfp+p$. 
In what follows, we take a sufficiently large positive constant $C_1'$.} We have 
\beas &&
E\big[|\hat{u}_n^{B,\alpha,\beta}|^p]
\\&\leq&
E\bigg[\bigg(\int_{\bbU^\beta_n} {\colorb \bbZ^\beta_n(u; \alpha)} \varpi(\theta^*+n^{-\beta}u)du\bigg)^{-1}
\int_{\bbU^\beta_n}|u|^p {\colorb \bbZ^\beta_n(u; \alpha)} \varpi(\theta^*+n^{-\beta}u)du\bigg]
\nn\\&&
\hspace{150pt}\quad(\text{Jensen's inequality, }p\geq1)
\\&\leq&
{\colorb C(\varpi)} \sum_{r=1}^\infty(r+1)^p 
\bigg\{\hspace{-2.8mm}\bigg\{
E \Bigg[ \bigg(\int_{\bbU^\beta_n} {\colorb \bbZ^\beta_n(u; \alpha)} du\bigg)^{-1}
\int_{\{u;r<|u|\leq r+1\}\cap\bbU^\beta_n} {\colorb \bbZ^\beta_n(u; \alpha)} du 
\\&&
\times\>1_{\bigg\{\int_{\{u;r<|u|\leq r+1\}\cap\bbU^\beta_n}
{\colorb \bbZ^\beta_n(u; \alpha)} du>\frac{C_1'}{r^{D-\sfp+1}}
\bigg\}} \Bigg]
\\&&
+
E\bigg[\bigg(\int_{\bbU^\beta_n} {\colorb \bbZ^\beta_n(u; \alpha)} du\bigg)^{-1}
\int_{\{u;r<|u|\leq r+1\}\cap\bbU^\beta_n} {\colorb \bbZ^\beta_n(u; \alpha)} du 
\>1_{\bigg\{\int_{\{u;r<|u|\leq r+1\}\cap{\tred\bbU_n^\beta}}
{\colorb \bbZ^\beta_n(u; \alpha)} du\leq\frac{C_1'}{r^{D-\sfp+1}}
\bigg\}}\bigg]
\bigg\}\hspace{-2.8mm}\bigg\}
\\&&
+{\colorb C(\varpi)}\quad\text{(The last term is for }r=0.\>
\text{The integrand is not greater than one.} )
\\&\leq&
{\colorb C(\varpi)}\sum_{r=1}^\infty(r+1)^p 
\bigg\{\hspace{-2.8mm}\bigg\{
P\bigg[\int_{\{u;r<|u|\leq r+1\}\cap \bbU^\beta_n}
{\colorb \bbZ^\beta_n(u; \alpha)} du>\frac{C_1'}{r^{D-\sfp+1}}\bigg]
\nn\\&&\hspace{100pt}
+\frac{C_1'}{r^{D-\sfp+1}}
E\bigg[\bigg(\int_{ \bbU^\beta_n } {\colorb \bbZ^\beta_n(u; \alpha)} du\bigg)^{-1}
\bigg]
\bigg\}\hspace{-2.8mm}\bigg\}
+{\colorb C(\varpi)}
\\&\leq&
{\colorb C(\varpi)}\sum_{r=1}^\infty(r+1)^p 
\bigg\{\hspace{-2.8mm}\bigg\{
P\bigg[
\sup_{u\in {\colorb \bbV^\beta_n(r)}}{\colorb \bbZ^\beta_n(u; \alpha)})>\frac{C_1}{r^{D}}\bigg]
+
\frac{C_1'}{r^{D-\sfp+1}}
E\bigg[\bigg(\int_{{\colorb \bbU^\beta_n}} {\colorb \bbZ^\beta_n(u; \alpha)} du\bigg)^{-1}\bigg]
\bigg\}\hspace{-2.8mm}\bigg\}+{\colorb C(\varpi)}
\\&\simleq&
\sum_{r=1}^\infty r^{-(L-p)}
+\sum_{r=1}^\infty r^{-(D-p-\sfp+1)}E\bigg[\bigg(\int_{{\colorb \bbU^\beta_n}} {\colorb \bbZ^\beta_n(u; \alpha)} du\bigg)^{-1}\bigg]
+{\colorb C(\varpi)}. 
\\&<&\infty
\eeas
{\colorr\noindent 
by Theorem \ref{300217-10}, suppose that 
\bea\label{300501-1}
E\bigg[\bigg(\int_{{\colorb \bbU^\beta_n}} {\colorb \bbZ^\beta_n(u; \alpha)} du\bigg)^{-1}\bigg]
&<& 
\infty.
\eea
However, one can show (\ref{300501-1}) by using Lemma 2 
of Yoshida 
\cite{yoshida2011}. 
\qed\halflineskip
}

\subsection{Proof of Theorem \ref{300217-10}}\label{300430-2}
We will prove Theorem \ref{300217-10} by Theorem 2 of Yoshida 
\cite{yoshida2011} 
{\rev
with the aid of the global filtering lemmas in Section \ref{300615-12}.}
Choose parameters $\eta$, $\beta_1$, $\rho_1$, $\rho_2$ and $\beta_2$ satisfying the following inequalities: 
\bea\label{300225-1} &&
0<\eta<1,\qquad
0<\beta_1<\half,\qquad 
0<\rho_1<\min\{1,\eta(1-\eta)^{-1},2{\colorg \beta_1}(1-\eta)^{-1}\},
\nn\\&&
2\eta<\rho_2,\qquad
\beta_2\geq0,\qquad
1-2\beta_2-\rho_2>0.
\eea

Let 
\beas 
\Delta_n(\alpha,\beta)&=&n^{-\beta}\partial_\theta\bbH^\beta_n(\theta^*;\alpha)
\yeq n^{-1+\beta}\partial_\theta\bbH_n(\theta^*;\alpha). 
\eeas
Let 
\beas 
\Gamma_n(\alpha)&=&-n^{-2\beta}\partial_\theta^2\bbH^\beta_n(\theta^*;\alpha)
\yeq -n^{-1}\partial_\theta^2\bbH_n(\theta^*;\alpha). 
\eeas
The ${\colorb \sfp \times \sfp}$ symmetric matrix $\Gamma^{(k)}
$ is defined by {the following formula:}
\beas 
\Gamma^{(k)}[u^{\otimes2}]
&=& 
\frac{1}{2T}
\int_0^T \text{Tr}\bigg(
(\partial_\theta S^{(k)}[u])(S^{(k)})^{-1}(\partial_\theta S^{(k)}[u])(S^{(k)})^{-1}(X_t,\theta^*)\bigg)
dt, 
\eeas
{\colorb where $u \in {\colorm \bbR^\sfp}$,} and $\Gamma$ by 
$
\Gamma=\sum_{k=1}^\sfk\Gamma^{(k)}$. 
We will need several lemmas. 
\begin{en-text}
We choose positive constants $\gamma_i$ ($i=1,2,3,4$) so that $\beta<\gamma_3<\gamma_2<\gamma_1<\gamma_0$ 
and that $\gamma_4<\gamma_2-\gamma_3$. 
Then 
we can choose parameters $\beta_1(\down0)$, $\beta_2(\up1/2)$, 
$\rho_2(\down0)$, $\eta(\down0)$ and $\rho_1(\down0)$ so that 
$\max\{2\beta\beta_1,\beta(1-2\beta_2)\}<\gamma_3$. 
Then there is {\colorg an} $\ep\in(\max\{2\beta\beta_1,\beta(1-2\beta_2)\},\gamma_3)$. 
\end{en-text}
{\colorg
We choose positive constants $\gamma_i$ ($i=1,2$) so that $\beta<\gamma_2<\gamma_1<\gamma_0$.  
Then 
we can choose parameters $\beta_1(\down0)$, $\beta_2(\up1/2)$, 
$\rho_2(\down0)$, $\eta(\down0)$ and $\rho_1(\down0)$ so that 
$\max\{2\beta\beta_1,\beta(1-2\beta_2)\}<\gamma_2$. 
Then there is {\colorg an} $\ep\in(\max\{2\beta\beta_1,\beta(1-2\beta_2)\},\gamma_2)$. 
}
\begin{lemma}\label{300221-A1(1)}
For every $p\geq1$, 
\beas
\sup_{n\in\bbN}E\bigg[\bigg(n^{-2\beta}\sup_{\theta\in\Theta}\big|\partial_\theta^3\bbH_n^\beta(\theta;\alpha)
\big|\bigg)^p
\bigg] &<& \infty. 
\eeas
\end{lemma}
\proof
We have 
$\bbH_{n}(\theta;\alpha) 
= \bbH_{n}^\circ(\theta;\alpha) +\bbM^\circ(\theta;\alpha)+\bbR^\circ(\theta;\alpha)$, 
where 
\beas 
\bbH_{n}^\circ(\theta;\alpha) 
&=& 
-\half\sum_{k=1}^\sfk\sum_{j\in\calj^{(k)}_n(\alpha^{(k)})}p(\alpha^{(k)})^{-1}\bigg\{
S^{(k)}(X_\tjm,\theta)^{-1}\big[S^{(k)}(X_\tjm,\theta^*)\big]
\\&&
+\log\det S^{(k)}(X_\tjm,\theta)\bigg\},
\eeas
\beas 
\bbM_{n}^\circ(\theta;\alpha) 
&=& 
-\half\sum_{k=1}^\sfk\sum_{j\in\calj^{(k)}_n(\alpha^{(k)})}
h^{-1}S^{(k)}(X_\tjm,\theta)^{-1}
\big[q^{(k)}(\alpha^{(k)})^{-1}\big(\sigma^{(k)}(X_\tjm,\theta^*)\Delta_jw^{(k)}\big)^{\otimes2}
\\&&
-hp(\alpha^{(k)})^{-1}S^{(k)}(X_\tjm,\theta^*)\big]
\eeas
and
\beas 
\bbR^\circ_{n}(\theta;\alpha) 
&=& 
-\half\sum_{k=1}^\sfk\sum_{j\in\calj^{(k)}_n(\alpha^{(k)})}
q^{(k)}(\alpha^{(k)})^{-1}h^{-1}S^{(k)}(X_\tjm,\theta)^{-1}
\nn\\&&\hspace{150pt}\cdot
\big[\big(\Delta_jY^{(k)}\big)^{\otimes2}K^{(k)}_{n,j}
-\big(\sigma^{(k)}(X_\tjm,\theta^*)\Delta_jw^{(k)}\big)^{\otimes2}
\big].
\eeas
Apply Lemma \ref{300217-8} to $\partial_\theta^i\bbR_{n}^\circ(\theta;\alpha)$ ($i=0,...,3$) 
to obtain 
\beas 
\sum_{i=0}^3\bigg\|\sup_{\theta\in\Theta}\big|\partial_\theta^in^{-1}\bbR_{n}^\circ(\theta;\alpha)\big|\bigg\|_p
&<&\infty
\eeas
for every $p>1$. 
Moreover, 
we apply Sobolev's inequality, Lemma {\fred\ref{300217-5} (ii)} and Lemma \ref{300217-6} (ii). 
Then it is sufficient to show that 
\bea\label{300326-1} 
\sum_{i=0}^4\sup_{\theta\in\Theta}\bigg\{\big\|\partial_\theta^i {\colorb n^{-1}} \bbH_{n}^\times(\theta;\alpha)\big\|_p
+\big\|\partial_\theta^i {\colorb n^{-1}} \bbM_{n}^\times(\theta;\alpha)\big\|_p
\bigg\}
&<&
\infty
\eea
for proving the lemma, where 
$\bbH_{n}^\times(\theta;\alpha)$ and $\bbM_{n}^\times(\theta;\alpha)$ 
are 
defined by the same formula as 
$\bbH_{n}^\circ(\theta;\alpha)$ and $\bbM_{n}^\circ(\theta;\alpha)$
, respectively,  
but with $\widetilde{\calj}^{(k)}_n(\alpha^{(k)})$ in place of $\calj^{(k)}_n(\alpha^{(k)})$. 
However, (\ref{300326-1}) is obvious. 
\qed\halflineskip

\begin{lemma}\label{300221-A1(2)}
For every $p\geq1$, 
\beas
\sup_{n\in\bbN}E\bigg[\bigg(
n^{2\beta\beta_1}\big|\Gamma_n(\alpha)-\Gamma%
\big|\bigg)^p
\bigg] &<& \infty. 
\eeas
\end{lemma}
\proof 
Consider the decomposition 
$\Gamma_{n}(\alpha) =\Gamma_n^*+M_n^*+R_n^*$ with 
\beas 
\Gamma_n^*
&=& 
\frac{1}{2n}\sum_{k=1}^\sfk\sum_{j\in\calj^{(k)}_n(\alpha^{(k)})}
p(\alpha^{(k)})^{-1}\bigg\{
\partial_\theta^2\log\det S^{(k)}(X_\tjm,\theta^*)
\nn\\&&\hspace{150pt}
+\big(\partial_\theta^2(S^{(k)}\>^{-1})\big)(X_\tjm,\theta^*)\big[
S(X_\tjm,\theta^*)
\big]
\bigg\},
\eeas
\beas 
M_n^*
&=& 
\frac{1}{2n}\sum_{k=1}^\sfk\sum_{j\in\calj^{(k)}_n(\alpha^{(k)})}
\big(\partial_\theta^2(S^{(k)}\>^{-1})\big)(X_\tjm,\theta^*)\bigg[
q^{(k)}(\alpha^{(k)})^{-1}
h^{-1}\big(\sigma^{(k)}(X_\tjm,\theta^*)\Delta_jw^{(k)}\big)^{\otimes2}
\\&&
-p(\alpha^{(k)})^{-1}S(X_\tjm,\theta^*)
\bigg]
\eeas
and 
\beas 
R_n^*
&=& 
{\colorb \frac{1}{2n}} \sum_{k=1}^\sfk\sum_{j\in\calj^{(k)}_n(\alpha^{(k)})}
q^{(k)}(\alpha^{(k)})^{-1} {\colorb h^{-1}} \big(\partial_\theta^2(S^{(k)}\>^{-1})\big)(X_\tjm,\theta^*)
\\&&\cdot
\bigg[\big(\Delta_jY^{(k)}\big)^{\otimes2}K^{(k)}_{n,j}
-\big(\sigma^{(k)}(X_\tjm,\theta^*)\Delta_jw^{(k)}\big)^{\otimes2}
\bigg].
\eeas

Since $2\beta\beta_1<\ {\colorb \gamma_2}$, we obtain 
\beas
\sup_{n\in\bbN}\big\|
n^{2\beta\beta_1}\big|\Gamma^*_n-\Gamma\big|
\big\|_p
&<& \infty
\eeas
by Lemma \ref{300217-9}, and also obtain  
\beas
\sup_{n\in\bbN}\big\|
n^{2\beta\beta_1}\big|R^*_n\big|
\big\|_p
&<& \infty
\eeas
by Lemma \ref{300217-8} for every $p>1$. 
Moreover, by {\fred Lemmas \ref{300217-5} (ii)} and \ref{300217-6} (ii) 
{\colorg applied to $2\beta\beta_1(<\gamma_2)$ for ``$\gamma_3$''}, 
we replace $\calj^{(k)}_n(\alpha^{(k)})$ in the expression of $M^*_n$ by $\widetilde{\calj}^{(k)}_n(\alpha^{(k)})$ 
and then apply the Burkholder-Davis-Gundy inequality to show 
\beas
\sup_{n\in\bbN}\big\|
n^{2\beta\beta_1}\big|M^*_n\big|
\big\|_p
&<& \infty
\eeas
for every $p>1$. 
This completes the proof. 
\qed\halflineskip

The following two lemmas are obvious under $[F3]$. 
\begin{lemma}\label{300221-A2}
For every $p\geq1$, there exists a constant $C_p$ such that 
\beas 
P\big[\lambda_{min}(\Gamma)<r^{-\rho_1}\big]&\leq& \frac{C_p}{r^p}
\eeas
for all $r>0$, 
where $\lambda_{min}(\Gamma)$ denotes the minimum eigenvalue of $\Gamma$. 
\end{lemma}
\halflineskip

\begin{lemma}\label{300221-A5}
For every $p\geq1$, there exists a constant $C_p$ such that 
\beas 
P\big[\chi_0<r^{-{\colorg (}\rho_2-2\eta{\colorg )}}\big]&\leq& \frac{C_p}{r^p}
\eeas
for all $r>0$. 
\end{lemma}
\halflineskip

\begin{lemma}\label{300221-A6(1)}
For every $p\geq1$, 
\beas 
\sup_{n\in\bbN}E\big[
\big|\Delta_n(\alpha,\beta)\big|^p\big]&<& \infty. 
\eeas
\end{lemma}
\proof 
We consider the decomposition 
$\Delta_n(\alpha,\beta)=
n^{-1+\beta}\partial_\theta\bbH_n(\theta^*;\alpha) =M_n^\vee+R_n^\vee$ with 
\beas 
M_n^\vee
&=& 
-\frac{n^{\beta}}{2{\colorg n}}\sum_{k=1}^\sfk\sum_{j\in\calj^{(k)}_n(\alpha^{(k)})}
\big(\partial_\theta(S^{(k)}\>^{-1})\big)(X_\tjm,\theta^*)
\\&&
\cdot
\bigg[
q^{(k)}(\alpha^{(k)})^{-1}
{\colorg h^{-1}}
\big(\sigma^{(k)}(X_\tjm,\theta^*)\Delta_jw^{(k)}\big)^{\otimes2}
-p(\alpha^{(k)})^{-1}S(X_\tjm,\theta^*)
\bigg]
\eeas
and 
\beas 
R_n^\vee
&=& 
-\frac{n^{\beta}}{2{\colorg n}}\sum_{k=1}^\sfk\sum_{j\in\calj^{(k)}_n(\alpha^{(k)})}
q^{(k)}(\alpha^{(k)})^{-1}{\colorg h^{-1}}
\\&&\times
\big(\partial_\theta(S^{(k)}\>^{-1})\big)(X_\tjm,\theta^*)
\bigg[\big(\Delta_jY^{(k)}\big)^{\otimes2}K^{(k)}_{n,j}
-\big(\sigma^{(k)}(X_\tjm,\theta^*)\Delta_jw^{(k)}\big)^{\otimes2}
\bigg].
\eeas
We see 
$
\sup_{n\in\bbN}\big\|
R_n^\vee(\alpha,\beta)\big\|_p< \infty
$
by Lemma \ref{300217-8}. 
Moreover 
$\sup_{n\in\bbN}\big\|
M_n^\vee(\alpha,\beta)\big\|_p< \infty
$
by Lemmas {\fred\ref{300217-5} (ii)} and \ref{300217-6} (ii) 
and the Burkholder-Davis-Gundy inequality. 
{\tred 
We note that symmetry between the components of $W^{(k)}_j$ is available. 
}
\qed

\halflineskip
\noindent
{\colorg 
As a matter of fact, $\Delta_n(\alpha,\beta)$ converges to $0$, as seen 
in the proof of Lemma \ref{300221-A6(1)}. 
The location shift of the random field $\bbZ^\beta_n(\cdot;\alpha)$ 
asymptotically vanishes. 
\halflineskip}

\begin{lemma}\label{300221-A6(2)}
For every $p\geq1$, 
\beas 
\sup_{n\in\bbN}E\bigg[\bigg(
\sup_{\theta\in\Theta}n^{\beta(1-2\beta_2)}
\big|\bbY_n(\theta;\alpha)-\bbY(\theta)\big|
\bigg)^p\bigg]&<& \infty.
\eeas
\end{lemma}
%
\proof 
In this situation, we use the decomposition 
\beas 
\bbY_n(\theta;\alpha)
&=&
\bbY_n^+(\theta;\alpha)+\bbM_n^+(\theta;\alpha)+\bbR_n^+(\theta;\alpha)
\eeas
with 
\beas 
{\colorg \bbY^+_n}(\theta;\alpha) 
&=& 
-\frac{1}{2n}\sum_{k=1}^\sfk\sum_{j\in\calj^{(k)}_n(\alpha^{(k)})}p(\alpha^{(k)})^{-1}\bigg\{
\text{Tr}\bigg(S^{(k)}(X_\tjm,\theta)^{-1}S^{(k)}(X_\tjm,\theta^*)-I_{\sfm_k}\bigg)
\\&&
+\log\frac{\det S^{(k)}(X_\tjm,\theta)}{\det S^{(k)}(X_\tjm,\theta^*)}\bigg\},
\eeas
\beas 
\bbM^+_{n}(\theta;\alpha) 
&=& 
-\frac{1}{2n}\sum_{k=1}^\sfk\sum_{j\in\calj^{(k)}_n(\alpha^{(k)})}
\bigg(S^{(k)}(X_\tjm,\theta)^{-1}-S^{(k)}(X_\tjm,\theta^*)^{-1}\bigg)
\\&&\cdot
\bigg[q^{(k)}(\alpha^{(k)})^{-1}h^{-1}\big(\sigma^{(k)}(X_\tjm,\theta^*)\Delta_jw^{(k)}\big)^{\otimes2}
-p(\alpha^{(k)})^{-1}S^{(k)}(X_\tjm,\theta^*)\bigg]
\eeas
and
\beas 
\bbR^+_{n}(\theta;\alpha) 
&=& 
-{\colorg\frac{1}{2n}}\sum_{k=1}^\sfk\sum_{j\in\calj^{(k)}_n(\alpha^{(k)})}
q^{(k)}(\alpha^{(k)})^{-1}
{\colorg h^{-1}}
\bigg(S^{(k)}(X_\tjm,\theta)^{-1}-S^{(k)}(X_\tjm,\theta^*)^{-1}\bigg)
\\&&\cdot
\bigg[\big(\Delta_jY^{(k)}\big)^{\otimes2}K^{(k)}_{n,j}
-\big(\sigma^{(k)}(X_\tjm,\theta^*)\Delta_jw^{(k)}\big)^{\otimes2}
\bigg]. 
\eeas
As assumed, $\beta(1-2\beta_2)<{\colorg \gamma_2}\leq1/2$. 
Lemma \ref{300217-8} gives 
\beas 
\sup_{n\in\bbN}E\bigg[\bigg(
\sup_{\theta\in\Theta}n^{\beta(1-2\beta_2)}
\big|\bbR^+_n(\theta;\alpha)\big|
\bigg)^p\bigg]&<& \infty
\eeas
for every $p>1$. 
Furthermore, Lemma \ref{300217-9} gives 
\beas 
\sup_{n\in\bbN}E\bigg[\bigg(
\sup_{\theta\in\Theta}n^{\beta(1-2\beta_2)}
\big|\bbY^+_n(\theta;\alpha)-\bbY(\theta)\big|
\bigg)^p\bigg]&<& \infty.
\eeas
On the other hand, 
Lemmas {\fred\ref{300217-5} (ii)} and \ref{300217-6} (ii) 
and the Burkholder-Davis-Gundy inequality 
{\colorg together with Sobolev's inequality deduce} 
\beas 
\sup_{n\in\bbN}E\bigg[\bigg(
\sup_{\theta\in\Theta}n^{\beta(1-2\beta_2)}
\big|\bbM^+_n(\theta;\alpha)\big|
\bigg)^p\bigg]&<& \infty
\eeas
for every $p>1$, which completes the proof. 
\qed\halflineskip

\noindent{\it Proof of Theorem \ref{300217-10}.} 
Now Theorem \ref{300217-10} follows from Theorem 2 of Yoshida 
\cite{yoshida2011} 
combined with 
Lemmas \ref{300221-A1(1)}, \ref{300221-A1(2)}, \ref{300221-A2}, 
\ref{300221-A5}, \ref{300221-A6(1)} and \ref{300221-A6(2)}.  
\qed\halflineskip


\begin{en-text}
\begin{remark}\rm 
Localization. Direct application of U-Y (2013) (before the first jump of $X$) to obtain $\chi_0>0$ a.s. 
Or direct proof of non-degeneracy of $\chi_0$ ($\chi_0^{-1}\in L^\inftym$) for jump process $X$. 
\end{remark}
\end{en-text}
\begin{en-text}
\subsection{Characteristic chart}
\bi
\im draw estimated values vs. $\alpha$
\ei
\end{en-text}

\section{Global filter with moving threshold}\label{300615-3}
\subsection{Quasi likelihood function with moving quantiles} 
Though the threshold method presented in the previous section removes 
jumps surely, it is conservative and does not attain 
{\rev the optimal rate of convergence that is attained by the QLA estimators 
(i.e. QMLE and QBE) 
in the case without jumps.}
On the other hand, it is possible to give more efficient estimators 
by aggressively taking bigger increments while it may {\tred cause} 
{\colorg miss-detection} of certain portion of jumps.  

Let $\delta_0\in(0,1/4)$ and $\delta_1^{(k)}\in(0,1/2)$. 
For simplicity, let $s_n^{(k)}=n-B^{(k)}\lfloor n^{\delta_1^{(k)}}\rfloor$ with positive constants $B^{(k)}$. 
Let $\alpha_n^{(k)}=1-s_n^{(k)}/n$ and $\alpha_n=(\alpha_n^{(1)},...,\alpha_n^{(\sfk)})$. 
Let 
\beas 
\calk^{(k)}_n&=&\big\{j\in\{1,...,n\};
V^{(k)}_j<V^{(k)}_{(s_n^{(k)})}\big\}
\eeas
where 
\beas 
V^{(k)}_j &=& |({\mathfrak S}^{(k)}_{n,j-1})^{-1/2}\Delta_jY^{(k)}|
\eeas
with some positive definite random matrix ${\mathfrak S}^{(k)}_{n,j-1}$, 
and $V^{(k)}_{(j)}$ is the $j$-th order statistic of $V^{(k)}_1,...,V^{(k)}_n$.

\def\scH{\stackrel{\circ}{\bbH}\!}
We consider a random field by removing increments of $Y$ including jumps from 
the full quasi-likelihood function. 
Define $\bbH_n(\theta)$ by 
\bea\label{300225-21} 
\bbH_n(\theta)
&=&
-\half\sum_{k=1}^\sfk\sum_{j\in\calk^{(k)}_n}\bigg\{
(q^{(k)}_n)^{-1}
h^{-1}S^{(k)}(X_\tjm,\theta)^{-1}\big[\big(\Delta_jY^{(k)}\big)^{\otimes2}\big]
K^{(k)}_{n,j}
\nn\\&&
+(p^{(k)}_n)^{-1}
\log\det S^{(k)}(X_\tjm,\theta)\bigg\}.
\eea
\begin{remark}\rm 
The truncation functional $K^{(k)}_{n,j}$ is given by (\ref{300211-5}). 
It is also reasonable to set it as 
\beas 
K^{(k)}_{n,j} &=& 
1_{\big\{V^{(k)}_j<C^{(k)}_*n^{-\frac{1}{4}-\delta_0}\big\}},
\eeas
where $C^{(k)}_*$ is an arbitrarily given positive constant. 
\end{remark}
\begin{remark}\rm 
The threshold is larger than $n^{-\half+0}$. 
The truncation {\colorg $K^{(k)}_{n,j}$} 
is for stabilizing the increments of $Y$, not for filtering. 
The factors ${\mathfrak S}^{(k)}_{n,j-1}$, $q^{(k)}_n$ and $p^{(k)}_n$ 
can freely be chosen if ${\mathfrak S}^{(k)}_{n,j-1}$ and its inverse are uniformly bounded in $L^\inftym$ 
and if $q^{(k)}_n$ and $p^{(k)}_n$ are sufficiently close to $1$. 
$\bar{S}^{(k)}_{n,j-1}$, $q^{(k)}(\alpha_n^{(k)})$ and $p(\alpha^{(k)}_n)$ are natural choices 
for ${\mathfrak S}^{(k)}_{n,j-1}$, $q^{(k)}_n$ and $p^{(k)}_n$, respectively. 
Asymptotic theoretically, the factors $(q^{(k)}_n)^{-1}$ and $(p^{(k)}_n)^{-1}$ can be replaced by $1$, 
and one can take ${\mathfrak S}^{(k)}_{n,j-1}=I_{{\rev \sf m_k}}$
{\tred; see Condition $[F2']$ below}. 
{\colorg Thus a} modification of $\bbH_n(\theta)$ is 
$\scH_n(\theta)$ defined by 
\beas 
\scH_n(\theta)
&=&
-\half\sum_{k=1}^\sfk\sum_{j\in\calk^{(k)}_n}\bigg\{
h^{-1}S^{(k)}(X_\tjm,\theta)^{-1}\big[\big(\Delta_jY^{(k)}\big)^{\otimes2}\big]
K^{(k)}_{n,j}
\\&&
+
\log\det S^{(k)}(X_\tjm,\theta)\bigg\}
\eeas
{\colorg with $\calk^{(k)}_n$ for $V^{(k)}_j=|\Delta_jY^{(k)}|$.}
The quasi-log likelihood function $\scH_n$ gives the same asymptotic results as $\bbH_n$.  
\end{remark}
\halflineskip

We denote by $\hat{\theta}^{M,\alpha_n}_n$ a QMLE of $\theta$ with respect to $\bbH_n$ given by (\ref{300225-21}). 
{\colorg We should remark that $\hat{\theta}^{M,\alpha_n}_n$ defined by $\bbH_n(\theta)$ 
can differ from $\hat{\theta}^{M,\alpha}_n$ previously defined by $\bbH_n(\theta;\alpha)$.} 
The {\bf quasi-Bayesian estimator} (QBE)  $\hat{\theta}^{B,\alpha_n}_n$ 
of $\theta$ is defined by 
\beas 
\hat{\theta}^{B,\alpha_n}_n
&=& 
\bigg[\int_\Theta\exp\big(\bbH_n(\theta)\big)\varpi(\theta)d\theta\bigg]^{-1}
\int_\Theta\theta\exp\big(\bbH_n(\theta)\big)\varpi(\theta)d\theta,
\eeas
where $\varpi$ is a continuous function on $\Theta$ satisfying 
$0<\inf_{\theta\in\Theta}\varpi(\theta)\leq\sup_{\theta\in\Theta}\varpi(\theta)<\infty$.

\subsection{Polynomial type large deviation inequality}
Let $\bbU_n=\{u\in\bbR^\sfp;\>\theta^*+n^{-1/2}u\in\Theta\}$. 
Let $\bbV_n(r)=\{u\in\bbU_n;\>|u|\geq r\}$. 
We define 
the quasi-likelihood ratio random field {\colorg $\bbZ_n$} by 
\beas 
\bbZ_n(u) &=& 
\exp\bigg\{\bbH_n(\theta^*+n^{-1/2}u)-\bbH_n(\theta^*)\bigg\}
\qquad(u\in\bbU_n).
\eeas

\bd
\im[[F2$'$\!\!]] 
{\bf (i)} The positive-definite measurable random matrices ${\mathfrak S}^{(k)}_{n,j-1}$ 
($k\in\{1,...,\sfk\},\>n\in\bbN,\>j\in\{1,...,n\}$) satisfy 
\beas 
\sup_{k\in\{1,...,\sfk\}\atop\>n\in\bbN,\>j\in\{1,...,n\}}\big(\|{\mathfrak S}^{(k)}_{n,j-1}\|_p
+\|({\mathfrak S}^{(k)}_{n,j-1})^{-1}\|_p
\big)&<&\infty
\eeas
for every $p>1$. 
\bd
\im[(ii)] Positive numbers $q^{(k)}_n$ and $p_n^{(k)}$ satisfy 
$|q^{(k)}_n-1|=o(n^{-1/2})$ 
and 
$|1-p_n^{(k)}|=o(n^{-1/2})$. 
\ed
\ed
\halflineskip

{\colorr
A polynomial type large deviation inequality is given by 
the following theorem, a proof of which is in Section \ref{300430-1}. }
\begin{theorem}\label{300224-10}
Suppose that $[F1]_4$, $[F2']$ and $[F3]$ are fulfilled. 
Let $c_0\in(1,2)$. 
Then, for every positive number $L$, 
there exists a constant $C(c_0,L)$ such that 
\beas 
P\bigg[\sup_{u\in\bbV_n(r)}\bbZ_n(u)\geq e^{-r^{c_0}}\bigg] &\leq& \frac{C(c_0,L)}{r^L}
\eeas
for all $r>0$ and $n\in\bbN$. 
\end{theorem}

The polynomial type large deviation inequality for $\bbZ_n$ in Theorem \ref{300224-10} ensures 
$L^\inftym$-boundedness of the QLA estimators. 
\begin{proposition}\label{300615-16}
Suppose that $[F1]_4$, $[F2']$ and $[F3]$ are satisfied. Then 
\beas 
\sup_{n\in\bbN}\big\|\sqrt{n}\big(\hat{\theta}^{{\sf A},\alpha_n}_n-\theta^*\big)\big\|_p\><\> \infty
\qquad({\sf A}=M,B)
\eeas 
for every $p>1$. 
\end{proposition}

\subsection{Proof of Theorem \ref{300224-10}}\label{300430-1}
\def\tH{\widetilde{\bbH}}
\def\daggerH{\bbH^\dagger}
{\tred Recall $\tY^{(k)}=Y^{(k)}-J^{(k)}$.} Let 
\beas 
\tH_n(\theta)
&=&
-\half\sum_{k=1}^\sfk\sum_{j=1}^n\bigg\{
h^{-1}S^{(k)}(X_\tjm,\theta)^{-1}[\big(\Delta_j\tY^{(k)}\big)^{\otimes2}\big]
+
\log\det S^{(k)}(X_\tjm,\theta)\bigg\}.
\eeas
\begin{lemma}\label{300225-11}
For every $p\geq1$, 
\bea\label{300223-2} 
\sum_{i=0}^4\sup_{\theta\in\Theta}
\bigg\|n^{-1/2}\partial_\theta^i\bbH_n(\theta)-n^{-1/2}\partial_\theta^i\tH_n(\theta)\bigg\|_p
&\to& 0
\eea
as $n\to\infty$. 
\end{lemma}
\proof
Let 
\beas 
{\mathfrak A}^{(k)}_n 
&=& 
{\tred 
\bigcup_{j=1}^n\bigg[\big\{ j\in(\calk^{(k)}_n)^c\big\}\cap\big\{\Delta_jN^{(k)}=0\big\}\bigg].
}
\eeas
Let 
\beas 
{\mathfrak B}^{(k)}_n 
&=& 
\bigcap_{j=1}^n\bigg[\big\{V^{(k)}_j\geq V^{(k)}_{(s_n)}\big\}\cup\big\{|\Delta_jJ^{(k)}|\leq n^{-\frac{1}{4}-\delta_0}\big\}\bigg].
\eeas
For $\omega\in{\mathfrak A}^{(k)}_n \cap ({\mathfrak B}^{(k)}_n)^c$, 
there exists $j(\omega)\in(\calk^{(k)}_n)^c$ such that $\Delta_{j(\omega)}N^{(k)}(\omega)=0$, and 
also there exists $j'(\omega)\in\{1,...,n\}$ such that 
$V^{(k)}_{j'(\omega)}(\omega)<V^{(k)}_{(s_n)}(\omega)$ and 
$|\Delta_{j'(\omega)}J^{(k)}(\omega)|>n^{-\frac{1}{4}-\delta_0}$. 
Then 
\beas &&
\bigg|({\mathfrak S}^{(k)}_{n,j'(\omega)-1})^{-1/2}\Delta_{j'(\omega)}J^{(k)}(\omega)\bigg|
- \bigg|({\mathfrak S}^{(k)}_{n,j'(\omega)-1}(\omega))^{-1/2}
 \Delta_{j'(\omega)}\tY^{(k)}(\omega)
 \bigg|
\\&\leq&
V^{(k)}_{j'(\omega)}(\omega)\><\>V^{(k)}_{j(\omega)}(\omega)
\yeq
 \bigg|({\mathfrak S}^{(k)}_{n,j(\omega)-1}(\omega))^{-1/2}
\Delta_{j(\omega)}\tY^{(k)}(\omega)
\bigg|
\eeas
and hence 
\beas 
n^{-\frac{1}{4}-\delta_0}
&\leq& 
2\big|{\mathfrak S}^{(k)}_{n,j'(\omega)-1}\big|^{1/2}
\max_{j=1,..,n}
 \bigg|({\mathfrak S}^{(k)}_{n,j-1}(\omega))^{-1/2}
\Delta_j\tY^{(k)}(\omega)
\bigg|
 \eeas
 where $\big|M\big|=\{\text{Tr}(MM^\star)\}^{1/2}$ for a matrix $M$. 
 Since 
 $\big\{h^{-1/2}\big|\Delta_j\tY^{(k)}\big|;\>j=1,...,n,\>n\in\bbN\big\}$
 is bounded in $L^\inftym$, 
 we obtain 
\beas 
P\big[{\mathfrak A}^{(k)}_n \cap ({\mathfrak B}^{(k)}_n)^c\big]
&=&
O(n^{-L})
\eeas
as $n\to\infty$ for every $L>0$. 
Moreover, $P[({\mathfrak A}^{(k)}_n)^c]=O(n^{-L})$ from the assumption for $N^{(k)}$ 
{\tred 
since 
\beas 
\big({\mathfrak A}^{(k)}_n\big)^c
&\subset&
\bigg\{\#\big\{j\in\{1,...,n\};\>\Delta_jN^{(k)}\not=0\big\}\ygeq n-s_n^{(k)}+1\bigg\}
\>\subset\>
\big\{N_T^{(k)}\geq B^{(k)}n^{\delta_1^{(k)}}\big\}.
\eeas
}\noindent
Thus 
\bea\label{300223-1}
P\bigg[\bigcap_{k=1}^\sfk
{\mathfrak B}^{(k)}_n
\bigg]
&=& 
1-O(n^{-L})
\eea
as $n\to\infty$ for every $L>0$.

Define {\colorg $\daggerH_n(\theta)$} by 
\beas 
\daggerH_n(\theta)
&=&
-\half\sum_{k=1}^\sfk\sum_{j\in\calk^{(k)}_n}\bigg\{
{\colorg (q^{(k)}_n)^{-1}}
h^{-1}S^{(k)}(X_\tjm,\theta)^{-1}[\big(
\Delta_jY^{(k)}-\Delta_jJ^{(k)}
\big)^{\otimes2}\big]
K^{(k)}_{n,j}
{\colorg 1_{\big\{|\Delta_jJ^{(k)}|\leq 1\big\}}}
\\&&
+{\colorg (p^{(k)}_n)^{-1}}
\log\det S^{(k)}(X_\tjm,\theta)\bigg\},
\eeas
{\colorg where 
the indicator function controls  
the moment outside of $\cap_{k=1}^\sfk{\mathfrak B}^{(k)}_n$. }
Then by (\ref{300223-1}), the cap {\colorg and $N_T\in L^\infm$}, we obtain 
\beas 
\sum_{i=0}^4\sup_{\theta\in\Theta}
\bigg\|n^{-1/2}\partial_\theta^i\bbH_n(\theta)-n^{-1/2}\partial_\theta^i\daggerH_n(\theta)\bigg\|_p
&\to& 0
\eeas
as $n\to\infty$ for every $p\geq1$. 
{\tred Indeed, 
we can estimate this difference of the two variables on the event 
${\mathfrak C}_n:=\cap_{k=1}^\sfk{\mathfrak B}^{(k)}_n$  
and on ${\mathfrak C}_n^c$, as follows. 
On ${\mathfrak C}_n$, 
$|\Delta_jJ^{(k)}|\leq n^{-1/4-\delta_0}1_{\{\Delta_jJ^{(k)}\not=0\}}$ 
whenever $j\in\calk_n^{(k)}$. The cap $K_{n,j}^{(k)}$ also offers 
the estimate $|\Delta_jY^{(k)}|<C^{(k)}_*n^{-1/4}$. 
On ${\mathfrak C}_n$, 
after removing the factor $1_{\big\{|\Delta_jJ^{(k)}|\leq 1\big\}}$ from 
the expression of $n^{-1/2}\partial_\theta^i\daggerH_n(\theta)$ with the help of 
$N_T\in L^\inftym$ and the $L^p$-estimate of $h^{-1}\big|\Delta_j\tY|^2$, 
we can estimate the cross term in the difference with  
\beas &&
n^{-1/2}\sum_{j\in\calk^{(k)}_n}\bigg|
h^{-1}S^{(k)}(X_\tjm,\theta)^{-1}[
\Delta_jY^{(k)}\otimes\Delta_jJ^{(k)}
\big]
K^{(k)}_{n,j}
\bigg|
\\&\leq&
\calm_n^{(k)}n^{-\delta_0}
\sum_{j=1}^n1_{\{\Delta_jJ^{(k)}\not=0\}}
\yleq
\bigg(n^{\delta_0/2}+\calm_n^{(k)}1_{\{\calm_n>n^{\delta_0/2}\}}\bigg)n^{-\delta_0}N_T
\eeas
for $\calm_n^{(k)}=\max_{j=1,...,n}  |S^{(k)}(X_\tjm,\theta)^{-1}|$,
as well as 
the term involving $\big(\Delta_jJ^{(k)}\big)^{\otimes2}$ and admitting a similar estimate. 
Estimation is much simpler on ${\mathfrak C}_n^c$ thanks to (\ref{300223-1}). 
The cap $1_{\big\{|\Delta_jJ^{(k)}|\leq 1\big\}}$ helps. 
}

We know that $\#(\calk^{(k)}_n)^c\sim B^{(k)}n^{\delta_1^{(k)}}$, and {\tred have} assumed that 
$|q^{(k)}_n-1|=o(n^{-1/2})$ 
and that 
$|1-p_n^{(k)}|=o(n^{-1/2})$. 
Then, {\colorg with (\ref{300223-1})}, it is easy to show 
\beas   
\sum_{i=0}^4\sup_{\theta\in\Theta}
\bigg\|n^{-1/2}\partial_\theta^i\daggerH_n(\theta)-n^{-1/2}\partial_\theta^i\tH_n(\theta)\bigg\|_p
&\to& 0,
\eeas
which implies (\ref{300223-2})
as $n\to\infty$ for every $p\geq1$. 
\qed\halflineskip

We choose parameters 
$\eta$, $\beta_1$, $\rho_1$, $\rho_2$ and $\beta_2$ satisfying (\ref{300225-1}) with 
$\beta_2>0$. 
Let 
\beas 
\Delta_n&=&n^{-1/2}\partial_\theta\bbH_n(\theta^*)
\quad
\text{and}
\quad
\Gamma_n\yeq-n^{-1}\partial_\theta^2\bbH_n(\theta^*). 
\eeas
Let 
\beas
\bbY_n(\theta) &=& n^{-1}\big\{\bbH_n(\theta)-\bbH_n(\theta^*)\big\}. 
\eeas

The following two estimates will play a basic role. 
\begin{lemma}\label{300225-UYL5}
Let $f\in C^{0,1}_\up\big(\bbR^\sfd\times\Theta;\bbR^{{\colorb \sfm_k}}\otimes\bbR^{{\colorb \sfm_k}}\big)$. 
Then under ${\colorg [F1]_0}$, 
\beas 
\sup_{n\in\bbN} E\bigg[\bigg(\sup_{\theta\in\Theta}\bigg|
{\colorg n^{\half-\ep}}
\sum_{j=1}^nf(X_\tjm,\theta)
{\colorg 
\bigg[\big(\Delta_j\tY^{(k)}\big)^{\otimes2}
-\big(\sigma^{(k)}(X_\tjm,\theta^*)\Delta_jw^{(k)}\big)^{\otimes2}\bigg]
}
\bigg|\bigg)^p\bigg]
&<&\infty
\eeas
for every $p>1$ {\colorg and $\ep>0$}. 
\end{lemma}
\proof 
{\colorg 
One can validate this lemma in a quite similar way as Lemma \ref{300217-8}. 
\qed\halflineskip
}

\begin{en-text}
{\colorg 
We expand $\Delta_j\tY^{(k)}$ with the formula 
\beas 
\Delta_j\tY^{(k)}
&=& 
\sigma^{(k)}(X_\tjm,\theta^*) \Delta_jw^{(k)}
+\int_\tjm^\tj\big\{\sigma^{(k)}(X_t,\theta^*)-\sigma^{(k)}(X_\tjm,\theta^*)\big\}dw^{(k)}_t
+\int_\tjm^\tj b^{(k)}_tdt
\\&=:&
\xi_{1,j}+\xi_{2,j}+\xi_{3,j}.
\eeas
}
Let $V^n_j(\theta)$ be random variables 
{\colorg depending on $\theta$}
such that \koko
\beas 
\sup_{n\in\bbN;\>j=1,...,n
\atop{\colorg \theta\in\Theta}
}\|\sqrt{n}V^n_j{\colorg (\theta)}\|_p=O(1)\quad \text{and} \quad 
\sup_{n\in\bbN
\atop{\colorg \theta\in\Theta}
}\bigg\|\max_{j=1,...,n}|\sqrt{n}V^n_j{\colorg (\theta)}|\bigg\|_p=O(n^{1/2-\ep_p})
\eeas
for every $p>1$ and some $\ep_p>0$ depending on $p$. 
Then, by the Burkholder-Davis-Gundy inequality and $[F1]_1$ (i), we have 
{\colorg 
\beas&&
\bigg\|\sqrt{n}\sum_{j=1}^n V^n_j(\theta)
\xi_{2,j}\bigg\|_p
\\&\leq&
\sum_{j=1}^n\big\|\sqrt{n}V^n_j(\theta)\big\|_{2p}
\bigg\|\bigg|\int_\tjm^\tj\big\{\sigma^{(k)}(X_t,\theta^*)-\sigma^{(k)}(X_\tjm,\theta^*)\big\}dw^{(k)}_t\bigg|\bigg\|_{2p}
\\&\simleq&
\sum_{j=1}^n\big\|\sqrt{n}V^n_j(\theta)\big\|_{2p}
\bigg( h^{p-1}
\int_\tjm^\tj E\big[\big|\sigma^{(k)}(X_t,\theta^*)-\sigma^{(k)}(X_\tjm,\theta^*)\big|^p\big]dt\bigg)^{1/p}\koko
\\&\simleq&
n^{p/2}E\bigg[\sum_{j=1}^n|V^n_j(\theta)|^p\int_\tjm^\tj
\big|\sigma^{(k)}(X_t,\theta^*)-\sigma^{(k)}(X_\tjm,\theta^*)\big|^pdt\bigg]
\\&\simleq&
O(1)+
n^{p/2}E\bigg[\sum_{j=1}^n|V^n_j(\theta)|^p\int_\tjm^\tj
\big|\sigma^{(k)}(X_t,\theta^*)-\sigma^{(k)}(X_\tjm,\theta^*)\big|^pdt
1_{\{\Delta_jN>0\}}
\bigg]
\\&&\koko
\bigg\|\sqrt{n}\sum_{j=1}^n V^n_j(\theta)
\int_\tjm^\tj\big\{\sigma^{(k)}(X_t,\theta^*)-\sigma^{(k)}(X_\tjm,\theta^*)\big\}dw^{(k)}_t\bigg\|_p
\\&\leq&
\sum_{j=1}^n\big\|\sqrt{n}V^n_j\big\|_{2p}
\bigg\|\bigg|\int_\tjm^\tj\big\{\sigma^{(k)}(X_t,\theta^*)-\sigma^{(k)}(X_\tjm,\theta^*)\big\}dw^{(k)}_t\bigg|
1_{\{\Delta_jN^X=0\}}\bigg\|_{2p}
\\&&
+
\bigg\|\max_{j=1,...,n}|\sqrt{n}V^n_j|\bigg\|_{4p}
\bigg\|\max_{j=1,...,n}\bigg|
\int_\tjm^\tj\big\{\sigma^{(k)}(X_t,\theta^*)-\sigma^{(k)}(X_\tjm,\theta^*)\big\}dw^{(k)}_t\bigg|\bigg\|_{4p}
\big\|E\big[N^X_T\big]\big\|_{2p}
\\&\leq&
O(n\times 1\times n^{-1})+O(n^{1/2-\ep_p}\times n^{-1/2+\ep_p}\times 1)
\\&=&
O(1)
\eeas
}
as $n\to\infty$ 
for every $p\geq2$. 
{\colorg Here 
to estimate 
$\int_\tjm^\tj\big\{\sigma^{(k)}(X_t,\theta^*)-\sigma^{(k)}(X_\tjm,\theta^*)\big\}dw^{(k)}_t$ 
on the event $\{\Delta_jN^X=0\}$, 
introducing the stopping time of the first jump 
}
Then we can show the lemma in the same way as Lemma 5 of Uchida and Yoshida 
\cite{uchida2013quasi}. 
\qed\halflineskip
\end{en-text}

\begin{lemma}\label{300225-UYL4}
Let $p>1$ and $\ep>0$. Let $f\in C^{1,1}_\up(\bbR^\sfd\times\Theta{\colorg ;\bbR})$. 
Suppose that ${\colorg [F1]_{ 0}}$ is satisfied. 
Then 
\beas
\sup_{n\in\bbN}E\bigg[\bigg(
\sup_{\theta\in\Theta}n^{\half-\ep}\bigg|
\frac{1}{n}\sum_{j=1}^nf(X_\tjm,\theta)-\frac{1}{T}\int_0^Tf(X_t,\theta)dt
\bigg|\bigg)^p\bigg]
&<&
\infty.
\eeas
\end{lemma}
\proof 
Let $p>1$. By taking an approach similar to the proof of Lemma \ref{300225-UYL5}, we obtain 
\beas&&
\sup_{\theta\in\Theta}
n^{\half-\ep}\bigg\|
{\colorg h}\sum_{j=1}^nf(X_\tjm,\theta)-\int_0^Tf(X_t,\theta)dt
\bigg\|_p
\\&\leq&
\sup_{\theta\in\Theta}n^{\half-\ep}\sum_{j=1}^n
\bigg\|\bigg|\int_\tjm^\tj\big\{f(X_t,\theta)-f(X_\tjm,\theta)\big\}dt\bigg|
1_{\{\Delta_jN^X=0\}}\bigg\|_p
\\&&
+
\sup_{\theta\in\Theta}n^{\half-\ep}\bigg\|\max_{j=1,...,n}\bigg|
\int_\tjm^\tj\big\{{\colorg f}(X_t,\theta)-{\colorg f}(X_\tjm,\theta)\big\}{\colorg dt}\bigg|\bigg\|_{2p}
\big\|E\big[N^X_T\big]\big\|_{2p}
\\&\leq&
O(n^{\half-\ep}\times n\times n^{-1.5})+o(n^{1/2-\ep}\times n^{-1/2+\ep}\times 1)
\\&=&
o(1)
\eeas
as $n\to\infty$. 
We also have the same estimate for $\partial_\theta f$ in place of $f$. 
Then the Sobolev inequality implies the result. 
\qed\halflineskip

We have the following estimates. 
\begin{lemma}\label{300225-A1(1)}
For every $p\geq1$, 
\beas
\sup_{n\in\bbN}E\bigg[\bigg(n^{-1}\sup_{\theta\in\Theta}\big|\partial_\theta^3\bbH_n(\theta)
\big|\bigg)^p
\bigg] &<& \infty. 
\eeas
\end{lemma}
\proof Applying Lemma \ref{300225-11} and Sobolev's inequality, 
{\colorg one can prove the lemma in a fashion similar to Lemma \ref{300221-A1(1)}.} 
\qed\halflineskip

\begin{lemma}\label{300225-A1(2)}
For every $p\geq1$, 
\beas
\sup_{n\in\bbN}E\big[\big(
n^{\beta_1}\big|\Gamma_n-\Gamma%
\big|\big)^p
\big] &<& \infty. 
\eeas
\end{lemma}
%
\proof 
Thanks to Lemma \ref{300225-11}, it is sufficient to show that 
\bea
\sup_{n\in\bbN}E\big[\big(
n^{\beta_1}\big|\widetilde{\Gamma}_n-\Gamma%
\big|\big)^p
\big] &<& \infty 
\eea
where 
\beas 
\widetilde{\Gamma}_n &=& -n^{-1}\partial_\theta^2\tH_n(\theta^*)
\eeas
Now {\colorg taking a similar way as Lemma \ref{300221-A1(2)},
one can} prove the desired inequality by applying 
Lemmas \ref{300225-UYL5} and \ref{300225-UYL4} as well as the Burkholder-Davis-Gundy inequality. 
\qed

\begin{en-text}
\begin{lemma}\label{300225-A2}
For every $p\geq1$, there exists a constant $C_p$ such that 
\beas 
P\big[\lambda_{min}(\Gamma)<r^{-\rho_1}\big]&\leq& \frac{C_p}{r^p}
\eeas
for all $r>0$. 
\end{lemma}

\begin{lemma}\label{300225-A5}
For every $p\geq1$, there exists a constant $C_p$ such that 
\beas 
P\big[\chi_0<r^{-\rho_2-2\eta}\big]&\leq& \frac{C_p}{r^p}
\eeas
for all $r>0$. 
\end{lemma}
\end{en-text}

\begin{lemma}\label{300225-A6(1)}
For every $p\geq1$, 
$\ds
\sup_{n\in\bbN}E\big[
\big|\Delta_n\big|^p\big]\><\> \infty. 
$
\end{lemma}
\proof 
{\colred 
By Lemma \ref{300225-11}, it suffices to show 
\bea\label{200225-12} 
\sup_{n\in\bbN}E\big[
\big|\widetilde{\Delta}_n\big|^p\big]&<& \infty
\eea
for 
\begin{en-text}
\bea\label{300626-11}
\widetilde{\Delta}_n&=&n^{-1/2}\partial_\theta\tH_n(\theta^*)
\nn\\&=&
\frac{1}{2\sqrt{n}}\sum_{k=1}^\sfk\sum_{j=1}^n
\big(
(S^{(k)})^{-1}(\partial_{\theta}S^{(k)})(S^{(k)})^{-1}\big)(X_\tjm,\theta^*)
\big[D^{(k)}_j\big]
\eea
where 
\beas 
D^{(k)}_j 
&=& 
h^{-1}\big(\Delta_j\tY^{(k)}\big)^{\otimes2}-S^{(k)}(X_\tjm,\theta^*).
\eeas
\koko
Let 
\end{en-text}
\bea\label{300626-11}
\widetilde{\Delta}_n&=&n^{-1/2}\partial_\theta\tH_n(\theta^*)
\yeq
\frac{1}{{\tred 2}\sqrt{n}}\sum_{k=1}^\sfk\sum_{j=1}^n
f_\tjm
\big[D^{(k)}_j\big]
\eea
where 
\beas 
f_\tjm&=&
{\tred \big((S^{(k)})^{-1}(\partial_{\theta}S^{(k)})(S^{(k)})^{-1}\big)(X_\tjm,\theta^*)}
\eeas
and
\beas 
D^{(k)}_j 
&=& 
h^{-1}\big(\Delta_j\tY^{(k)}\big)^{\otimes2}-S^{(k)}(X_\tjm,\theta^*).
\eeas
\begin{en-text}
By Lemma \ref{300225-11}, it sufficient to verify
\bea\label{0109060436} 
\sup_{n\in\bbN}E\big[
\big|\widetilde{\Delta}_n\big|^p\big]&<& \infty.
\eea
\end{en-text}

We have $N^X_T\in L^\infm$ and 
\beas
\bigg\|\max_{j=1,...,n}\big|f_\tjm\big[D^{(k)}_j\big]\big|\bigg\|_p=O(n^{1/4})
\eeas
for every $p>1$. 
Therefore  
\beas 
\bigg\|n^{-1/2}\sum_{j=1}^nf_\tjm\big[D^{(k)}_j\big]\bigg\|_p
&=&
\bigg\|n^{-1/2}\sum_{j=1}^n1_{\{\Delta_jN^X=0\}}f_\tjm\big[D^{(k)}_j\big]\bigg\|_p
+o(1)
\eeas
for every $p>1$. 
In this situation, it suffices to show that 
\bea\label{0109060315} 
\bigg\|n^{-1/2}\sum_{j=1}^n1_{\{\Delta_jN^X=0\}}f_\tjm\big[D^{(k)}_j\big]\bigg\|_p
&=&
O(1)
\eea
as $n\to\infty$ 
for every $p>1$. 
%

Now, we have the equality 
\beas 
1_{\{\Delta_jN^X=0\}}\Delta_j\tY^{(k)}
&=&
1_{\{\Delta_jN^X=0\}}\big(\Xi_{1,j}+\Xi_{2,j}+\Xi_{3,j}\big),
\eeas
where 
\beas 
\Xi_{1,j}&=&\sigma^{(k)}(X_\tjm,\theta^*) \Delta_jw^{(k)},
\\
\Xi_{2,j}&=&\int_\tjm^\tj\big\{
\sigma^{(k)}(X_\tjm+\widetilde{X}_t-\widetilde{X}_\tjm,\theta^*)
-\sigma^{(k)}(X_\tjm,\theta^*)\big\}dw^{(k)}_t,
\\
\Xi_{3,j}&=&\int_\tjm^\tj b^{(k)}_tdt.
\eeas
Define $C(x,y)$ by 
\beas 
C(x,y)&=&\bigg|\int_0^1\partial_x\sigma^{(k)}(x+r(y-x),\theta^*)dr\bigg|. 
\eeas
Then, by the same reason as in (\ref{0109060315}), and by It\^o's formula 
and the Burkholder-Davis-Gundy inequality, 
\beas &&
\bigg\|n^{-1/2}\sum_{j=1}^n1_{\{\Delta_jN^X=0\}}h^{-1}f_\tjm\big[
\Xi_{1,j}\otimes\Xi_{2,j}
\big]\bigg\|_p
\\&=&
\bigg\|n^{-1/2}\sum_{j=1}^nh^{-1}f_\tjm\big[
\Xi_{1,j}\otimes\Xi_{2,j}
\big]\bigg\|_p+o(1)
\\&\simleq&
\bigg\|n^{-1/2}\sum_{j=1}^nh^{-1}|f_\tjm|
\big|\sigma^{(k)}(X_\tjm,\theta^*)\big|
\\&&\times
\int_\tjm^\tj
\big|\sigma^{(k)}(X_\tjm+\widetilde{X}_t-\widetilde{X}_\tjm,\theta^*)
-\sigma^{(k)}(X_\tjm,\theta^*)\big|dt
\bigg\|_p+O(1)
\eeas
and the last expression is not greater than 
\beas&&
\bigg\|n^{-1/2}\sum_{j=1}^nh^{-1}|f_\tjm|
\big|\sigma^{(k)}(X_\tjm,\theta^*)\big|\int_\tjm^\tj
C(X_\tjm,\widetilde{X}_t-\widetilde{X}_\tjm)
\big|\widetilde{X}_t-\widetilde{X}_\tjm
\big|dt
\bigg\|_p+O(1)
\\&\simleq&
n^{-1/2}\sum_{j=1}^nh^{-1}\int_\tjm^\tj
\bigg\||f_\tjm|
\big|\sigma^{(k)}(X_\tjm,\theta^*)\big|
C(X_\tjm,\widetilde{X}_t-\widetilde{X}_\tjm)
\big|\widetilde{X}_t-\widetilde{X}_\tjm
\big|
\bigg\|_pdt+O(1)
\\&\simleq&
n^{-1/2}\sum_{j=1}^n
\sup_{t\in[\tjm,\tj]}\big\|\widetilde{X}_t-\widetilde{X}_\tjm\big\|_{2p}
\sup_{t\in[\tjm,\tj]\atop j=1,...,n}\bigg\||f_\tjm|
\big|\sigma^{(k)}(X_\tjm,\theta^*)\big|
C(X_\tjm,\widetilde{X}_t-\widetilde{X}_\tjm)
\bigg\|_{2p}
\nn\\&&
+O(1)
\\&=&
O(1)
\eeas
for $p>1$ since 
$\big\|\widetilde{X}_t-\widetilde{X}_\tjm\big\|_{2p}\leq C_{2p}n^{-1/2}$ and 
$\sup_{t\in[0,T]}\|X_t\|_p+\sup_{t\in[0,T]}\|\widetilde{X}_t\|_p<\infty$ 
by the continuity of the mapping $t\mapsto \widetilde{X}_t\in L^{p}$ 
for every $p>1$. 
In a similar manner, we obtain 
\beas 
\bigg\|n^{-1/2}\sum_{j=1}^n1_{\{\Delta_jN^X=0\}}h^{-1}f_\tjm\big[
\Xi_{i_1,j}\otimes\Xi_{i_2,j}
\big]\bigg\|_p
&=&
O(1)
\eeas
for every $p>1$ and $(i_1,i_2)\in\{1,2,3\}^2\setminus\{(1,1)\}$. 
 Finally, for $(i_1,i_2)=(1,1)$, 
\beas&& 
\bigg\|n^{-1/2}\sum_{j=1}^n1_{\{\Delta_jN^X=0\}}f_\tjm\big[
h^{-1}\Xi_{1,j}\otimes\Xi_{1,j}-S^{(k)}(X_\tjm,\theta^*)
\big]\bigg\|_p
\\&=&
\bigg\|n^{-1/2}\sum_{j=1}^nf_\tjm\big[
h^{-1}\Xi_{1,j}\otimes\Xi_{1,j}-S^{(k)}(X_\tjm,\theta^*)
\big]\bigg\|_p
+o(1)
\\&=&
O(1)
\eeas
by the Burkholder-Davis-Gundy inequality. 
Therefore we obtained (\ref{0109060315}) and hence 
(\ref{200225-12}). 
}
\qed\halflineskip

\begin{lemma}\label{300225-A6(2)}
For every $p\geq1$, 
\beas 
\sup_{n\in\bbN}E\bigg[\bigg(
\sup_{\theta\in\Theta}n^{\half-\beta_2}
\big|\bbY_n(\theta)-\bbY(\theta)\big|
\bigg)^p\bigg]&<& \infty.
\eeas
\end{lemma}
\proof 
We use Lemmas \ref{300225-11}, \ref{300225-UYL5} and \ref{300225-UYL4} besides 
the Burkholder-Davis-Gundy inequality and Sobolev's inequality. 
Then the proof is similar to {\colorg Lemma \ref{300221-A6(2)} and also to} 
Lemma 6 of Uchida and Yoshida 
\cite{uchida2013quasi}. 
\qed\halflineskip

\noindent{\it Proof of Theorem \ref{300224-10}.} 
The result follows from Theorem 2 of Yoshida 
\cite{yoshida2011} 
with the aid of Lemmas \ref{300221-A2}, \ref{300221-A5}, 
\ref{300225-A1(1)}, \ref{300225-A1(2)}, \ref{300225-A6(1)} and \ref{300225-A6(2)}. 
\qed\halflineskip


\subsection{Limit theorem and convergence of moments}
In this section, asymptotic mixed normality of the QMLE and QBE will be established. 
\bd
\im[[F1$'$\!\!]]$_\kappa$ 
Conditions $(ii)$, $(iii)$ and $(iv)$ of $[F1]_\kappa$ are satisfied in addition to 
\bd\im[(i)] 
the process $X$ has a representation 
{\colorg 
\beas 
X_t &=& X_0+\int_0^t\tilde{b}_sds
+\int_0^t\tilde{a}_sd\tilde{w}_s+J^X_t
\qquad(t\in[0,T])
\eeas
where $J^X=(J^X_t)_{t\in[0,T]}$ is a \cadlag adapted pure jump process, 
$\tilde{w}=(\tilde{w}_t)_{t\in[0,T]}$ is an $\sfr_1$-dimensional $\F$-Wiener process, 
$\tilde{b}=(\tilde{b}_t)_{t\in[0,T]}$ is a $\sfd$-dimensional \cadlag adapted process
and $\tilde{a}=(\tilde{a}_t)_{t\in[0,T]}$ is a  
progressively measurable processes taking values in 
$\bbR^\sfd\otimes\bbR^{\sfr_1}$. 
Moreover,  
\beas 
\|X_0\|_p+\sup_{t\in[0,T]}\big(\|\tilde{b}_t\|_p
+\|\tilde{a}_t\|_p
{\colorg +\|J^X_t\|_p}
\big)
&<& \infty
\eeas
for every $p>1$. 
}
\ed
\ed
{\colorg 
The Wiener process $\tilde{w}$ is possibly correlated with $w$. 
}

Recall that $\hat{\theta}^{B,\alpha_n}_n$ denotes the quasi-Bayesian estimator (QBE) of $\theta$ with respect to $\bbH_n$ 
defined by (\ref{300225-21}). 
We extend the probability space $(\Omega,\calf,P)$ so that 
a $\sfp$-dimensional standard Gaussian random vector $\zeta$ independent of $\calf$ 
{\tred is} defined on the extension $(\overline{\Omega},\overline{\calf},\overline{P})$. 
Define a random field $\bbZ$ on $(\overline{\Omega},\overline{\calf},\overline{P})$ by 
\beas 
\bbZ(u) &=& 
\exp\bigg(\Delta[u]-\half\Gamma[u^{\otimes2}]\bigg)
\qquad(u\in\bbR^\sfp)
\eeas
where 
$\Delta[u]=\Gamma^{1/2}[\zeta,u]$. 
We write $\hat{u}^{{\sf A},\alpha_n}_n=\sqrt{n}\big(\hat{\theta}^{{\sf A},\alpha_n}_n-\theta^*\big)$ for ${\sf A}\in\{M,B\}$.

{\colorg 
Let $B(R)=\{u\in\bbR^\sfp;\>|u|\leq R\}$ for $R>0$. 
Equip the space $C(B(R))$ of continuous functions on $B(R)$ 
with the sup-norm. Denote by $d_s(\calf)$ the $\calf$-stable convergence.  

\begin{lemma}\label{300623-5}
Suppose that $[F1']_4$, $[F2']$ and $[F3]$ are fulfilled. Then 
\bea\label{300624-30}
\bbZ_n|_{B(R)}\to^{d_s(\calf)}\bbZ|_{B(R)}
\qquad\text{in  }C(B(R))
\eea
as $n\to\infty$ for every $R>0$. 
\end{lemma}
\proof 
Fix $k\in\{1,...,\sfk\}$. 
Let 
\beas 
\widetilde{D}^{(k)}_j &=& 
\big(\Delta_j\tY^{(k)}\big)^{\otimes2}
-\big(\sigma^{(k)}(X_\tjm,\theta^*)\Delta_jw^{(k)}\big)^{\otimes2}
\eeas
and let 
$f_\tjm=
\big((S^{(k)})^{-1}(\partial_{\theta}S^{(k)})(S^{(k)})^{-1}\big)(X_\tjm,\theta^*)$.
We will show 
\bea\label{300624-1}
\bigg\|\sum_{j=1}^nn^{1/2}f_\tjm[\widetilde{D}^{(k)}_j]\bigg\|_p&\to&0
\eea
for every $p>1$. 
Let 
\beas &&
{\sf B}_j \yeq \int_\tjm^\tj b^{(k)}_sds,\quad
{\sf C}_j \yeq \sigma^{(k)}(X_\tjm,\theta^*)\Delta_jw^{(k)},\quad
\\&&
{\sf D}_j \yeq \int_\tjm^\tj \big(\sigma^{(k)}(X_s,\theta^*)-\sigma^{(k)}(X_\tjm,\theta^*)\big)dw_s,\quad \
{\sf E}_j \yeq \int_\tjm^\tj \sigma^{(k)}(X_s,\theta^*)dw_s. 
\eeas
Then 
\beas 
\widetilde{D}^{(k)}_j
&=&
({\sf B}_j)^{\otimes2} 
+\big\{{\sf B}_j\otimes{\sf E}_j + {\sf E}_j\otimes{\sf B}_j \big\}
+\big\{{\sf C}_j\otimes{\sf D}_j +{\sf D}_j\otimes{\sf C}_j+{\sf D}_j\otimes{\sf D}_j\big\}.
\eeas

It is easy to see 
\bea\label{300624-7} 
\bigg\|\sum_{j=1}^nn^{1/2}f_\tjm[{\sf B}_j^{\otimes2}]\bigg\|_p
&\to&
0.
\eea

For $p>2$, we have 
\beas 
\bigg\|\sum_{j=1}^nn^{1/2}f_\tjm[{\sf B}_j\otimes{\sf E}_j]\bigg\|_p
&\leq&
\bigg\|\sum_{j=1}^nn^{1/2}f_\tjm[hb_\tjm\otimes{\sf E}_j]\bigg\|_p
\\&&
+
\bigg\|\sum_{j=1}^nn^{1/2}f_\tjm\bigg[\int_\tjm^\tj\big(b_s-b_\tjm\big)ds
\otimes{\sf E}_j\bigg]\bigg\|_p
\\&\simleq&
\bigg\|\sum_{j=1}^nn^{-1}|f_\tjm|^2|b_\tjm|^2 |{\sf E}_j|^2\bigg\|_{p/2}^{1/2}
\\&&
+
\bigg\|\sum_{j=1}^nn^{1/2}f_\tjm\bigg[\int_\tjm^\tj\big(b_s-b_\tjm\big)ds
\otimes{\sf E}_j\bigg]\bigg\|_p
\\&\leq&
\bigg\{\sum_{j=1}^nn^{-1}\big\||f_\tjm|\big\|_{3p}^2
\big\||b_\tjm|\big\|_{3p}^2 \big\||{\sf E}_j|\big\|_{3p}^2\bigg\}^{1/2}
\\&&
+
\bigg\|\sum_{j=1}^nn^{1/2}\big|f_\tjm\big|\big|{\sf E}_j\big|\int_\tjm^\tj\big|b_s-b_\tjm\big|ds
\bigg\|_p
\eeas
by the Burkholder-Davis-Gundy inequality and H\"older's inequality. 
Therefore 
\bea\label{300624-8} 
\bigg\|\sum_{j=1}^nn^{1/2}f_\tjm[{\sf B}_j\otimes{\sf E}_j]\bigg\|_p
&\to&
0
\eea
since 
\beas
I_n\>:=\>
\bigg\|\sum_{j=1}^nn^{1/2}\big|f_\tjm\big|\big|{\sf E}_j\big|\int_\tjm^\tj\big|b_s-b_\tjm\big|ds
\bigg\|_p
&\to&
0.
\eeas
Indeed, for any $\ep>0$, there exists a number $\delta>0$ such that 
$P\big[w'(b,\delta)>\ep\big]<\ep$, where $w'(x,\delta)$ is 
the modulus of continuity defined by 
\beas 
w'(x,\delta) 
&=& 
\inf_{(s_i)\in\cals_\delta}\max_{i}\sup_{r_1,r_2\in[s_{i-1},s_i)}|x(r_1)-{\rev x}(r_2)|,
\eeas
where $\cals_\delta$ is the set of sequences $(s_i)$ such that 
$0=s_0<s_1<\cdots<s_v= T$ and $\min_{i=1,...,{\tred v-1}}(s_i-s_{i-1})>\delta$. 
Then 
\beas 
I_n
&\leq&
\bigg\|\sum_{j=1}^nn^{1/2}\big|f_\tjm\big|\big|{\sf E}_j\big|
\bigg\|_p\ep h
+\bigg\|\max_{j=1,...,n}V_j
\bigg\|_p\frac{T}{\delta}
+\bigg\|\sum_{j=1}^n
V_j
\bigg\|_{2p}P\big[w'(b,\delta)>\ep\big]^{\frac{1}{2p}}
\\&\simleq&
\ep
+\bigg(n^{-1/2}+\sum_{j=1}^n\big\|V_j1_{\{V_j>n^{-1/2}\}}\big\|_p\bigg)
\frac{T}{\delta}
+
\ep^{\frac{1}{2p}}
\eeas
for $n>T/\delta$, 
where 
\beas 
V_j &=& 
n^{1/2}\big|f_\tjm\big|\big|{\sf E}_j\big|\int_\tjm^\tj\big(\big|b_s\big|+\big|b_\tjm\big|\big)|ds.
\eeas
Thus we obtain $\lim_{n\to\infty}I_n=0$ and hence (\ref{300624-8}).

It\^o's formula gives 
\beas 
\sigma^{(k)}(X_t,\theta^*)-\sigma^{(k)}(X_\tjm,\theta^*)
&=& 
\int_\tjm^t\bigg(\partial_x\sigma^{(k)}(X_s,\theta^*)[\tilde{b}_s]
+\half\partial_x^2\sigma^{(k)}(X_s,\theta^*)\big[\tilde{a}_s\tilde{a}_s^\star\big]\bigg)
ds
\\&&
+\int_\tjm^t\partial_x\sigma^{(k)}(X_{s-},\theta^*)[\tilde{a}_sd\tilde{w}_s]
\\&&
+\int_{(\tjm,t]}\big(\sigma^{(k)}(X_{s},\theta^*)-\sigma^{(k)}(X_{s-},\theta^*)\big)dN^X_s
\\&=:&
{\sf b}_j(t)+{\sf a}_j(t)+{\sf d}_j(t)
\eeas
for $t\in[\tjm,\tj]$. 
With It\^o's formula, one can show 
\beas
\bigg\|\sum_{j=1}^nn^{1/2}f_\tjm\bigg[
{\sf C}_j\otimes\int_\tjm^\tj{\sf a}_j(s)dw_s
\bigg]\bigg\|_p
&\to&
0.
\eeas
Obviously 
\beas
\bigg\|\sum_{j=1}^nn^{1/2}f_\tjm\bigg[
{\sf C}_j\otimes\int_\tjm^\tj{\sf b}_j(s)dw_s
\bigg]\bigg\|_p
&\to&
0.
\eeas
Moreover, for 
$\hat{V}_j=n^{1/2}\big|f_\tjm\big|
\big|{\sf C}_j\big|\big|\int_\tjm^\tj{\sf d}_j(s)dw_s
\big|$, we have 
\bea\label{300626-1}
\bigg\|\sum_{j=1}^nn^{1/2}f_\tjm\bigg[
{\sf C}_j\otimes\int_\tjm^\tj{\sf d}_j(s)dw_s
\bigg]\bigg\|_p
&\leq&
\bigg\|\max_{j=1,...,n}\hat{V}_jN^X_T
\bigg\|_p
\nn\\&\leq&
n^{-1/4}\big\|N^X_T
\big\|_p
+
P\bigg[\max_{j=1,...,n}\hat{V}_j>n^{-1/4}\bigg]^{\frac{1}{2p}}\big\|N^X_T
\big\|_{2p}
\nn\\&\to&
0. 
\eea
Therefore 
\bea\label{300624-10} 
\bigg\|\sum_{j=1}^nn^{1/2}f_\tjm[{\sf C}_j\otimes{\sf D}_j]\bigg\|_p
&\to&
0.
\eea
Similarly to (\ref{300626-1}), we know 
\beas
\bigg\|\sum_{j=1}^nn^{1/2}f_\tjm\bigg[
\bigg(\int_\tjm^\tj{\sf d}_j(s)dw_s\bigg)^{\otimes2}
\bigg]\bigg\|_p
&\to&
0
\eeas
and {\tred also} 
\bea\label{300624-11} 
\bigg\|\sum_{j=1}^nn^{1/2}f_\tjm[{\sf D}_j\otimes{\sf D}_j]\bigg\|_p
&\to&
0.
\eea
From 
(\ref{300624-7}), (\ref{300624-8}), (\ref{300624-10}), (\ref{300624-11}) and 
symmetry, we obtain (\ref{300624-1}). 
In particular, (\ref{300624-1}) {\tred and} (\ref{300626-11}) 
give the approximation 
\beas 
\widetilde{\Delta}_n
&\equiv& 
n^{-1/2}\partial_\theta\widetilde{\bbH}_n(\theta^*)
\\&=&
\frac{1}{2\sqrt{n}}\sum_{k=1}^\sfk\sum_{{\tred j=1}}^nf_\tjm
\bigg[h^{-1}\big(\sigma^{(k)}(X_\tjm,\theta^*)\Delta_jw^{(k)}\big)^{\otimes2}
-S^{(k)}(X_\tjm,\theta^*)\bigg]
+o_p(1), 
\eeas
and so $\widetilde{\Delta}_n
\to^{d_s(\calf)}
\Gamma^\half\zeta$ as $n\to\infty$. 
Furthermore, Lemma \ref{300225-11} ensures 
\bea\label{300624-20} 
\Delta_n
&\to^{d_s(\calf)}&
\Gamma^\half\zeta
\eea
as $n\to\infty$.

Let $R>0$. Then there exists $n(R)$ such that 
for all $n\geq n(R)$ and all $u\in B(R)$, 
\bea\label{300624-21} 
\log\bbZ_n(u) &=& 
\Delta_n[u]
+\frac{1}{2n}\partial_\theta^2\bbH_n(\theta^*)[u^{\otimes2}]
+r_n(u),
\eea
where
\beas 
r_n(u) 
&=&
\int_0^1(1-s)\big\{n^{-1}\partial_\theta^2\bbH_n(\theta^\dagger_n(su))[u^{\otimes2}]
-n^{-1}\partial_\theta^2\bbH_n(\theta^*)[u^{\otimes2}]\big\}ds
\eeas
with $\theta^\dagger_n(u)=\theta^*+n^{-1/2}u$. 
Combining (\ref{300624-20}), Lemmas \ref{300225-A1(2)} and \ref{300225-A1(1)} 
with the representation (\ref{300624-21}), we conclude 
the finite-dimensional stable convergence 
\bea\label{300624-22}
\bbZ_n &\to^{d_{s\text{-}f}(\calf)}& \bbZ
\eea
as $n\to\infty$. 
Since Lemma \ref{300225-A1(1)} validates the tightness of $\{\bbZ_n|_{B(R)}\}_{n\geq n(R)}$, we obtain the functional stable convergence (\ref{300624-30}). 
\qed\halflineskip
}

\begin{theorem}\label{300430-8}
Suppose that $[F1']_4$, $[F2']$ and $[F3]$ are fulfilled. Then  
\beas 
E\big[f\big(\hat{u}^{{\sf A},\alpha_n}_n\big)\Phi\big] &\to& \bbE\big[f\big(\Gamma^{-1/2}\zeta\big)\Phi\big]
\eeas
as $n\to\infty$ for ${\sf A}\in\{M,B\}$, 
any continuous function $f$ of at most polynomial growth, and 
any $\calf$-measurable random variable $\Phi\in\cup_{p>1}L^p$. 
\end{theorem}
\proof 
{\colorg 
To prove the result for ${\sf A}=M$, we apply Theorem 5 of \cite{yoshida2011} 
with the help of Lemma \ref{300623-5} and Proposition \ref{300615-16}. 
For the case  ${\sf A}=B$, we obtain the convergence 
\beas 
\int_{\bbU_n}f(u)\bbZ_n(u)\varpi(\theta^*+n^{-1/2}u)du
\to^{d_s(\calf)}\int_{\bbR^\sfp}f(u)\bbZ(u)\varpi(\theta^*)du
\eeas
for any continuous function of at most polynomial growth,  
by applying Theorem 6 of \cite{yoshida2011}. 
For that, we use Lemma \ref{300623-5} and Theorem \ref{300224-10}. 
Estimate with Lemma 2 of \cite{yoshida2011} ensures 
Condition (i) of Theorem 8 of \cite{yoshida2011}, 
which proves the stable convergence as well as moment convergence. 
\qed\halflineskip
}

\section{{\colorg Efficient one-step estimators}}\label{300430-10}
{\rev 
In Section \ref{300615-3}, the asymptotic optimality was established 
for the QMLE $\hat{\theta}^{M,\alpha_n}_n$ and the QBE $\hat{\theta}^{B,\alpha_n}_n$ having 
a moving threshold specified by $\alpha_n$ converging to $0$. 
However, 
in practice 
for fixed $n$, these estimators are essentially the same as 
the $\alpha$-QMLE and $\alpha$-QBE for a fixed $\alpha$ though 
they gained some freedom of choice of ${\mathfrak S}^{(k)}_{n,j-1}$, $p^{(k)}_n$ and $q^{(k)}_n$ 
in the asymptotic theoretical context.}

It was found {\rev in Section \ref{300615-13}} that the $\alpha$-QMLE $\hat{\theta}^{M,\alpha}_n$ 
and the $(\alpha,\beta)$-QBE $\hat{\theta}_n^{B,\alpha,\beta}$ 
based on a fixed $\alpha$-threshold are consistent. 
However they have pros and cons. 
They are expected to remove jumps completely but they are conservative and  
the rate of convergence is not optimal. 
In this section, 
{\rev as the second approach to optimal estimation,}
we try to recover efficiency by combining 
{\rev these less optimal estimators}
with the aggressive random field $\bbH_n$ 
given by (\ref{300225-21}), 
{\rev expecting to keep high precision of jump detection by the fixed $\alpha$ filters.}

Suppose that $\kappa\in\bbN
$ satisfies $\kappa>1+(2\gamma_0)^{-1}$. 
We assume $[F1']_{\kappa{\tred\vee4}}$, ${\fred[F2]}$, $[F2']$ and $[F3]$. 
According to {\tred Proposition} \ref{300225-31}, $\hat{\theta}^{M,\alpha}_n$ attains $n^{-\beta}$-consistency 
for any $\beta\in\big(2^{-1}(\kappa-1)^{-1},\gamma_0\big)$, 
{\tred 
and then $\beta(\kappa-1)>1/2$.
} 
For $\theta^*\in\Theta$, there exists an open ball $B(\theta^*)\subset\Theta$ around $\theta^*$. 
If $\partial_\theta^2\bbH_n(\theta_0)$ is invertible, then 
Taylor's formula gives 
\beas 
\theta_1-\theta_0 &=& 
\big(\partial_\theta^2\bbH_n(\theta_0)\big)^{-1}
\big[\partial_\theta\bbH_n(\theta_1)-\partial_\theta\bbH_n(\theta_0)\big]
+\sum_{i=2}^{\kappa-2}A_{1,i}(\theta_0)\big[(\theta_1-\theta_0)^{\otimes i}\big]
\\&&
+A_{1,\kappa-1}(\theta_1,\theta_0)\big[(\theta_1-\theta_0)^{\otimes(\kappa-1)}\big]
\eeas
for $\theta_1,\theta_0\in B(\theta^*)$. 
{\tred The second term on the right-hand side reads $0$ when $\kappa=3$.} 
Here $A_{1,i}$ ($i=2,...,\kappa-2$) are written by $\big(\partial_\theta^2\bbH_n(\theta_0)\big)^{-1}$ 
and $\partial_\theta^i\bbH_n(\theta_0)$ ($i={\tred3},...,\kappa-1$), 
{\tred respectively,} 
and 
$A_{1,{\tred\kappa-1}}(\theta_0,\theta_1)$ is by $\big(\partial_\theta^2\bbH_n(\theta_0)\big)^{-1}$ 
and $\partial_\theta^\kappa\bbH_n(\theta)$ ($\theta\in B(\theta^*)$). 
Let 
\bea\label{300625-5} 
F(\theta_1,\theta_0)
&=&
\ep(\theta_0)
+\sum_{i=2}^{\kappa-2}A_{1,i}(\theta_0)\big[(\theta_1-\theta_0)^{\otimes i}\big], 
\eea
{\colorg 
where 
\beas 
\ep(\theta_0)
&=&
-\big(\partial_\theta^2\bbH_n(\theta_0)\big)^{-1}[\partial_\theta\bbH_n(\theta_0)],
\eeas
i.e., 
$\ep(\theta_0)[u]=-\big(\partial_\theta^2\bbH_n(\theta_0)\big)^{-1}[\partial_\theta\bbH_n(\theta_0),u]$ for $u\in\bbR^\sfp$. 
}
We write 
{\colorg $\sum_{i=2}^{\kappa-2}A_{1,i}(\theta_0)\big[F(\theta_1,\theta_0)^{\otimes i}\big]$ 
in the form}
\beas 
\sum_{i=2}^{\kappa-2}A_{1,i}(\theta_0)\big[F(\theta_1,\theta_0)^{\otimes i}\big]
&=&
{\colorg A_2}(\theta_0)
+\sum_{i_1+i_2\geq3}
A_{2,i_1,i_2}(\theta_0)\big[\ep(\theta_0)^{\otimes i_1},(\theta_1-\theta_0)^{\otimes i_2}\big] 
\eeas
{\colorg with 
\beas 
A_2(\theta_0) 
&=& 
\sum_{i=2}^{\kappa-2}A_{1,i}(\theta_0)\big[\ep(\theta_0)^{\otimes i}\big].
\eeas
}
Next we write
\beas 
\sum_{i_1+i_2\geq3}
A_{2,i_1,i_2}(\theta_0)\big[\ep(\theta_0)^{\otimes i_1},F(\theta_1,\theta_0)^{\otimes i_2}\big]
&=&
{\colorg A_3}(\theta_0)
+\sum_{i_1+i_2\geq4}
A_{3,i_1,i_2}(\theta_0)\big[\ep(\theta_0)^{\otimes i_1},(\theta_1-\theta_0)^{\otimes i_2}\big] 
\eeas
{\colorg 
with 
\beas 
A_3(\theta_0)
&=& 
\sum_{i_1+i_2\geq3}
A_{2,i_1,i_2}(\theta_0)\big[\ep(\theta_0)^{\otimes (i_1+i_2)}\big]. 
\eeas
}
\begin{en-text}
Repeating this procedure, we obtain the formula of the form 
\beas 
F(\theta_1,\theta_0) 
&=& 
\sum_{i=1}^{\kappa-2}A_i(\theta_0)
+\sum_{i_1+i_2\geq\kappa-1}
A_{\kappa-2,i_1,i_2}(\theta_1,\theta_0)\big[\ep(\theta_0)^{\otimes i_1},(\theta_1-\theta_0)^{\otimes i_2}\big]. 
\eeas
\end{en-text}
{\colorg\noindent
Repeat this procedure up to 
\beas&& 
\sum_{i_1+i_2\geq\kappa-2}
A_{\kappa-3,i_1,i_2}(\theta_0)\big[\ep(\theta_0)^{\otimes i_1},F(\theta_1,\theta_0)^{\otimes i_2}\big]
\\&=&
A_{\kappa-2}(\theta_0)
+\sum_{i_1+i_2\geq\kappa-1}
A_{\kappa-2,i_1,i_2}(\theta_0)\big[\ep(\theta_0)^{\otimes i_1},(\theta_1-\theta_0)^{\otimes i_2}\big] 
\eeas
with 
\beas 
A_{\kappa-2}(\theta_0)
&=& 
\sum_{i_1+i_2\geq\kappa-2}
A_{\kappa-3,i_1,i_2}(\theta_0)\big[\ep(\theta_0)^{\otimes (i_1+i_2)}\big]. 
\eeas
Let $A_1(\theta_0)=\ep(\theta_0)$. Thus, 
the sequence of $\bbR^\sfp$-valued random functions 
\beas 
A_i(\theta_0)\qquad(i=1,...,\kappa-2)
\eeas
are defined on $\{\theta_0\in\Theta;\>\partial_\theta^2\bbH_n(\theta_0)$ is invertible$\}$. 
For example, when $\kappa=4$, 
\beas 
A_1(\theta_0) &=& 
-\big(\partial_\theta^2\bbH_n(\theta_0)\big)^{-1}[\partial_\theta\bbH_n(\theta_0)],
\\
A_2(\theta_0) &=& 
-\half\big(\partial_\theta^2\bbH_n(\theta_0)\big)^{-1}
\big[\partial_\theta^3\bbH_n(\theta_0)[A_1(\theta_0)^{\otimes2}]\big]. 
\eeas

}

Let 
\beas 
{\mathfrak M}_n &=& 
\bigg\{\hat{\theta}_n^{M,\alpha}\in\Theta,\>
\det\partial_\theta^2\bbH_n(\hat{\theta}_n^{M,\alpha}){\colorg \not=0,\>
\hat{\theta}_n^{M,\alpha}
+\sum_{i=1}^{\kappa-2}A_i(\hat{\theta}_n^{M,\alpha})\in\Theta}\bigg\}.
\eeas
Define $\check{\theta}_n^{M,\alpha}$ by 
\beas 
\check{\theta}_n^{M,\alpha}&=&
\l\{\begin{array}{lll}
\hat{\theta}_n^{M,\alpha}
+\sum_{i=1}^{\kappa-2}A_i(\hat{\theta}_n^{M,\alpha})&\text{on}&{\mathfrak M}_n
\y
\theta_*&\text{on}&{\mathfrak M}_n^c
\end{array}
\r.
\eeas
where $\theta_*$ is an arbitrary value in $\Theta$. 

On the event ${\mathfrak M}^0_n:=\{\hat{\theta}^{M,\alpha_n}_n,\>\hat{\theta}_n^{M,\alpha}\in B(\theta^*)\}\cap{\mathfrak M}_n$, 
the QMLE $\hat{\theta}^{M,\alpha_n}_n$ for $\bbH_n$ satisfies
\bea\label{300625-1} 
\hat{\theta}^{M,\alpha_n}_n-\hat{\theta}_n^{M,\alpha}
&=& 
F(\hat{\theta}^{M,\alpha_n}_n,\hat{\theta}_n^{M,\alpha})
+
A_{1,\kappa-1}(\hat{\theta}^{M,\alpha_n}_n,\hat{\theta}_n^{M,\alpha})
\big[(\hat{\theta}^{M,\alpha_n}_n-\hat{\theta}_n^{M,\alpha})^{\otimes(\kappa-1)}\big].
\eea
{\colorg 
Let 
\beas 
{\mathfrak M}'_n
&=&
\bigg\{\hat{\theta}^{M,\alpha_n}_n,\>\hat{\theta}_n^{M,\alpha}\in B(\theta^*),\>
|\det{\tred n^{-1}}\partial_\theta^2\bbH_n(\hat{\theta}_n^{M,\alpha})|\geq2^{-1}\det\Gamma,\>
\hat{\theta}_n^{M,\alpha}
+\sum_{i=1}^{\kappa-2}A_i(\hat{\theta}_n^{M,\alpha})\in\Theta\bigg\}.
\eeas
Then the estimate 
\bea\label{300625-2}
\bigg\|\bigg\{
\hat{\theta}^{M,\alpha_n}_n-\hat{\theta}_n^{M,\alpha}
-
A_1(\hat{\theta}_n^{M,\alpha})
-
\sum_{i=2}^{\kappa-2}A_{1,i}(\hat{\theta}_n^{M,\alpha})\big[(\hat{\theta}^{M,\alpha_n}_n-\hat{\theta}_n^{M,\alpha})^{\otimes i}\big]
\bigg\}1_{{\mathfrak M}'_n}
\bigg\|_p
&=&
O(n^{-\beta(\kappa-1)})
\nn\\&&
\eea
for every $p>1$ 
follows from the representation (\ref{300625-1}), 
Propositions \ref{300225-31} and \ref{300615-16} and 
Lemma \ref{300221-A2}. 
Moreover, 
Lemmas \ref{300221-A2}, \ref{300225-A1(2)} and \ref{300225-A1(1)} 
together with $L^p$-boundedness of the estimation errors 
yield 
$P[({\mathfrak M}'_n)^c]=O(n^{-L})$ for every $L>0$. 

Now on the event ${\mathfrak M}^0_n$, we have 
\beas&&
\sum_{i=2}^{\kappa-2}A_{1,i}(\hat{\theta}_n^{M,\alpha})
\big[\big(\hat{\theta}^{M,\alpha_n}_n-\hat{\theta}_n^{M,\alpha}\big)^{\otimes i}\big]
\\&=&
\sum_{i=2}^{\kappa-2}A_{1,i}(\hat{\theta}_n^{M,\alpha})
\bigg[\bigg(
F(\hat{\theta}^{M,\alpha_n}_n,\hat{\theta}_n^{M,\alpha})
+
A_{1,\kappa-1}(\hat{\theta}^{M,\alpha_n}_n,\hat{\theta}_n^{M,\alpha})
\big[(\hat{\theta}^{M,\alpha_n}_n-\hat{\theta}_n^{M,\alpha})^{\otimes(\kappa-1)}\big]
\bigg)^{\otimes i}\bigg].
\eeas
Therefore it follows from (\ref{300625-2}) that 
\beas&&
\bigg\|\bigg\{
\hat{\theta}^{M,\alpha_n}_n-\hat{\theta}_n^{M,\alpha}
-
A_1(\hat{\theta}_n^{M,\alpha})
-
A_2(\hat{\theta}_n^{M,\alpha})
\nn\\&&
-
\sum_{i_1+i_2\geq3}
A_{2,i_1,i_2}(\hat{\theta}_n^{M,\alpha})\big[\ep(\hat{\theta}_n^{M,\alpha})^{\otimes i_1},(\hat{\theta}^{M,\alpha_n}_n-\hat{\theta}_n^{M,\alpha})^{\otimes i_2}\big] 
\bigg\}1_{{\mathfrak M}'_n}
\bigg\|_p
\nn\\&=&
O(n^{-\beta(\kappa-1)})
\eeas
for every $p>1$. 
Inductively, 
\beas 
\bigg\|\bigg\{
\hat{\theta}^{M,\alpha_n}_n-\hat{\theta}_n^{M,\alpha}
-\sum_{i=1}^{\kappa-2}
A_i(\hat{\theta}_n^{M,\alpha})
\bigg\}1_{{\mathfrak M}'_n}
\bigg\|_p
&=&
O(n^{-\beta(\kappa-1)}).
\eeas
Consequently, 
using boundedness of $\Theta$ on  $({\mathfrak M}'_n)^c$, we obtain 
}
\beas
\big\|\hat{\theta}^{M,\alpha_n}_n-\check{\theta}^{M,\alpha}_n\big\|_p
&=& 
O(n^{-\beta(\kappa-1)}){\tred\yeq o(n^{-1/2})}
\eeas
and this implies 
\beas 
\big\|\check{\theta}^{M,\alpha}_n-\theta^*\big\|_p
&=& 
O(n^{-1/2})
\eeas
for every $p>1$. 
{\tred 
We note that $\beta$ in the above argument is a working parameter chosen so that 
$\beta>2^{-1}(\kappa-1)^{-1}$. 
}

{\tred 
Next, we will consider a Bayesian estimator as the initial estimator. 
We are supposing that $\kappa>1+(2\gamma_0)^{-1}$, and furthermore 
we suppose $\beta$ satisfies $\beta\in(2^{-1}(\kappa-1)^{-1},\gamma_0)$. 
Remark that this $\beta$ is the parameter involved in 
the estimator $\hat{\theta}_n^{B,\alpha,\beta}$, not a working parameter. 
}
Let 
\beas 
{\mathfrak B}_n &=& 
\bigg\{\hat{\theta}_n^{B,\alpha,\beta}\in\Theta,\>
\det\partial_\theta^2\bbH_n(\hat{\theta}_n^{B,\alpha,\beta}){\colorg \not=0, \>
\hat{\theta}_n^{B,\alpha,\beta}
+\sum_{i=1}^{\kappa-2}A_i(\hat{\theta}_n^{B,\alpha,\beta})
\in\Theta
}
\bigg\}.
\eeas
Define $\check{\theta}_n^{B,\alpha,\beta}$ by 
\beas 
\check{\theta}_n^{B,\alpha,\beta}&=&
\l\{\begin{array}{lll}
\hat{\theta}_n^{B,\alpha,\beta}
+\sum_{i=1}^{\kappa-2}A_i(\hat{\theta}_n^{B,\alpha,\beta})&\text{on}&{\mathfrak B}_n
\y
\theta_*&\text{on}&{\mathfrak B}_n^c.
\end{array}
\r.
\eeas
Then we obtain
\beas\
\big\|\hat{\theta}^{M,\alpha_n}_n-\check{\theta}^{B,\alpha,\beta}_n\big\|_p
&=& 
O(n^{-\beta(\kappa-1)})
{\tred \yeq o(n^{-1/2})}
\eeas
and
\beas 
\big\|\check{\theta}^{B,\alpha,\beta}_n-\theta^*\big\|_p
&=& 
O(n^{-1/2})
\eeas
for every $p>1$. 

Write $\check{u}^{\sf A}_n=\sqrt{n}\big(\check{\theta}^{\sf A}_n-\theta^*\big)$ 
for ${\sf A}=$``$M,\alpha$'' and ``$B,\alpha,\beta$''. 
Thus, we have obtained the following result 
{\colorg from Theorem \ref{300430-8} for $\hat{\theta}_n^{M,\alpha_n}$.} 

\begin{en-text}
\begin{theorem}
Suppose that $\beta<\gamma_0$ and that an integer $\kappa$ satisfies $\kappa>1+(2\gamma_0)^{-1}$. 
Suppose that  $[F1']_{\kappa{\tred\vee4}}$, 
${\fred[F2]}$, $[F2']$ and $[F3]$ are fulfilled. Then  
\beas 
E\big[f\big(\check{u}^{\sf A}_n\big)\Phi\big] &\to& \bbE\big[f\big(\Gamma^{-1/2}\zeta\big)\Phi\big]
\eeas
as $n\to\infty$ for ${\sf A}=$``$M,\alpha$'' and ``$B,\alpha,\beta$'', 
any continuous function $f$ of at most polynomial growth, and 
any $\calf$-measurable random variable $\Phi\in\cup_{p>1}L^p$. 
\end{theorem}
\end{en-text}
{\tred 
\begin{theorem}
Suppose that  $[F1']_{\kappa{\tred\vee4}}$, 
${\fred[F2]}$, $[F2']$ and $[F3]$ are fulfilled. 
Let  $f$ be any continuous function of at most polynomial growth, and 
let $\Phi$ be any $\calf$-measurable random variable in $\cup_{p>1}L^p$. 
Suppose that an integer $\kappa$ satisfies $\kappa>1+(2\gamma_0)^{-1}$. 
Then 
\bd
\im[$(a)$] 
$\ds E\big[f\big(\check{u}^{M,\alpha}_n\big)\Phi\big] \to\bbE\big[f\big(\Gamma^{-1/2}\zeta\big)\Phi\big]$
as $n\to\infty$. 
\im[$(b)$]
$\ds E\big[f\big(\check{u}^{B,\alpha,\beta}_n\big)\Phi\big] \to\bbE\big[f\big(\Gamma^{-1/2}\zeta\big)\Phi\big]$
as $n\to\infty$, 
suppose that $\beta\in(2^{-1}(\kappa-1)^{-1},\gamma_0)$. 
\ed
\end{theorem}
}

\begin{en-text}
\section{Limit theorems}
\bd
\im[[F1$^\sharp$\!\!]] 
Conditions $(ii)$, $(iii)$, $(iv)$ and $(v)$ of $[F1]$ are satisfied in addition to \vspace{-2mm}
\bd\im[(i)] 
the process $X$ has a representation 
\beas 
X_t &=& X_0+\int_0^t\tilde{b}_sds+\int_0^ta_sdw_s+\int_0^t\tilde{a}_sd\tilde{w}_s
\qquad(t\in[0,T])
\eeas
where $\tilde{w}=(\tilde{w}_t)_{t\in[0,T]}$ is an $\sfr_1$-dimensional Wiener process possibly correlated with $w$, 
$\tilde{b}=(\tilde{b}_t)_{t\in[0,T]}$, $a=(a_t)_{t\in[0,T]}$ and $\tilde{a}=(\tilde{a}_t)_{t\in[0,T]}$ are 
progressively measurable processes taking values in $\bbR^\sfd$, $\bbR^\sfd\otimes\bbR^\sfr$ an 
$\bbR^\sfd\otimes\bbR^{\sfr_1}$, respectively, such that 
\beas 
\|X_0\|_p+\sup_{t\in[0,T]}\big(\|\tilde{b}_t\|_p+\|a_t\|_p+\|\tilde{a}\|_p\big)
&<& \infty
\eeas
for very $p>1$. 
\ed
\ed
\end{en-text}
\begin{en-text}
\bi
\im 
\beas
n^{-1}\bbH_n(\theta;\alpha) - n^{-1}\bbH_n(\theta^*;\alpha)
&\to^p&
\bbY(\theta)
\eeas

\im score
\beas 
\partial_{\theta}\bbH_{n}(\theta;\alpha) 
&=& 
\half\sum_{k=1}^\sfk\sum_{j\in\calj^{(k)}_n(\alpha^{(k)})}
\big(
(S^{(k)})^{-1}(\partial_{\theta}S^{(k)})(S^{(k)})^{-1}
\big)(X_\tjm,\theta)
\big[D^{(k)}_j\big]
\eeas
where 
\beas 
D^{(k)}_j 
&=& 
r^{(k)}_n(\alpha^{(k)})^{-1}h^{-1}\big(\Delta_jY^{(k)}\big)^{\otimes2}
-p(\alpha^{(k)})^{-1}S^{(k)}(X_\tjm,\theta)
\eeas

\im score
\beas 
\partial_{\theta}\bbH_{n}(\theta) 
&=& 
\half\sum_{j\in\calj_n}\bigg\{
S^{-1}(\partial_{\theta}S)S^{-1}(X_\tjm,\theta)
\big[\big(\Delta_jY\big)^{\otimes2}\big]
-\big(S^{-1}(\partial_{\theta}S)S^{-1}\big)[S](X_\tjm,\theta)\bigg\}
\eeas

\koko 
\im Conjecture. 
\beas 
n^{-1}\partial_{\theta}\bbH_{n}(\theta) 
&=^a&
\frac{1}{2n}\sum_{j\in\calj_n}\bigg\{
S^{-1}(\partial_{\theta}S)S^{-1}(X_\tjm,\theta)
\big[S(X_\tjm,\theta^*)\big]
-\big(S^{-1}(\partial_{\theta}S)S^{-1}\big)[S](X_\tjm,\theta)\bigg\}
\\&\sim&
\half\int_\cala 
S^{-1}(\partial_{\theta}S)S^{-1}(x,\theta)
\big[S(x,\theta^*)-S(x,\theta)\big]\nu(dx)
\eeas
\begin{en-text}
\subsection{Convergence of the empirical distribution of $h^{-1/2}\Delta_jY$}
To show the conjectures in the previous section, 
\bi
\im empirical distribution
\beas 
n^{-1}\sum_{j=1}^n\delta_{h^{-1/2}\Delta_jY}
&\to^{p!}&
\int N_\sfm(0,s)P^{S(X_0,\theta^*)}(ds)
\eeas

\im joint
\beas 
n^{-1}\sum_{j=1}^n\delta_{(X_\tjm,h^{-1/2}\Delta_jY)}(dx,dy)
&\to^{p!}&
\pi(dx)\int N_\sfm(0,s)(dy)P^{S(x,\theta^*)}(ds)
\eeas
\ei
\end{en-text}

\section{Localization}\label{300228-1}
{\colorr 
In the preceding sections, we established 
asymptotic properties of the estimators, in particular, $L^p$-estimates for them. 
{\colorb Though} it was thanks to $[F3]$, verifying it is not straightforward. 
An analytic criterion and a geometric criterion are known to insure Condition $[F3]$ 
when $X$ is a non-degenerate diffusion process 
(Uchida and Yoshida 
\cite{uchida2013quasi}). 
It is possible to give similar criteria even for jump-diffusion processes but 
we do not pursue this problem here. 
Instead, it is also possible to relax $[F3]$ 
in order to only obtain stable convergences. 

We will work with  
\bd\im[[F3$^\flat$\!\!]] $\chi_0>0$ a.s.
\ed
in place of $[F3]$. 

Let $\ep>0$. Then there exists a $\delta>0$ such that 
$P[A_\delta]\geq1-\ep$ for $A_\delta=\{\chi_0>\delta\}$. 
Define $^\delta\bbH_n(\theta;\alpha)$ by 
\beas 
^\delta\bbH_n(\theta;\alpha)_\omega 
&=&
\l\{\begin{array}{ll}
\bbH_n(\theta;\alpha)_\omega&(\omega\in A_\delta)\y
-n|\theta-\theta^*|^2&(\omega\in A_\delta^c).
\end{array}\r.
\eeas
The way of modification of $\bbH_n$ on $A_\delta^c$ is not essential in the following argument. 
Let 
\beas 
^\delta\bbZ^\beta_n(u;\alpha) &=& 
\exp\bigg\{\>^\delta\bbH^\beta_n\big(\theta^*+n^{-\beta}u;\alpha\big)-\>^\delta\bbH^\beta_n\big(\theta^*;\alpha\big)\bigg\}
\qquad(u\in\bbU^\beta_n)
\eeas
{\colorg 
for $\>^\delta\bbH_n^\beta(\theta;\alpha)=n^{-1+2\beta}\>^\delta\bbH_n(\theta;\alpha)$. 
}
The random field $^\delta\bbY_n(\theta;\alpha)$ is defined by 
\beas
^\delta\bbY_n(\theta;\alpha) &=& n^{-2\beta}\big\{\>^\delta\bbH_n^\beta(\theta;\alpha)-\>^\delta\bbH_n^\beta(\theta^*;\alpha)\big\}
\yeq n^{-1}\big\{\>^\delta\bbH_n(\theta;\alpha)-\>^\delta\bbH_n(\theta^*;\alpha)\big\}. 
\eeas
The limit of $^\delta\bbY_n(\theta;\alpha)$ is now 
\beas 
^\delta\bbY(\theta) 
&=& 
\bbY(\theta)1_{A_\delta}-|\theta-\theta^*|^21_{A_\delta^c}.
\eeas
The corresponding key index is 
\beas 
^\delta\chi_0 &=&  \inf_{\theta\not=\theta^*}\frac{-\>^\delta\bbY(\theta)}{|\theta-\theta^*|^2}.
\eeas
Then Condition $[F3]$ holds for $^\delta\chi_0$
under the conditional probability given $A_\delta$, that is, 
\beas 
P\big[\>^\delta\chi_0<r^{-1}\big|A_\delta\big] &\leq& C_{L,\delta}\>r^{-L}\quad(r>0)
\eeas
for every $L>0$. 
Now it is not difficult to follow the proof of Propositions \ref{300225-31} 
and \ref{300226-1} 
to obtain 
\beas 
\sup_{n\in\bbN}\bigg\{
E\big[\big|n^\beta\big(\hat{\theta}^{M,\alpha}_n-\theta^*\big)\big|^p1_{A_\delta}\big]
+
E\big[\big|n^\beta\big(\hat{\theta}^{B,\alpha,\beta}_n-\theta^*\big)\big|^p1_{A_\delta}\big]
\bigg\}
\><\> \infty
\eeas
for every $p>1$ and every $\beta<\gamma_0$, 
under $[F1]_4$ and ${\fred[F2]}$ in addition to $[F3^\flat]$. 
Thus we obtained the following results. 
\begin{proposition}\label{300430-5}
Suppose that $[F1]_4$, ${\fred[F2]}$ and $[F3^\flat]$ are satisfied. Then 
$\ds 
n^\beta\big(\hat{\theta}^{M,\alpha}_n-\theta^*\big)\yeq O_p(1)$ 
and 
$\ds 
n^\beta\big(\hat{\theta}^{B,\alpha,\beta}_n-\theta^*\big)\yeq O_p(1)$ 
as $n\to\infty$ 
for every $\beta<\gamma_0$. 
\end{proposition}

In a similar way, we can obtain the stable convergence of 
the estimators with moving $\alpha$, as a counterpart to  
Theorem \ref{300430-8}. 
\begin{theorem}\label{300430-9}
Suppose that $[F1']_4$, $[F2']$ and $[F3^\flat]$ are fulfilled. Then  
\beas 
\hat{u}^{{\sf A},\alpha_n}_n&\to^{d_s}& \Gamma^{-1/2}\zeta
\eeas
as $n\to\infty$ for ${\sf A}\in\{M,B\}$. 
\end{theorem}

Moreover, a modification of the argument in Section \ref{300430-10} gives 
the stable convergence of the one-step estimators. 
\begin{en-text}
\begin{theorem}
Suppose that $\beta<\gamma_0$ and that an integer $\kappa$ satisfies $\kappa>1+(2\gamma_0)^{-1}$. 
Suppose that $[F1']_\kappa$, ${\fred[F2]}$, $[F2']$ and $[F3^\flat]$ are fulfilled. Then  
\beas 
\check{u}^{\sf A}_n&\to^{d_s}& \Gamma^{-1/2}\zeta
\eeas
as $n\to\infty$ for ${\sf A}=$``$M,\alpha$'' and ``$B,\alpha,\beta$''. 
\end{theorem}
\end{en-text}
{\tred 
\begin{theorem}
Suppose that  $[F1']_{\kappa{\tred\vee4}}$, 
${\fred[F2]}$, $[F2']$ and $[F3^\flat]$ are fulfilled. 
Suppose that an integer $\kappa$ satisfies $\kappa>1+(2\gamma_0)^{-1}$. 
Then 
\bd
\im[$(a)$] 
$\ds \check{u}^{M,\alpha}_n\to^{d_s}\Gamma^{-1/2}\zeta$
as $n\to\infty$. 
\im[$(b)$]
$\ds \check{u}^{B,\alpha,\beta}_n\to^{d_s}\Gamma^{-1/2}\zeta$
as $n\to\infty$, 
suppose that $\beta\in(2^{-1}(\kappa-1)^{-1},\gamma_0)$. 
\ed
\end{theorem}
}

Suppose that the process $X$ satisfies the stochastic integral equation 
\beas 
X_t &=& X_0+\int_0^t\tilde{b}(X_s)ds
+\int_0^t\tilde{a}(X_s)d\tilde{w}_s+J^X_t
\qquad(t\in[0,T])
\eeas
with a finitely active jump part $J^X$ with $\Delta J^X_0=0$. 
The first jump time $T_1$ of $J^X$ satisfies $T_1>0$ a.s. 
Suppose that $X'$ is a solution to 
\beas 
X'_t &=& X_0+\int_0^t\tilde{b}(X'_s)ds
+\int_0^t\tilde{a}(X'_s)d\tilde{w}_s
\qquad(t\in[0,T])
\eeas 
and that $X'=X^{T_1}$ on $[0,T_1)$ for the stopped process $X^{T_1}$ of $X$ at $T_1$. 
This is the case where the stochastic differential equation has 
a unique strong solution. 
Furthermore, suppose that 
the key index $\chi_{0,\ep}$ defined for $(X'_t)_{t\in[0,\ep]}$ is non-degenerate 
for every $\ep>0$ 
in that $\sup_{r>0}r^LP[\chi_{0,\ep}<r^{-1}]<\infty$ for every $L>0$. 
Then on the event $\{T_1>\ep\}$, we have 
positivity of $\chi_0$. This implies Condition $[F3^\flat]$. 
To verify non-degeneracy of $\chi_{0,\ep}$, we may apply 
a criterion in Uchida and Yoshida 
\cite{uchida2013quasi}. 
}

\begin{en-text}
\begin{acknowledgements}
The authors thank the reviewers for their valuable comments 
to improve this article. 
\end{acknowledgements}

\newpage\begin{Large}
\noindent
\centerline{\bf Global jump filters and quasi-likelihood analysis for volatility}\vspace{5mm}
\centerline{Haruhiko Inatsugu and Nakahiro Yoshida}\vspace{3mm}
\end{Large}
\centerline{Graduate School of Mathematical Sciences, University of Tokyo}
\centerline{Japan Science and Technology Agency CREST}
\begin{Large}
\vspace{1cm} \\
\noindent
{\bf \underline{Supplementary materials}}
\end{Large}
\end{en-text}

\section{Simulation Studies}\label{300615-6}

\subsection{Setting of simulation}\label{3103031434}

In this section, we numerically 
investigate the performance of the global threshold estimator. 
We use the following one-dimensional Ornstein-Uhlenbeck process with jumps 
\begin{equation}
dX_t = - \eta X_t dt + \sigma d{\rev w}_t + dJ_t\qquad (t\in[0,1])
\end{equation}
{\rev starting from $X_0$.} 
Here ${\rev w=(w_t)_{t\in[0,1]}}$ is a one-dimensional Brownian motion and $J$ is a one-dimensional compound Poisson process defined by 
$$
J_t = \sum_{i=1}^{N_t} \xi_i, \qquad \xi_i \sim \mathcal{N}(0, {\rev\varepsilon^2}), 
$$
where $\varepsilon > 0$ and $N = {\rev(N_t)_{t\in[0,1]}}$ is a Poisson process with intensity $\lambda > 0$. 
The {\rev parameters} $\eta$, $\varepsilon$, and $\lambda$ are nuisance parameters, 
whereas $\sigma$ is unknown to be estimated from the 
discretely observed data ${\rev(X_{t_i^n})_{i=0,1,...,n}}$. 

There are already several parametric estimation methods for stochastic differential equations with jumps.  
Among them, Shimizu and Yoshida 
\cite{ShimizuYoshida2006} proposed a local threshold method for optimal parametric estimation.
They used method of jump detection by comparing each increment $|{\rev\Delta_i X}|$ with 
$h_n^{\rho}$, where $h_n = t_i^n - t_{i-1}^n$ is the time interval and $\rho \in (0, 1/2)$.  
More precisely, an increment ${\rev\Delta_i X}$ satisfying $|{\rev\Delta_i X}| > h_n^{\rho}$ is 
regarded as being driven by {\rev the} compound Poisson jump part, 
and is removed when constructing the likelihood function of the continuous part. 
The likelihood function of the continuous part is defined by 
$$
l_n(\sigma) = \sum_{i=1}^n \left[ -\frac{1}{2 \sigma^2 h_n } |\bar{X}_i^n|^2 - \frac{1}{2} \log \sigma^2  \right] \mathbf{1}_{ \{ |\Delta X_i| \leq h_n^{\rho} \} }, 
$$
where $\bar{X}_i^n = X_{t_i^n} - X_{t_{i-1}^n} + \eta X_{t_{i-1}^n} h_n$. 
Obviously, the jump detection scheme is essentially different from our approach in this paper. 
They do not use any other increments to determine whether an increment {\rev has} a jump or not. 
Our approach, however, {\rev uses all the increments.}

Shimizu and Yoshida 
{\rev \cite{ShimizuYoshida2006}} proved that this estimator is consistent as the sample size $n$ tends to infinity; 
that is, asymptotic property of {\rev the} local and {\rev the} global threshold approaches are the same from the viewpoint of consistency. 
However, precision of jump detection may be different in the case of (large but) finite samples.  
Comparison of two approaches is the main purpose of this section. 

{\rev In our setting, however, we assume that the jump size is normally distributed, the case of which is 
not dealt with in Shimizu and Yoshida 
\cite{ShimizuYoshida2006}. 
In their original paper, they assume that the jump size must be bounded away from zero. 
Ogihara and Yoshida 
\cite{OgiharaYoshida2011} accomodated {\rev a restrictive 
assumption on} the distribution of jump size. They proved 
that the local threshold estimator works well 
{\rev under this assumption} 
by using some elaborate arguments. 
Hence, the local estimator can be used in our setting and thus we can compare 
its estimates with the global threshold estimator. 

Note that, we do not impose too restrictive assumption about the 
distribution of jump sizes in our paper: we only assume natural moment conditions on 
the number of jumps. Versatility in this sense can be regarded as 
the advantage of our approach. }

The setting of the simulation is as follows. 
The initial value is $X_0 = 1$. 
The true value of the unknown parameter $\sigma$ is 0.1. 
Other parameters are all known and given by $\eta = 0.1$, $\varepsilon = 0.05$, and $\lambda = 20$. 
The sample size  is $n = 1,000$ {\colred in Section 6.2 to see the accuracy of the jump detection of our filter and 
$n = 5,000$ {\colred in Section \ref{202006211336} and} 
thereafter to compare the estimates of each estimator. We} assume the equidistant case, so that $h_n = 1/n = 0.001$
 {\colred and $h_n = 0.0002$}. Since the time horizon is now finite and 
$\eta$ is not consistently estimable, we set $\eta$ in $l_n(\sigma)$ at the true value $0.1$, that is the most preferable value for {\rev the estimator in Shimizu and Yoshida 
\cite{ShimizuYoshida2006}}.

{\rev In applying the global estimator, we need to set several tuning parameters. 
we set $C_{*}^{(k)} = 1$ for the truncation function $K_{n,j}^{(k)}$ in (\ref{300211-5}), 
that is used for the definition of $\alpha$-quasi-log likelihood function. 
For the one-step global estimator, we use the parameter $C_{*}^{(k)} = 1$ and 
$\delta_0 = 1/5$ for the truncation function 
$K_{n,j}^{(k)} = 1_{ \big\{ V_j^{(k)} < C_{*}^{(k)}n^{-\frac{1}{4}-\delta_0} \big\}}$. 
Moreover, we set $\delta_1^{(k)} = 4/9$ so that $p_n^{(k)} = (n - \lfloor n^{4/9} \rfloor) / n$ 
in the definition of the moving threshold quasi-likelihood function in (\ref{300225-21}). }

\begin{en-text}
{\rev We conduct a Monte Carlo simulation by generating 100 sample paths, calculating 
the estimates for each path, and then averaging the estimates. 
Moreover, we calculate the standard deviation of the estimates to see the stablity of our estimator. 
}
\end{en-text}

{\rev Figure 1 shows a sample path of $(X,J)$.}
The left panel is the sample path of $X$ and the right panel is its jump part {\rev $J$}. Note that the jump part is not 
observable and thus we {\rev need} to discriminate the jump from the sample path of $X$.  
 

\begin{figure}
\begin{center}
    \begin{tabular}{c}

      \begin{minipage}{0.5\hsize}
        \begin{center}
          \includegraphics[keepaspectratio, scale=0.55]{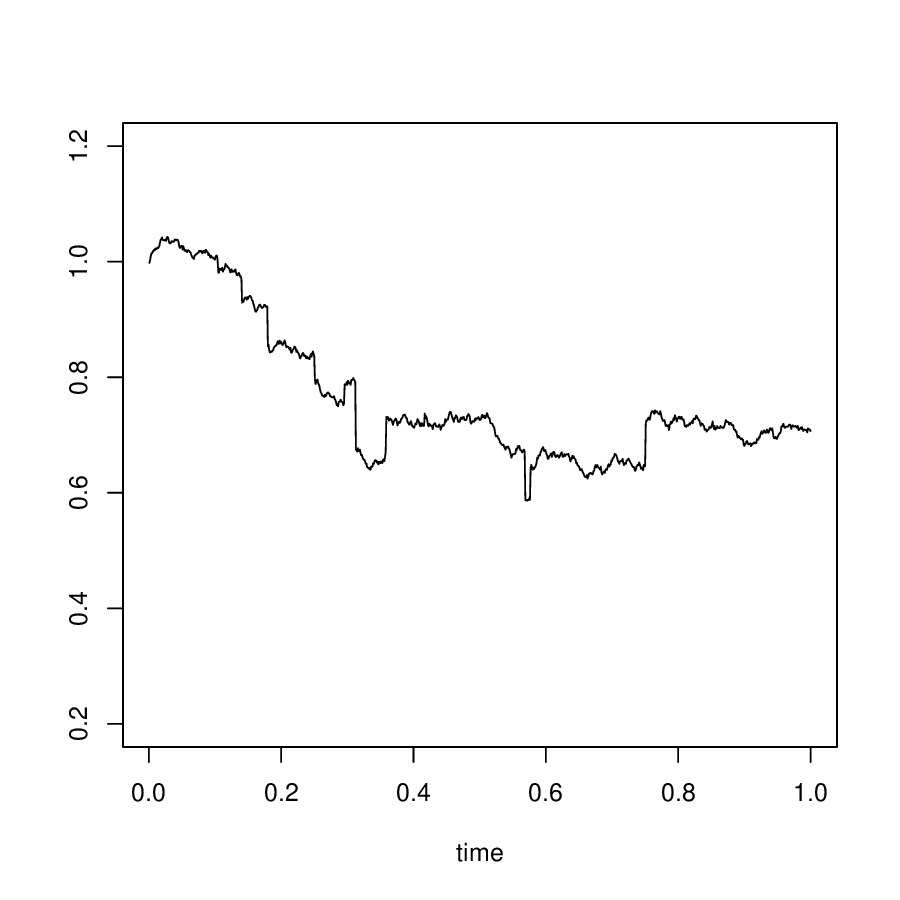}
          {(a) Sample path of $X$}
        \end{center}
      \end{minipage}

      \begin{minipage}{0.5\hsize}
        \begin{center}
          \includegraphics[keepaspectratio, scale=0.55]{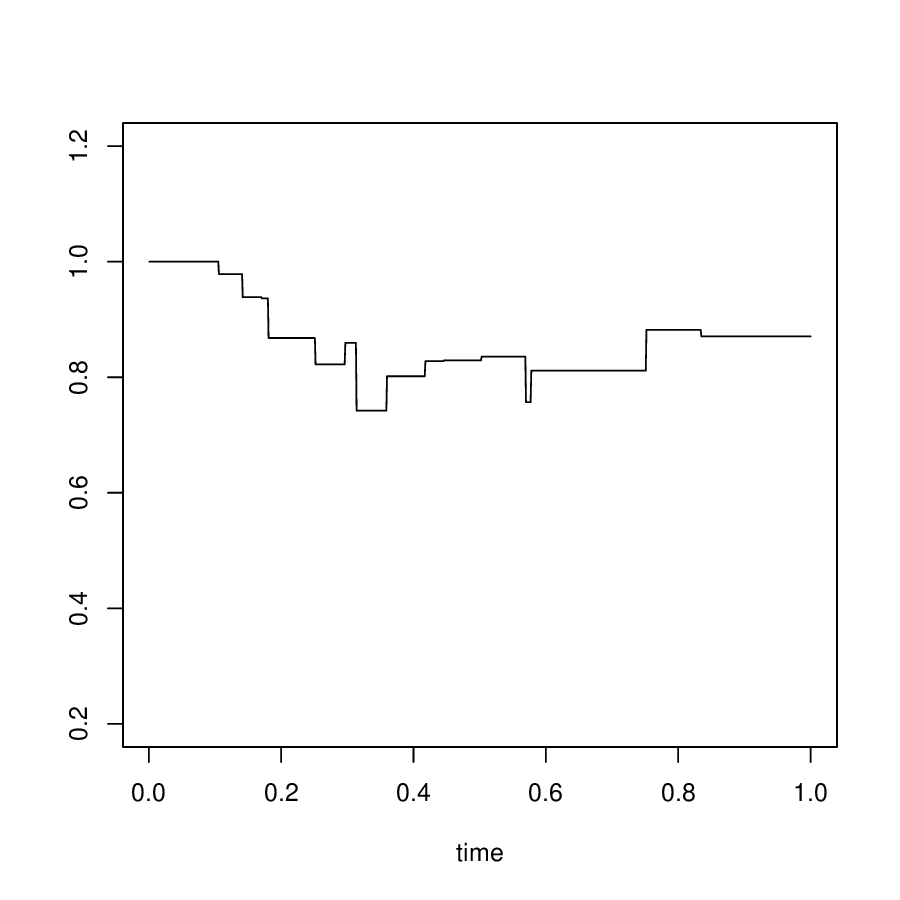}
          {(b) Sample path of the jump part {\rev $J$} of $X$}
        \end{center}
      \end{minipage}

    \end{tabular}
    \caption{Sample paths of $X$ and its jump part}

\end{center}
\end{figure}

\subsection{Accuracy of jump detection}

Before comparing the results of parameter estimation, we check the accuracy of jump detection of 
each estimation procedure. If there are too many misjudged increments, the estimated {\rev value} 
{\rev can have} a significant bias. Hence it is important how accurately we can eliminate jumps from the observed 
data $X$.

\subsubsection{Local threshold method} 

First, we check the accuracy of jump detection of the local threshold method.
Figure {\rev 2} shows the results of jump detection by the local threshold method of Shimizu and Yoshida 
\cite{ShimizuYoshida2006} for 
$\rho = 1/3$ {\rev in} panel (a) and $\rho = 1/2$ {\rev in} panel (b). 
{\rev The red vertical lines indicate the jump detected by each estimator, 
whereas 
{\colred the triangles on the horizontal axis}  
indicate the true jumps. As these {\colorg figures} show, the accuracy of the jump detection heavily depend on 
{\rev a} choice {\rev of} the tuning parameter $\rho$. For relatively small $\rho$ (say $\rho = {\colblue 1/3}$), 
we cannot {\colred completely} detect jumps: the estimator detects only one {\colred jump} for $\rho = {\colblue 1/3}$. 
On the other hand, in the case of {\colred (theoretically banned)} $\rho = 1/2$, 
the estimator detects the jumps better than 
the case of $\rho = {\colblue 1/3}$. 
Note that the case of $\rho = 1/2$} 
is not dealt {\rev with} in Shimizu and Yoshida 
\cite{ShimizuYoshida2006}, but it is useful for us 
to compare 
{\rev the local threshold method} 
with the global threshold method later and so we show the result of the exceptional case.  
%



\begin{figure}
    \begin{tabular}{c}

      \begin{minipage}{0.5\hsize}
        \begin{center}
          \includegraphics[keepaspectratio, scale=0.55]{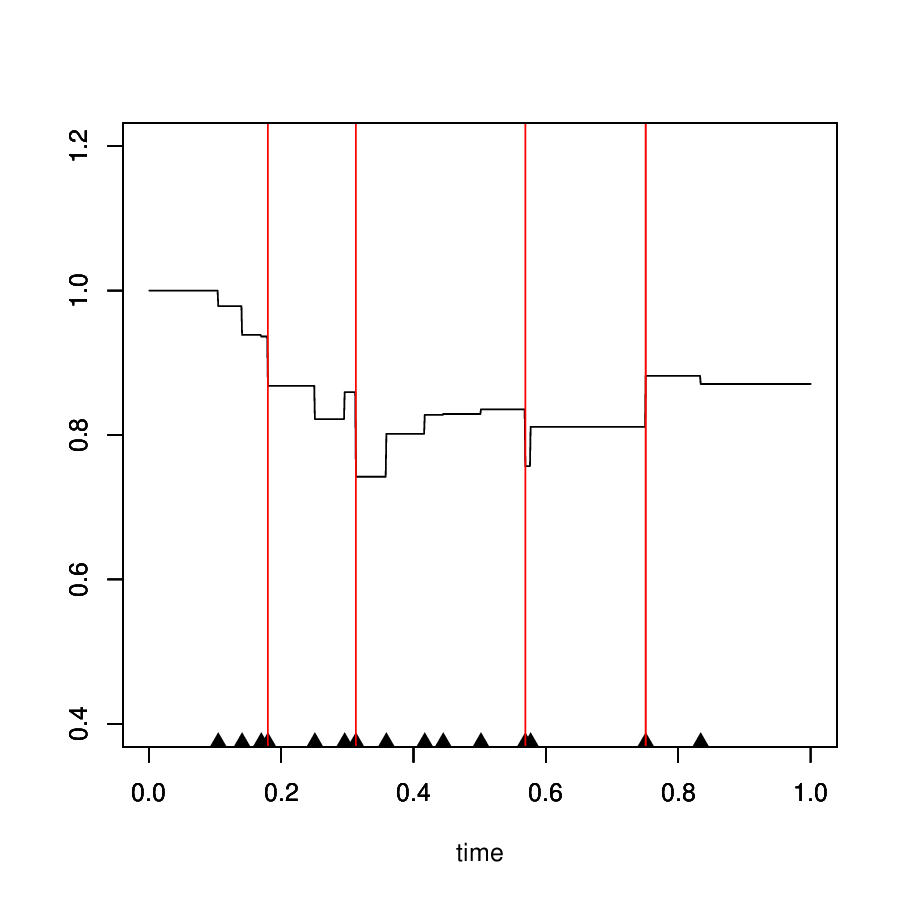}
          {(a) $\rho = \frac{1}{3}$}
        \end{center}
      \end{minipage}

      \begin{minipage}{0.5\hsize}
        \begin{center}
          \includegraphics[keepaspectratio, scale=0.55]{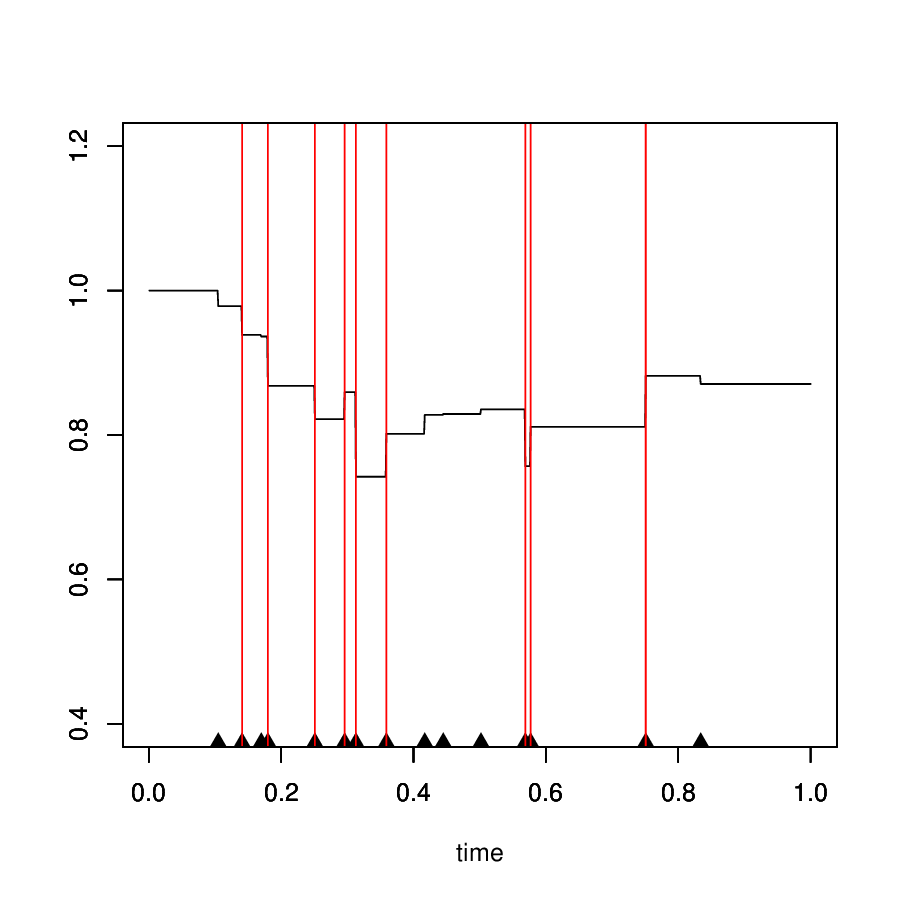}
          {(b) $\rho = \frac{1}{2}$}
        \end{center}
      \end{minipage}

    \end{tabular}
    \caption{Results of jump detection by local threshold method}

\end{figure}

\subsubsection{Global threshold method} 

Next, we discuss the jump detection by global threshold method. 
The accuracy of jump detection depends on the tuning parameter $\alpha \in (0,1)$, so 
we here show results of four cases, namely, the case $\alpha = 0.005, 0.010, 0.020, 0.050$. 

From the figures, we see that the too small $\alpha$ cannot detect jumps sufficiently, mistakenly judging some genuine 
jumps as increments driven by the continuous part{\rev , which is similar to the case of 
small $\rho$ of the Shimizu-Yoshida estimator. 
By setting $\alpha$ a little larger, the accuracy of jump detection increases, as shown in 
panels (b) and (c).}
On the other hand, too large $\alpha$ discriminate too many increments as 
jumps{\rev, as panel (d) shows. In this case, there are many increments that are 
regarded as jumps but are actually generated by the continuous part of the process only. 
These figures suggests that one should choose the tuning parameter $\alpha$ carefully to detect 
jumps appropriately. }

\begin{figure}
\begin{center}
    \begin{tabular}{c}

      \begin{minipage}{0.5\hsize}
        \begin{center}
           \includegraphics[keepaspectratio, scale=0.55]{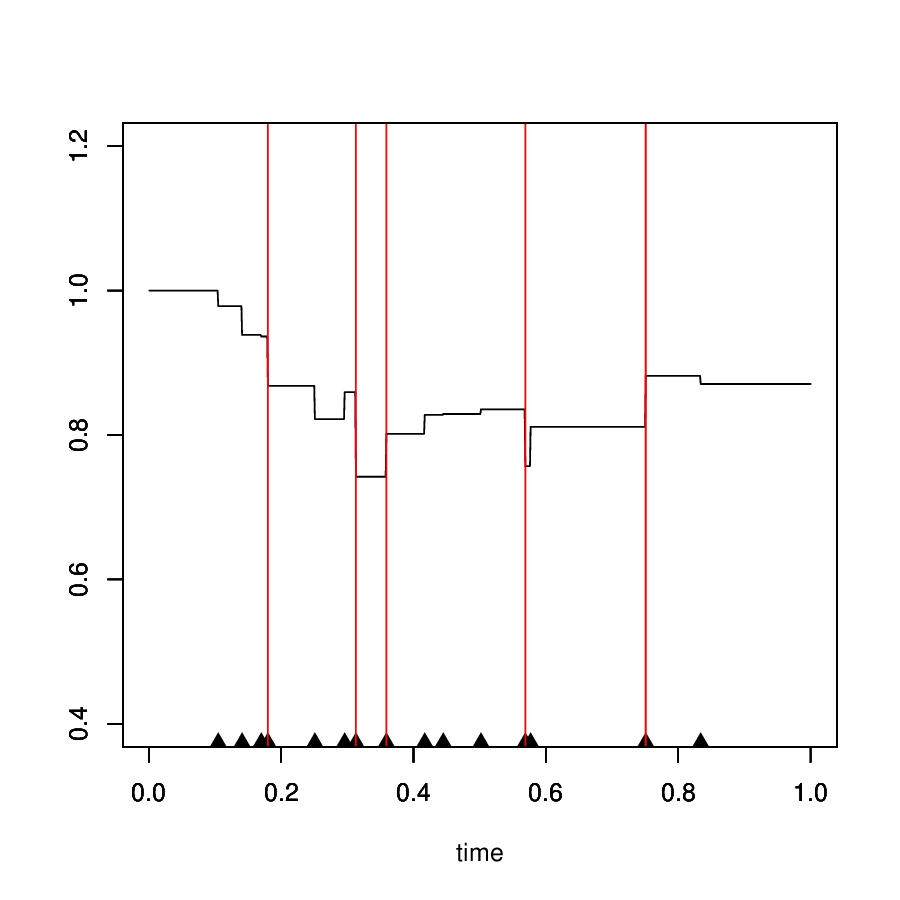}
          {(a) $\alpha= 0.005$}
        \end{center}
      \end{minipage}

      \begin{minipage}{0.5\hsize}
        \begin{center}
          \includegraphics[keepaspectratio, scale=0.55]{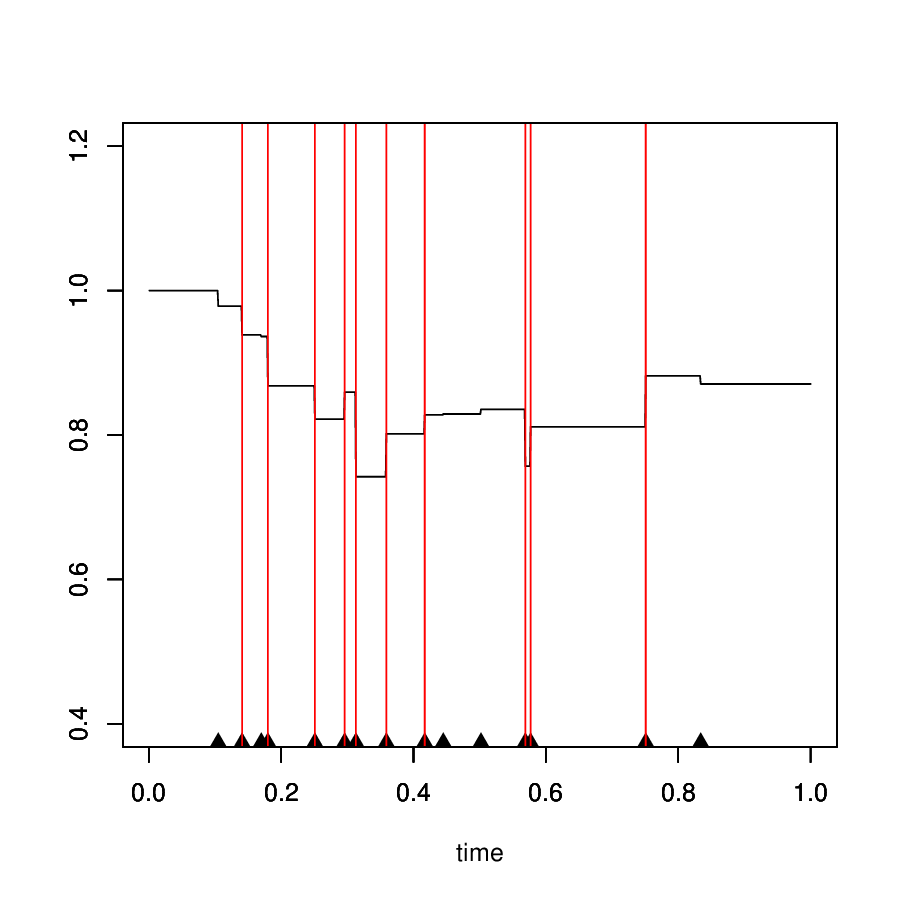} 
          {(b) $\alpha= 0.010$}
        \end{center}
      \end{minipage} \\

      \begin{minipage}{0.5\hsize}
        \begin{center}
          \includegraphics[keepaspectratio, scale=0.55]{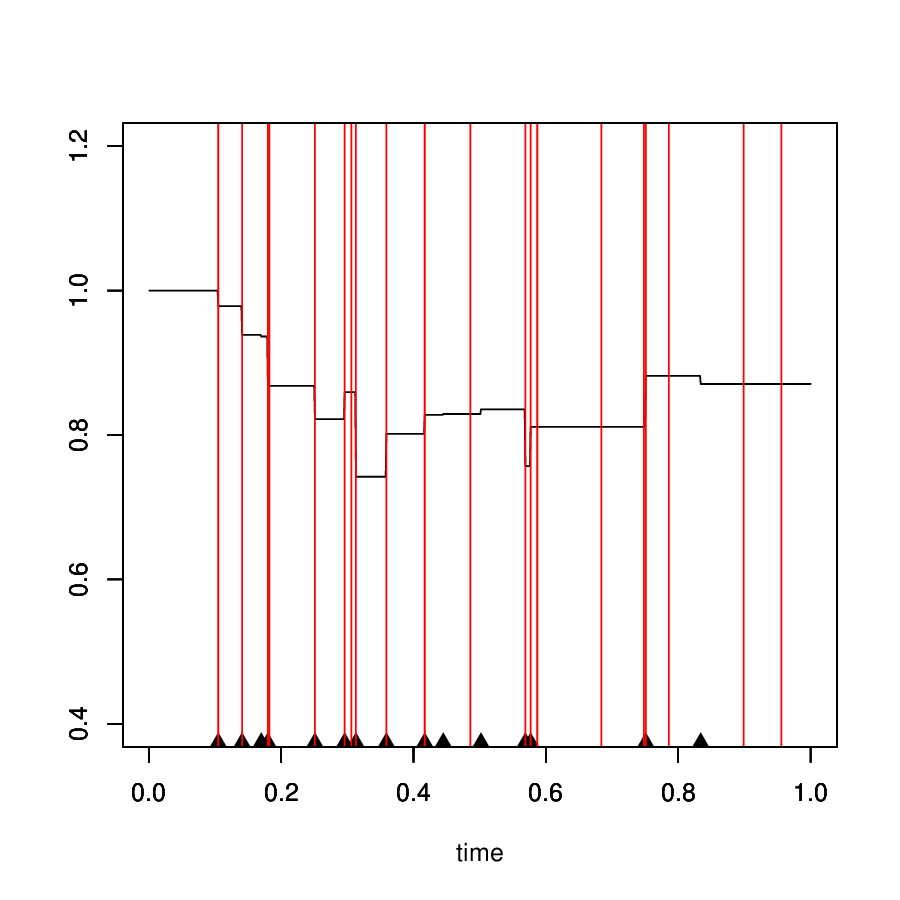} 
         {(c) $\alpha= 0.020$}
        \end{center}
      \end{minipage}
 
      \begin{minipage}{0.5\hsize}
        \begin{center}
          \includegraphics[keepaspectratio, scale=0.55]{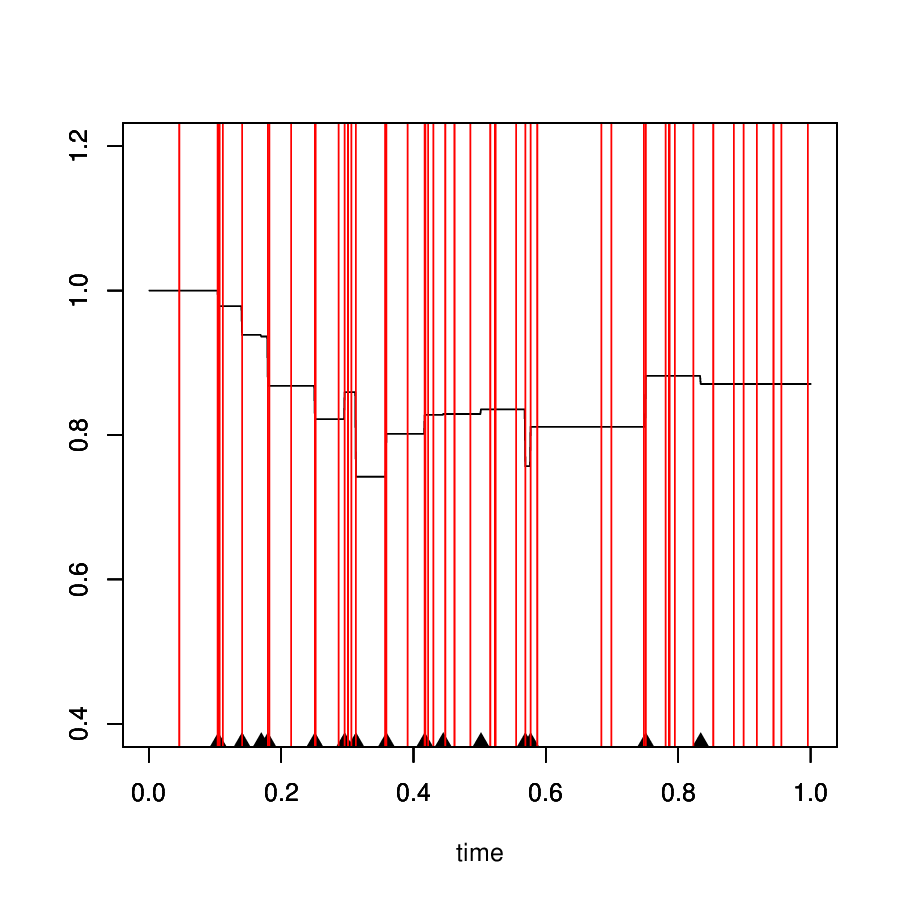} 
          {(d) $\alpha= 0.050$}
        \end{center}
      \end{minipage}

    \end{tabular}
    \caption{Results of jump detection by global threshold method}

\end{center}
\label{fig4}
\end{figure}

{\rev 
We show the false negative / positive ratio of jump detection in Table \ref{tb:ref-table}. Note that 
{\it false negative} means that our method did not judge an increment as a jump, despite it was actually 
driven by the compound Poisson jump part. The meaning of {\it false positive} is the opposite; that is, 
our method judged an increment which was not driven by the jump part as a jump.

The false negative ratio for small $\alpha$ tends to be large because in this case the estimator 
judges only big increments as jumps, and ignores some jumps of intermediate size. 
On the other hand, the false positive ratio for large $\alpha$ is high, since the estimator 
judges small increments as jumps, but almost increments are actually driven by the continous part.
From this table as well, we can infer that there should be some optimal range of $\alpha$ for jump detection. 
In any case, a large value of false negative may seriously bias  
the estimation, while a large value of false positive only decreases efficiency. 
Sensitivity of the local filter is also essentially observed by this experiment 
since each value of $\alpha$ of the global filter 
corresponds to a value of the threshold $Lh^\rho$ of the local filter. 

\begin{table}[!tbp]
  \caption{False Negative/Positive ratio of jump detection\label{tb:ref-table}} 
  \begin{center}
  \begin{tabular}{lrrrrrrrr}
  \hline\hline
  \multicolumn{1}{l}{alpha}&\multicolumn{1}{c}{0.005}&\multicolumn{1}{c}{0.01}&\multicolumn{1}{c}{0.015}&\multicolumn{1}{c}{0.02}&\multicolumn{1}{c}{0.025}&\multicolumn{1}{c}{0.05}&\multicolumn{1}{c}{0.1}&\multicolumn{1}{c}{0.25}\tabularnewline
  \hline
  False Negative&$73.333$&$40.000$&$26.667$&$26.667$&$26.667$&$26.667$&$26.667$&$20.000$\tabularnewline
  False Positive&$ 0.000$&$0.000$&$ 0.305$&$ 0.812$&$ 1.320$&$ 3.858$&$ 8.934$&$24.061$\tabularnewline
  \hline
  \end{tabular}\end{center}
  \end{table}
}

{\rev
\subsection{Comparison of the estimators}\label{202006211336} }
Next, we investigate the estimation results of {\rev the} global threshold method. 
{\colred In this section, we set the number of samples $n = 5,000$ to let the biases of the estimators as small as possible.}
Since the estimator depends on the parameter $\alpha$, we check the stability of 
the estimator with respect to the parameter $\alpha$. 
Remember that too small $\alpha$ is not able to detect jumps effectively, 
but too large $\alpha$ mistakenly eliminates small increments driven by the Brownian motion which should be 
used to construct the likelihood function of the continuous part. So there would be a suitable level $\alpha$. 

\begin{figure}%
  \begin{center}
  \includegraphics[keepaspectratio, scale=0.55]{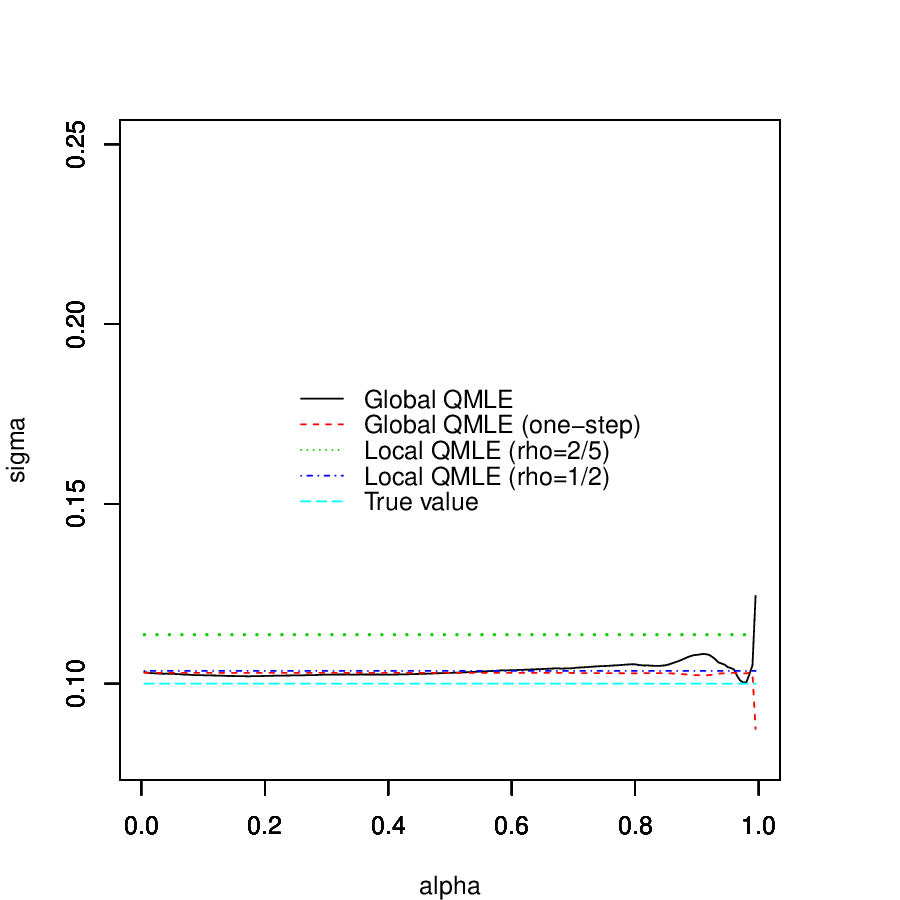}
  \end{center}
\caption{{\rev  
Comparison of estimators given a sample path
} }
\label{fig4}
\end{figure}

{\rev Figure \ref{fig4} compares the global QMLEs with the local QMLE with $\rho={\colred 2/5}$, as 
$\rho=1/2$ is theoretically prohibited, and 
suggests that the global methods are superior to the local methods.
Figure \ref{fig4} also compares the performance of the global threshold estimator $\hat{\sigma}^{M, \alpha}_n$ and the
one-step estimator $\check{\sigma}^{M, \alpha}_n$ with $\alpha$ ranging in $(0, 1)$, 
as well as that of the local filters.}
{\colorg Here we used $\kappa = 3$ to construct the one-step estimator; 
that is, the one-step estimator is given by $\check{\sigma}^{M, \alpha}_n = \hat{\sigma}^{M, \alpha}_n + A_1(\hat{\sigma}^{M, \alpha}_n)$, 
where the adjustment term $A_1$ is defined in Section \ref{300430-10}.}
{\colorg As the figure shows, for suitably small $\alpha$, 
both the estimate $\hat{\sigma}^{M, \alpha}_n$ {\rev and $\check{\sigma}^{M, \alpha}_n$ are} well close to $\sigma$. 
{\rev However}, as this figure indicates, the global threshold estimator may be somewhat
unstable with respect to the choice of $\alpha$. 
Although the global estimator with moving $\alpha$ and one-step global estimator are asymptotically equivalent, 
when we use the original global estimator, it would be recommended to use 
the one-step estimator as well and to try estimation for 
several $\alpha$'s in order to check the stability of the estimates.
%
}

{\rev To compare statistical properties of the estimators,} 
%
{\colorg we used the 100 outcomes of Monte Carlo simulation 
to calculate the average estimates, the root mean square error (RMSE), and the standard deviation of this 
experiment.}
Looking at the average values of the estimators shown in the Figure \ref{fig5} {\colorg (a)}, 
we see 
the global threshold {\rev estimators outperform} the local threshold estimator. 
It is concluded that the accuracy of the global estimator is not dependent on a sample path. 
High average accuracy can also be checked by RMSE. 
As shown in Figure \ref{fig5} (b), RMSEs {\rev of} the global estimators are smaller than those of 
the local estimators, except for the extreme choices of $\alpha$.

%




\begin{figure}%
      \begin{tabular}{c}
  
        \begin{minipage}{0.5\hsize}
          \begin{center}
            \includegraphics[keepaspectratio, scale=0.55]{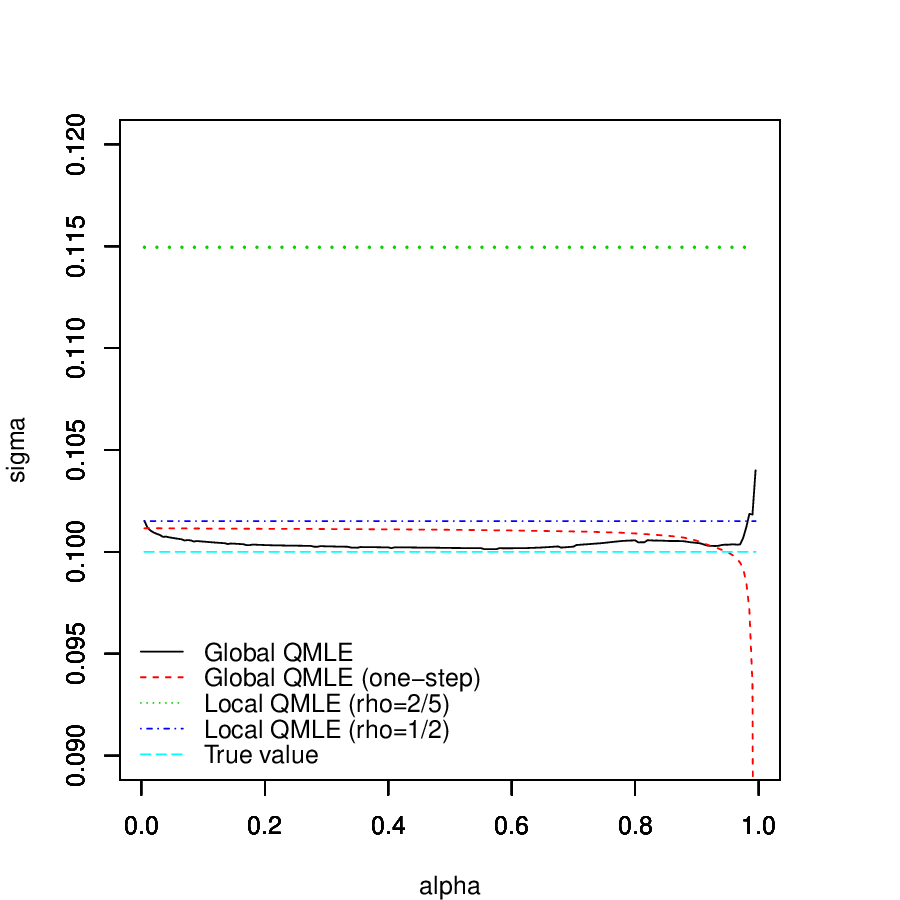}
            {(a) Average estimates}
          \end{center}
        \end{minipage}
  
        \begin{minipage}{0.5\hsize}
          \begin{center}
            \includegraphics[keepaspectratio, scale=0.55]{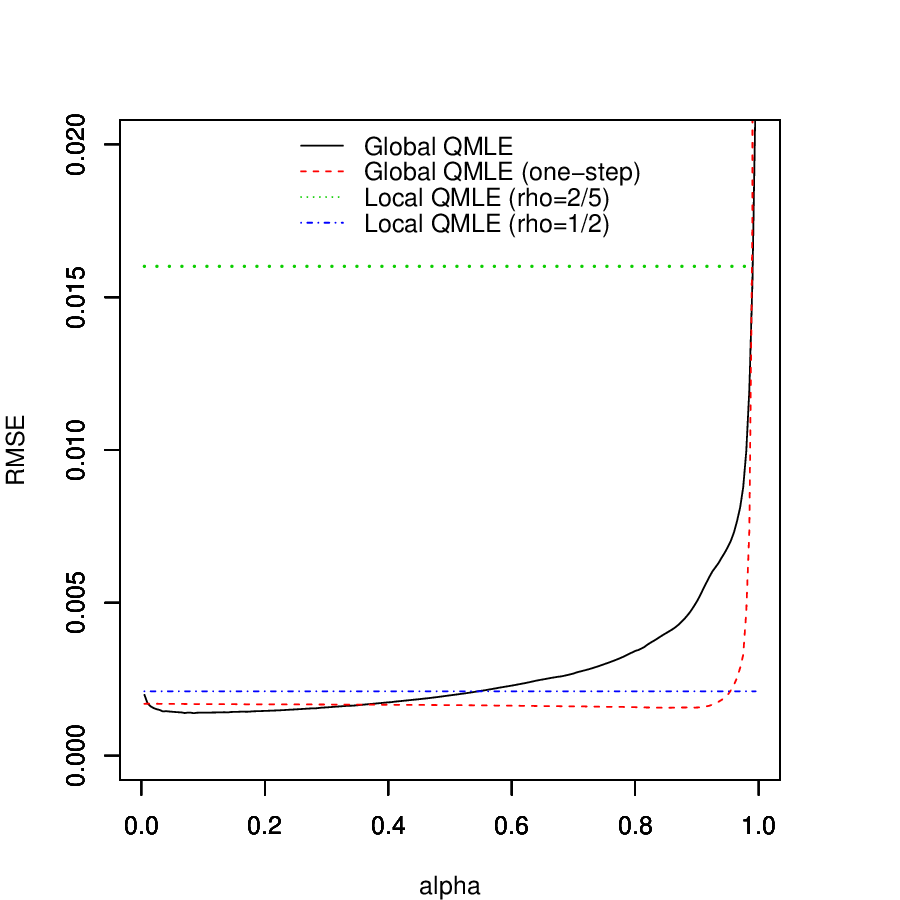}
            {(b) RMSE}
          \end{center}
        \end{minipage}

      \end{tabular}
      \caption{Results of jump detection by local threshold method: comparison of averaged results}
  
  \label{fig5}
  \end{figure}
%
%
%
%
%
%
%
%
Figure \ref{fig6} {\colred indicates the estimates for global QMLE estimator with standard error band. The standard 
errors are calculated by using 100 Monte carlo trials.} 
It shows that {\colred the global QMLE estimator works very well with or without one-step adjustment. 
We can see, however, the one-step adjusted estimator is robust against the choice of the tuning parameter $\alpha$. 
For large $\alpha$, the global threshold tends to eliminate increments that are not driven by the jump part of the underlying process, and this could result in the large standard deviation of the estimate. The one-step estimator works well for such large $\alpha$. 
\begin{en-text}
Also, the result shows that the one-step estimator might work well for extremely small $\alpha$. Without one-step 
adjustment, the estimator tends to be a little unstable for small $\alpha$, 
whereas the one-step estimator is robust in such a case. 
Note that we have used $n = 5,000$ in this simulation and, given the definition of our jump-detecting filter, this means 
even $\alpha = 0.005$ (which is used in the previous section) might be too large and so false-positive misdetection might occur highly frequently. However, the one-step estimator is successful in mitigating instability of estimates due to such false-positive misdetection.}
\end{en-text}
  
\begin{figure} 
      \begin{tabular}{c}
  
        \begin{minipage}{0.5\hsize}
          \begin{center}
            \includegraphics[keepaspectratio, scale=0.53]{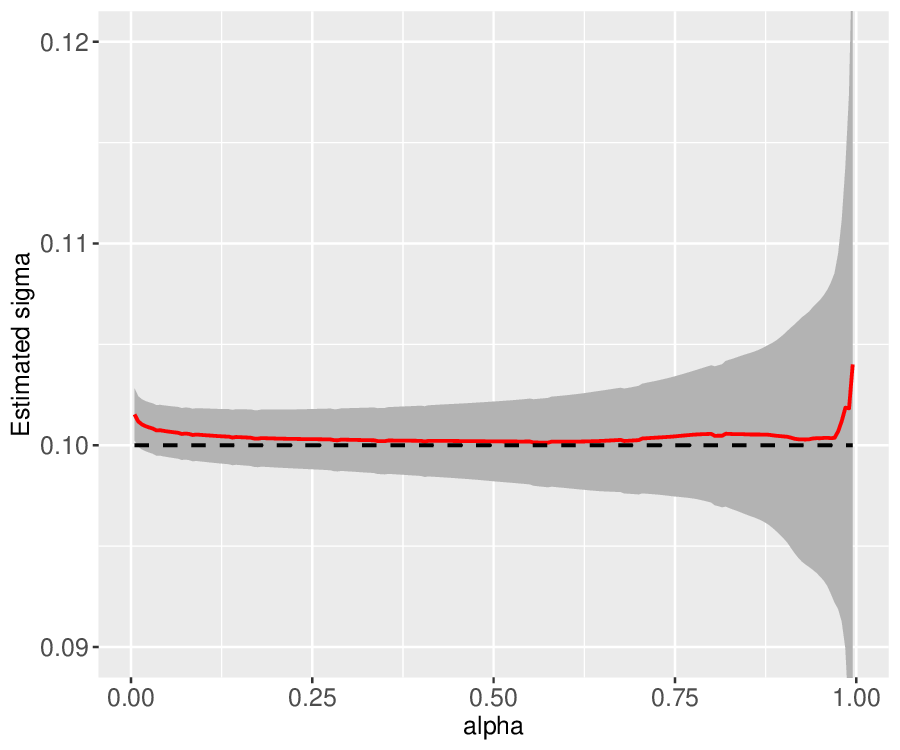}
           {(a) global QMLE}
          \end{center}
        \end{minipage}
  
        \begin{minipage}{0.5\hsize}
          \begin{center}
            \includegraphics[keepaspectratio, scale=0.53]{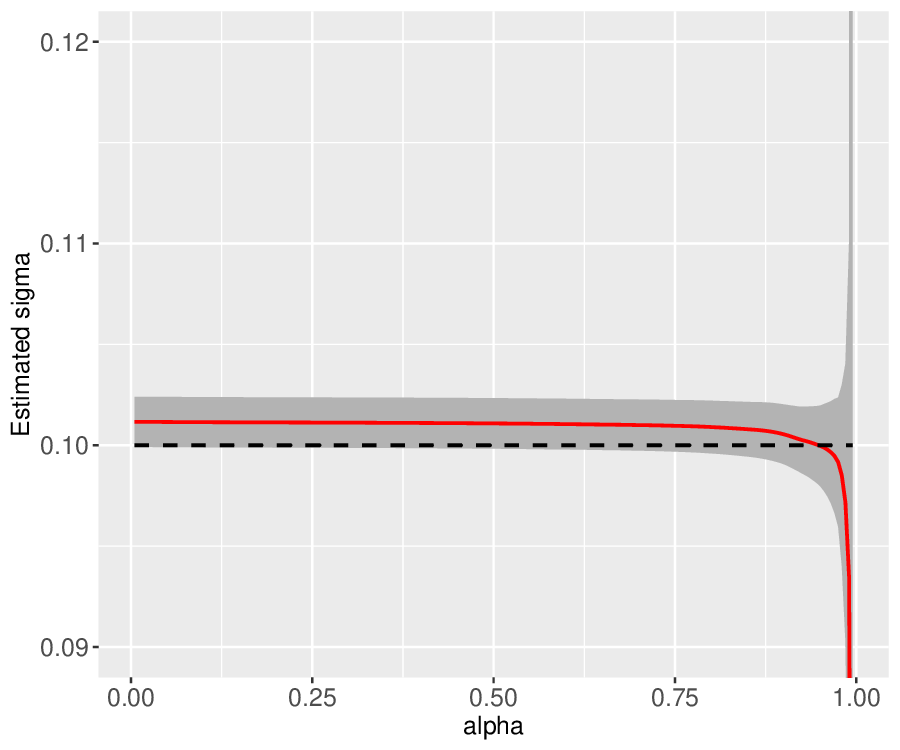}
            {(b) global QMLE (one-step)}  
          \end{center}
        \end{minipage}

      \end{tabular}
\caption{Estimation results of global QMLE estimator with standard error band}
\label{fig6}
\end{figure}

A suitably chosen $\alpha$ will yield a good 
estimate of the unknown parameter, although too small or too large $\alpha$ might 
tends to {\rev bias the} estimate. 
The global threshold estimator seems to generally be robust to the choice of 
the tuning parameters. 
The {\rev optimal} choice of $\alpha$  
{\rev depends on the situation}. Hence, it is desirable to use several values of $\alpha$ 
and to compare the results to determine the preferable value of $\alpha$ 
in using the global estimator. {\colred Moreover, it is worth considering of using one-step adjustment to 
get more robust estimates}.




{\rev 
The global filter sets a number for the critical value of the threshold though it is determined after observing the data. 
In this sense, the global filter looks similar to the local filter, that has a predetermined 
number as its threshold. 
However, the critical values used by the two methods are fairly different in practice. 
We consider the situation where, for some $n$, the local filter with threshold $Lh^\rho$ approximately 
performs as good as the global filter with $\alpha$. 
For simplicity, let us consider a one-dimensional case with $\sigma(x,\theta)=1$ constantly. 
Hence the critical value should approximately be near to the upper $\alpha/2$-quantile of $\Delta_jw$. 
Moreover, let $n=10^3$, $\rho={\colred 2/5}$ and $\alpha=0.1$. 
Then the constant $L$ in the threshold of the local filter should satisfy
$(10^{-3})^{-1/2}\times1.64=(10^{-3})^{-\rho}L$, namely, $L\sim3.27$ approximately. 
Since $L$ is a predetermined common constant for different numbers $n$, 
the critical value of the threshold of the local filter becomes $10^{-5\rho}L\sim{\colred 0.0327}$ when $n=10^5$, 
while the threshold of the global filter is about $(10^{-5})^{1/2}\times1.64\sim{0.00519}$. 
Some of jumps may not be detected by the local filter, 
since its critical value is not so small, compared with $\ep=0.05$. 
}

\ \\

{\rev
\subsection{Asymmetric jumps}
In the previous subsection, we assumed that the distribution of jump size 
{\tred was centered Gaussian  
and thus 
symmetric.} In a real situations, however, 
the distribution of 
the size of jumps might be not symmetric. For example, stock prices have an asymmetric distribution with heavier 
tail in negative price changes.  
In this subsection, we show that our global estimator performs well 
for jumps with asymmetric distribution. 

Although there are many asymmetric jumps in applications, we use just 
a normal distribution with a negative average 
because heavier tails would make jump detection easier. 
More precisely, we assume that the jump process $J$ is given by 
$$J_t = \sum_{i=1}^{N_t} \xi_t, \qquad \xi_i \sim \mathcal{N}(\mu, \varepsilon^2), $$ 
where $\mu = -0.01$ and $\varepsilon = 0.05$. 
In this setting, as shown in Figure \ref{fig7}, negative jumps appear more frequently than positive ones. 
}

  \begin{figure}%
        \begin{tabular}{c}
    
          \begin{minipage}{0.5\hsize}
            \begin{center}
              \includegraphics[keepaspectratio, scale=0.55]{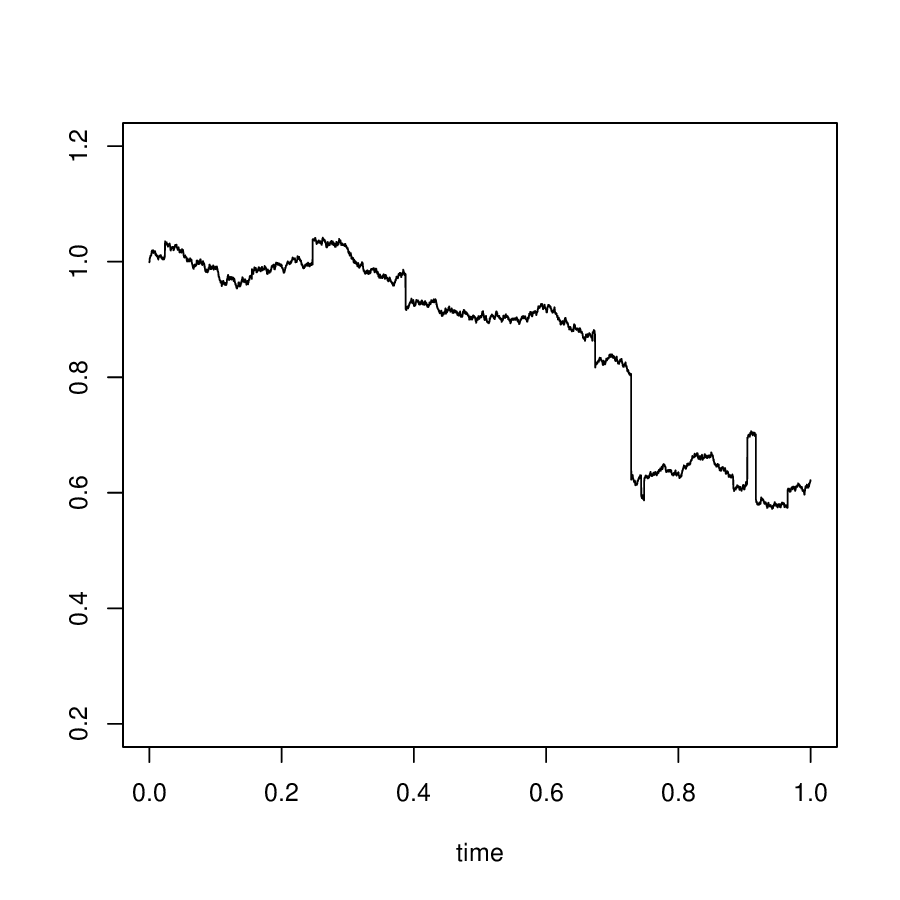}
              {(a) Sample path of $X$}
            \end{center}
          \end{minipage}
    
          \begin{minipage}{0.5\hsize}
            \begin{center}
              \includegraphics[keepaspectratio, scale=0.55]{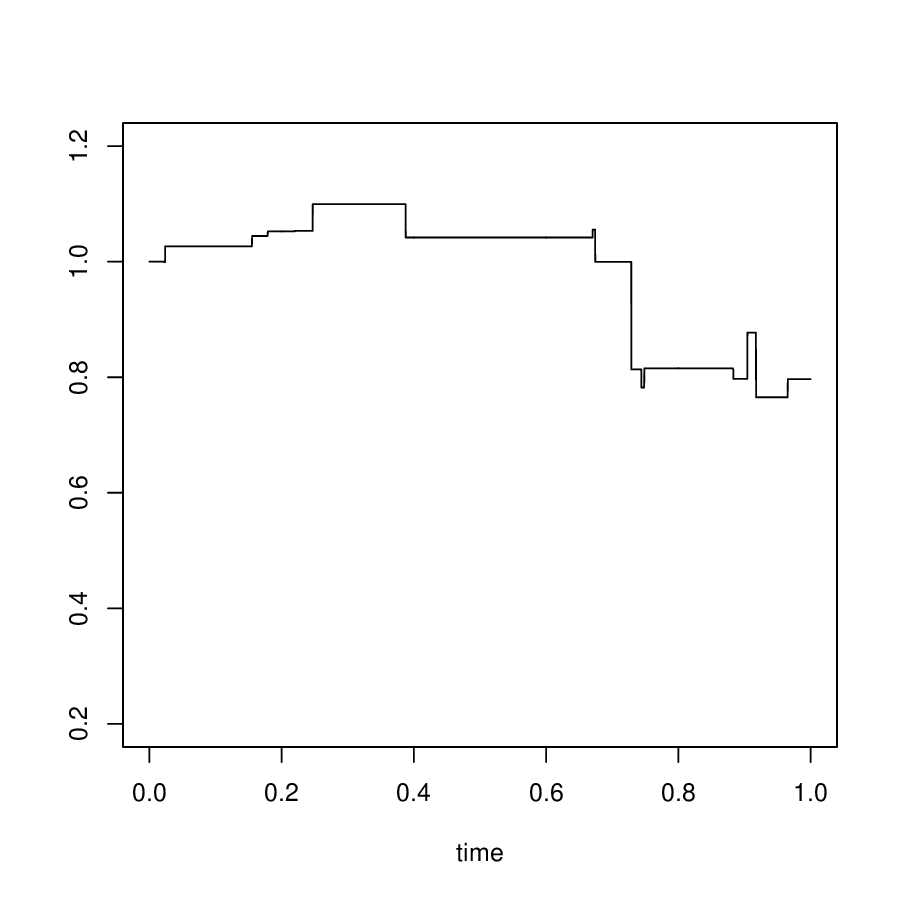}
              {(b) Sample path of the jump part $J$ of $X$}
            \end{center}
          \end{minipage}
  
        \end{tabular}
        \caption{Sample paths of $X$ and its jump part: in the case of asymmetric jump distribution}
     
    \label{fig7}
    \end{figure}

{\rev
As Figure \ref{fig8} shows, the global estimator performs well 
even in the case of asymmetric jumps. 
The estimates are well similarly to those in the case of symmetric jumps in 
the previous subsection. This example implies that out estimator will 
work very well under realistic circumstances, like financial time series where 
changes in asset prices have an symmetric distribution with heavy tail 
in negative price changes. 
}

\begin{figure}%
      \begin{tabular}{c}
  
        \begin{minipage}{0.5\hsize}
          \begin{center}
            \includegraphics[keepaspectratio, scale=0.55]{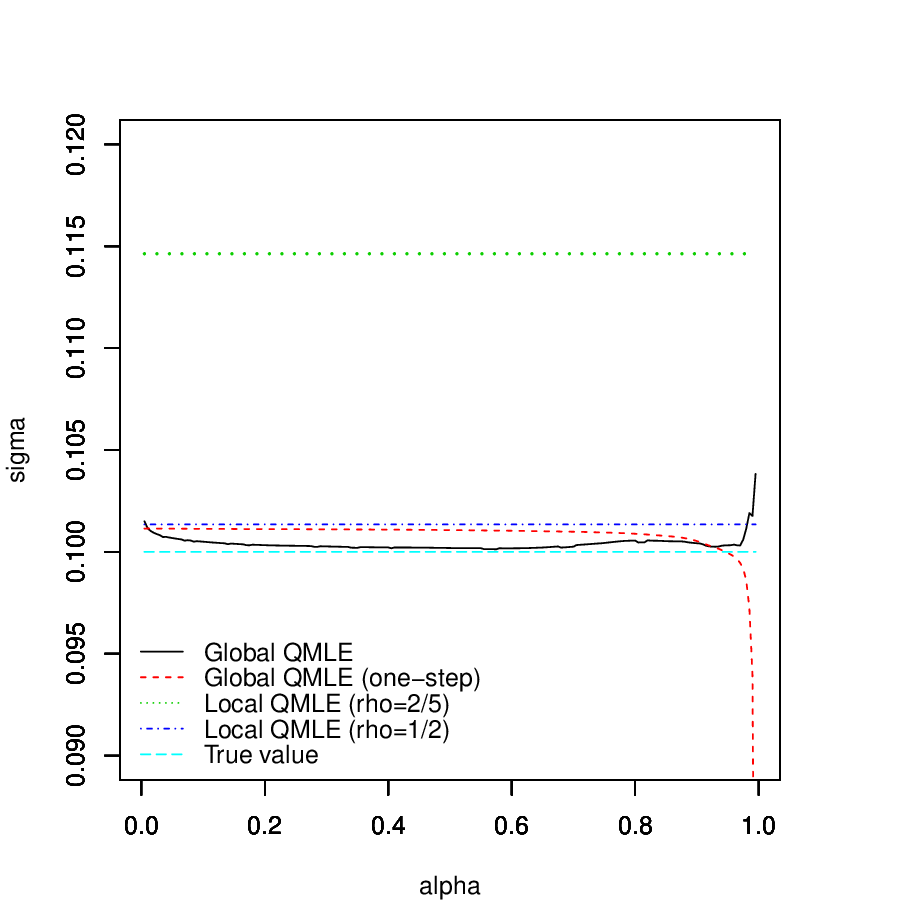}
            {(a) Average estimates}
          \end{center}
        \end{minipage}
  
        \begin{minipage}{0.5\hsize}
          \begin{center}
            \includegraphics[keepaspectratio, scale=0.55]{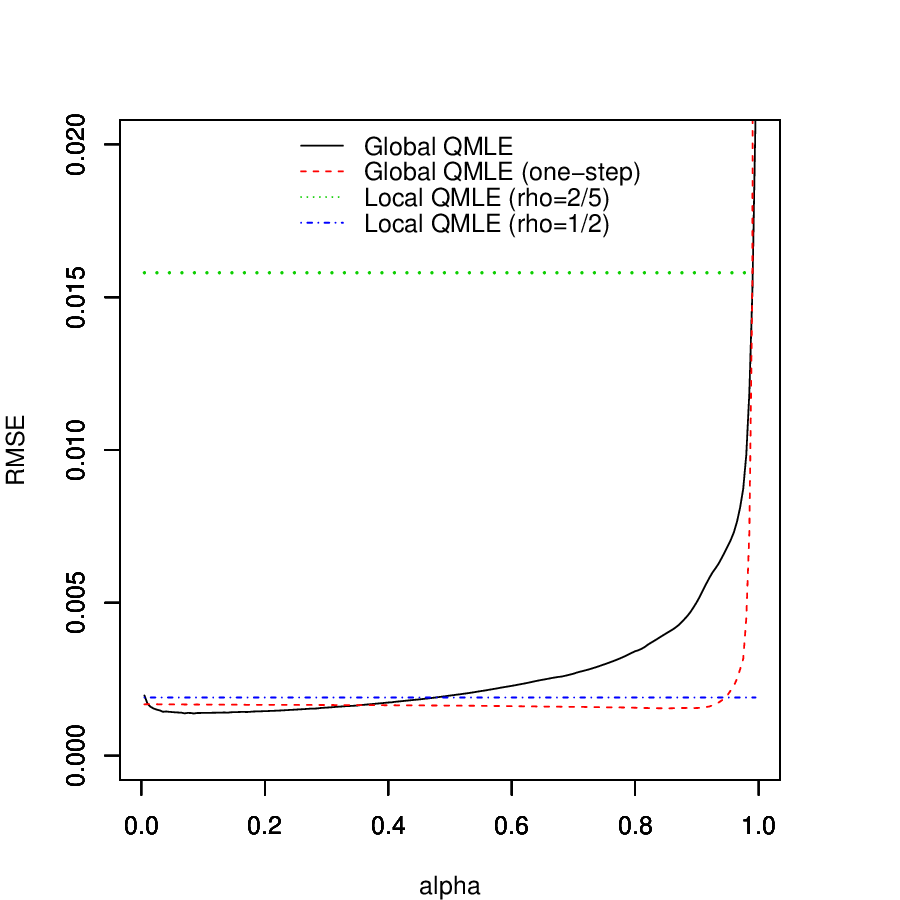}
            {(b) RMSE}
          \end{center}
        \end{minipage}

      \end{tabular}
      \caption{Results of jump detection: in the case of asymmetric jump disribution}
  
  \label{fig8}
  \end{figure}
%

{\rev
\subsection{Location-dependent diffusion coefficient}
Here we assume that the diffusion coefficient is given by $\sigma \sqrt{1+x^2}$, where 
$\sigma$ is an unknown positive parameter to be estimated. 
Other settings are entirely the 
same as those given in the Section \ref{3103031434}. In particular, we assume that 
the distribution of jump size is centered, contrary to the previous subsection. 

{\colred 
In this example, we have to set an estimator $\bars_{n,j-1}$ of the
volatility matrix, $\Big( \sigma \sqrt{1+X_{t_{j-1}^n}^2} \Big)^2$,
which satisfies the condition {\fred[F2]}(ii).
It is obvious that we can choose $\bars_{n,j-1} = 1+X_{t_{j-1}^n}^2$
to satisfy the condition.
}
The results are shown in Figure \ref{fig9}. 
Like in the case of constant coefficient, the global estimators perform well. 
Except for too small or large $\alpha$ for which the estimates are unstable and different from those of 
the case of constant diffusion coefficient, our estimators yield a good 
estimate even in the case of location-dependent diffusion coefficient. 
}

\begin{figure}%
        \begin{tabular}{c}
    
          \begin{minipage}{0.5\hsize}
            \begin{center}
              \includegraphics[keepaspectratio, scale=0.55]{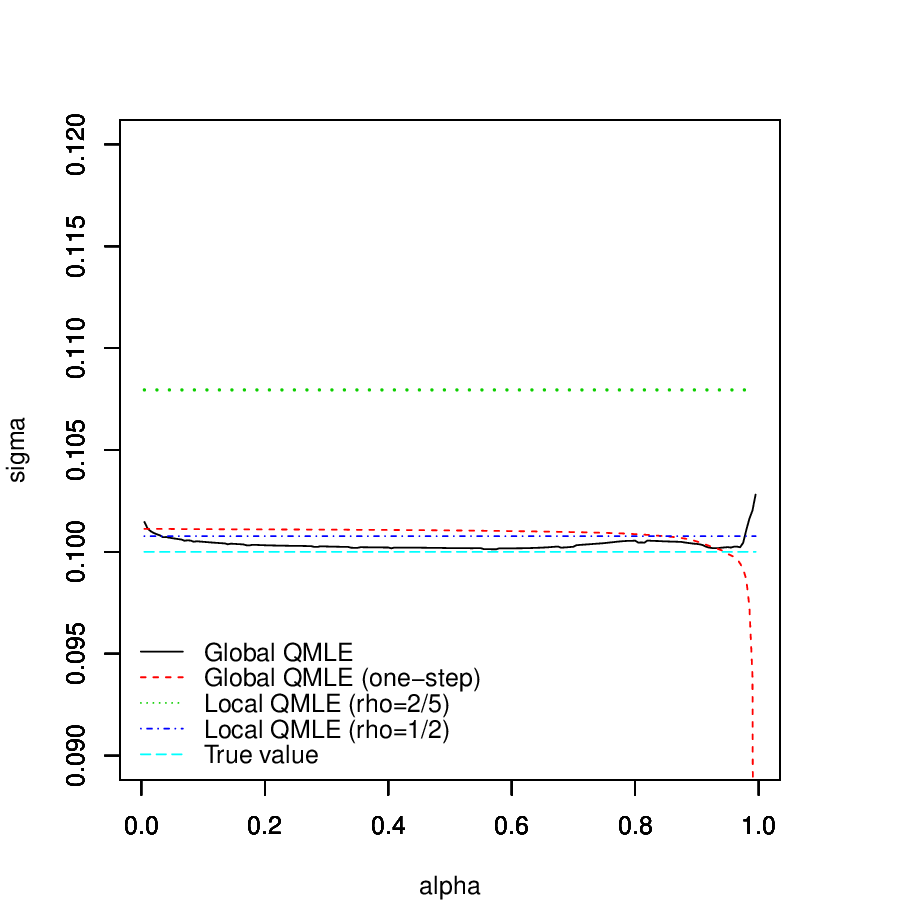}
              {(a) Average estimates}
            \end{center}
          \end{minipage}
    
          \begin{minipage}{0.5\hsize}
            \begin{center}
              \includegraphics[keepaspectratio, scale=0.55]{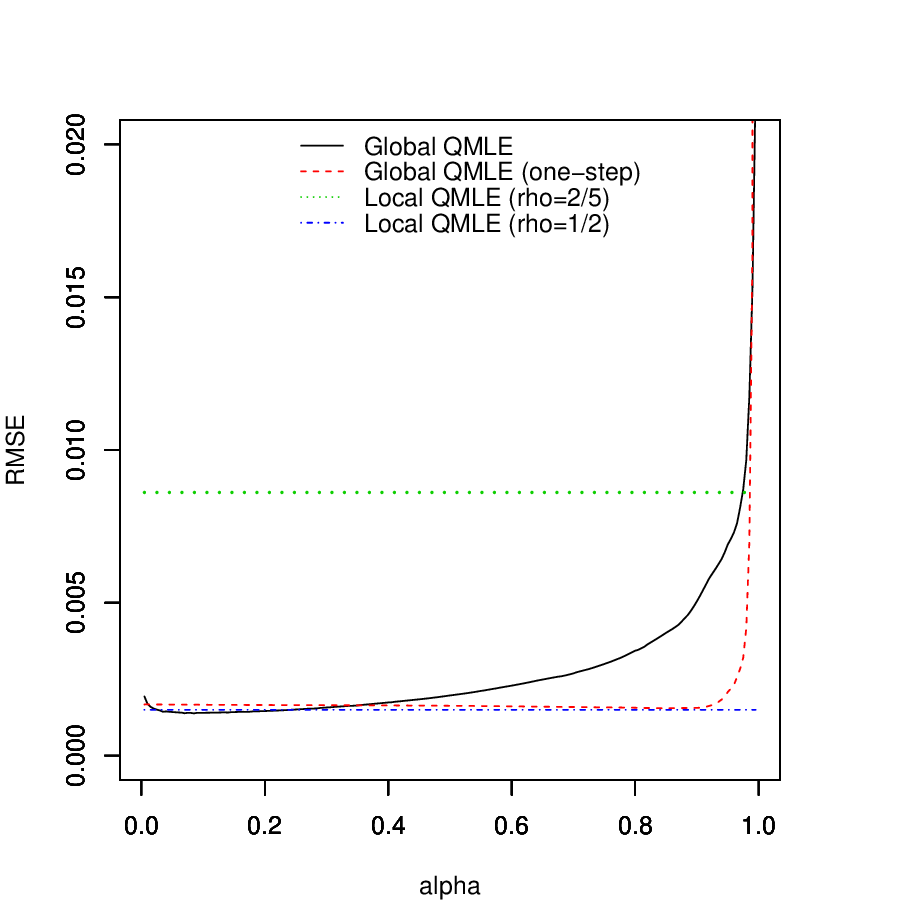}
              {(b) RMSE}  
            \end{center}
          \end{minipage}
  
        \end{tabular}
        \caption{Results of jump detection: a location-dependent diffusion coefficient}
    
    \label{fig9}
    \end{figure}

\section{Further topics and future work} 
{\rev 
In this paper, we payed main attention to removing jumps and to obtaining stable estimation of the diffusion parameter. 
The removed data consist of relatively large Brownian increments and the increments having jumps. 
Then it is possible to apply a suitable testing procedure to the removed data, 
e.g., the goodness-of-fit test for the cut-off normal distribution, in order to test existence of jumps. 

It is also possible to consider asymptotics where the intensity of jumps goes to infinity 
at a moderate rate that does not essentially change the argument of removing jumps. 
In such a situation, estimation of jumps becomes an issue. 
Probably, some central limit theorem holds for the error of the estimators of the structure of jumps. 
Furthermore, a statistical test of the existence of jumps will be possible in this framework. 
The ergodic case as $T\to\infty$ will be another situation where the parameters of jumps are estimable. 

The global filtering methods can apply to the realized volatility to estimate the integrated volatility. 
The superiority of the global filter to the several existing filtering methods 
used in this context is numerically observed as well as a mathematical proof. 
For details, see the forthcoming paper by the authors. 

The global jump filter was motivated by data analysis. 
This scheme is to be implemented on 
YUIMA, a comprehensive R package for statistical inference and simulation for stochastic processes.




\begin{en-text}
{\color{gray}

\koko 
Decompose $\widetilde{\Delta}_n$ according to the decomposition 
\bea\label{300626-12} 
D^{(k)}_j &=& 
h^{-1}\bigg\{\big(\Delta_j\tY^{(k)}\big)^{\otimes2}
-\big(\sigma^{(k)}(X_\tjm,\theta^*)\Delta_jw^{(k)}\big)^{\otimes2}\bigg\}
\nn\\&&
+\bigg\{h^{-1}\big(\sigma^{(k)}(X_\tjm,\theta^*)\Delta_jw^{(k)}\big)^{\otimes2}
-S^{(k)}(X_\tjm,\theta^*)\bigg\}
\eea
and apply Lemma \ref{300225-UYL5} and the Burkholder-Davis-Gundy inequality to obtain 
$L^\inftym$-boundedness of $\widetilde{\Delta}_n$. 
A similar procedure already appeared in the proof of Lemma \ref{300221-A6(1)}.
\qed\halflineskip
}
\end{en-text}

\bibliographystyle{spmpsci}      
\bibliography{bibtex-20180426-20180615}   

\end{document}